\documentclass[letterpaper]{emulateapj}
\usepackage{apjfonts}
\usepackage{lscape} 
\usepackage{amsmath,amsfonts}
\usepackage{graphicx}
\usepackage[dvips]{color}
\usepackage[breaklinks=true]{hyperref}
\usepackage{natbib} 
\bibpunct[,]{(}{)}{;}{a}{}{,} 

\newcommand{\SgrA}{Sgr~A$^\ast$}
\newcommand{\MBH}{\ensuremath{M_{\bullet}}}
\newcommand{\Msun}{\ensuremath{{\rm M}_{\odot}}}
\newcommand{\Mstar}{\ensuremath{m_{\ast}}}
\newcommand{\Nstar}{\ensuremath{N_{\ast}}}
\newcommand{\Rstar}{\ensuremath{r_{\ast}}}
\newcommand{\Rnb}{\ensuremath{R_{\rm NB}}}
\newcommand{\Tnb}{\ensuremath{T_{\rm NB}}}
\newcommand{\Tfp}{\ensuremath{T_{\rm FP}}}
\newcommand{\Npart}{\ensuremath{N_{\rm p}}}
\newcommand{\Mcl}{\ensuremath{M_{\rm cl}}}
\newcommand{\Rsun}{\ensuremath{{\rm R}_{\odot}}}
\newcommand{\Mpccube}{\ensuremath{M_{\odot}\,\mathrm{pc}^{-3}}}
\newcommand{\pc}{\ensuremath{\mathrm{pc}}}

\newcommand{\pccube}{\ensuremath{\mathrm{pc}^{-3}}}
\newcommand{\RS}{\ensuremath{R_{\mathrm{S}}}}
\newcommand{\kms}{\ensuremath{\mathrm{km\,s}^{-1}}}
\newcommand{\peryr}{\ensuremath{\mathrm{yr}^{-1}}}
\newcommand{\Vrel}{\ensuremath{V_{\rm rel}}}
\newcommand{\Vstar}{\ensuremath{V_\ast}}
\newcommand{\trh}{\ensuremath{t_{\rm rh}}}

\newcommand{\tcoll}{\ensuremath{t_{\rm coll}}}

\newcommand{\trlx}{\ensuremath{t_{\rm rlx}}}
\newcommand{\trperi}{\ensuremath{t_{\rm r,p}}}

\newcommand{\Rtd}{\ensuremath{R_{\rm t.d.}}}
\newcommand{\Rplunge}{\ensuremath{R_{\rm plunge}}}
\newcommand{\Rcoll}{\ensuremath{R_{\rm coll}}}

\newcommand{\Rh}{\ensuremath{R_{\rm h}}}
\newcommand{\GCoulomb}{\ensuremath{\gamma_{\rm c}}}

\newcommand{\tLA}{\ensuremath{t_{\rm LA}}}
\newcommand{\Rinfl}{\ensuremath{R_{\rm infl}}}
\newcommand{\Porb}{\ensuremath{P_{\rm orb}}}

\newcommand{\Rperi}{\ensuremath{R_{\rm peri}}}

\newcommand{\Rloss}{\ensuremath{R_{\rm loss}}}
\newcommand{\Rcr}{\ensuremath{R_{\rm cr}}}
\newcommand{\ThLC}{\ensuremath{\theta_{\rm LC}}}
\newcommand{\Mstavrg}{\ensuremath{\langle\Mstar\rangle}}

\newcommand{\MESSY}{\sc ME(SSY)$\ast\ast$2}

\newcommand{\NBODY}{\sc Nbody}
\newcommand{\NBFOUR}{\sc Nbody4}

\newcommand{\tento}[2]{\ensuremath{{#1}\times 10^{#2}}}

\shorttitle{Mass segregation in galactic nuclei}
\shortauthors{Freitag, Amaro-Seoane \& Kalogera}

\begin{document}


\title{Stellar remnants in galactic nuclei: mass segregation.}


\author{Marc Freitag\altaffilmark{1,2}\thanks{e-mail: freitag@ast.cam.ac.uk},
Pau Amaro-Seoane\altaffilmark{3} and
Vassiliki Kalogera\altaffilmark{1}}

\altaffiltext{1}{Department of Physics and Astronomy, 
Northwestern University, Evanston, IL 60208, USA}
\altaffiltext{2}{Institute of Astronomy, University of Cambridge, Madingley Road, CB3~0HA Cambridge, UK}
\altaffiltext{3}{Max Planck Intitut f\"ur Gravitationsphysik (Albert-Einstein-Institut), D-14476 Potsdam, Germany}




\begin{abstract}

The study of how stars distribute themselves around a massive black
hole (MBH) in the center of a galaxy is an important prerequisite for
the understanding of many galactic-center processes. 
These include the observed overabundance of point X-ray
sources at the Galactic center, the prediction of rates and
characteristics of tidal disruptions of
extended stars by the MBH and of inspirals of compact stars into the
MBH, the latter being events of high importance for the future 
space borne gravitational wave interferometer LISA. In relatively small galactic
nuclei, hosting MBHs with masses in the range $10^5-10^7\,\Msun$, the
single most important dynamical process is 2-body relaxation. It
induces the formation of a steep density cusp around the MBH and
strong mass segregation, as more massive stars lose energy to lighter
ones and drift to the central regions. Using a spherical stellar
dynamical Monte-Carlo code, we simulate the long-term relaxational
evolution of galactic nucleus models with a spectrum of stellar
masses. Our focus is the concentration of stellar black holes to the
immediate vicinity of the MBH. We quantify this mass segregation for a
variety of galactic nucleus models and discuss its astrophysical
implications. Special attention is given to models developed 
to match the conditions in the Milky Way nucleus; we examine 
the presence of compact objects in connection to recent high-resolution 
X-ray observations.
\end{abstract}


\section{Introduction}

Massive black holes (MBHs), with masses ranging from a few
$10^4\,\Msun$ to a few $10^9\,\Msun$ are probably present in the
centers of most galaxies. The most compelling line of evidence is
based on measurements of the kinematics of gas and stars in the
central regions of nearby galaxies \citep[e.g.,][ for recent
reviews]{Barth04,Kormendy04,Richstone04,FF05}. The inferred
masses of the central dark objects correlate with different properties
of the host galaxy, probably most tightly and most fundamentally with
the overall velocity dispersion of the spheroidal stellar component of
the galaxy ($M-\sigma$ relation, see
\citealt{Gebhardt00,FM00,TremaineEtAl02,NFD05}). The observational
statistics are dominated by systems in which $\MBH>10^7\,\Msun$
because kinematic detection of such massive objects is easier to
achieve. However, if the $M-\sigma$ relation extends to lower masses, a
possibility supported by observations of low-luminosity active
galactic nuclei \citep{GH04,BGH05}, and most correspondingly small
spheroids harbor MBHs, the nuclear MBHs in the $10^5-10^7\,\Msun$
range, of special interest for the present work, may reach a density
of order $10^{-2}\,{\rm Mpc}^{-3}$ in the local universe
\citep{AR02,ShankarEtAl04}.

Our own Milky Way (MW) is the galaxy for which we have the strongest
observational evidence for the presence of a central
MBH. Spectroscopic and astrometric measurements of the motion of stars
in the vicinity of the radio source {\SgrA} indeed indicate that they
are orbiting a dark mass concentration of some $\tento{3-4}{6}\,\Msun$
whose average density must exceed $\tento{3}{19}\,\Mpccube$. This high
density is incompatible with a stable cluster of any known less
massive astronomical
objects~\citep{GenzelEtAl03,SchoedelEtAl03,GhezEtAl05}, and therefore
the presence of a central black hole of the above mass is well
accepted. For a comprehensive review of the possible interactions
between the {\SgrA} MBH and the surrounding stars and their
observational consequences, see \citet{Alexander05}.

Mass segregation is thought to bring thousands of stellar BHs in the
innermost pc around {\SgrA}
\citep{Morris93,MEG00}. This central overpopulation may have a 
variety of consequences. We note that the compact objects probably
dominate the stellar mass density in a region ($R\lesssim 0.1\,$pc) in
which the ``S'' stars are
confined~\citep{GenzelEtAl03,SchoedelEtAl03,GhezEtAl05}. From infrared
photometry and spectroscopy, these stars appear to be main sequence
(MS) objects with masses of order $4-16\,\Msun$ and therefore younger
than 100~Myr
\citep{GGBLMM02,GhezEtAl03b,EisenhauerEtAl05}. This apparent youth in
a environment where normal stellar formation is made impossible by the
strong tidal forces is an unsolved enigma.

The compact stars may collide and merge with MS stars and giants,
hence creating unusual objects, once suggested to be the S-stars
themselves \citep{Morris93}, or increase the rotation rate of extended
stars through multiple tidal interactions
\citep{AK01}. It has also been proposed that S-stars are young stars, formed at 
$\gtrsim 1$\,pc from the MBH, whose orbital eccentricities were
increased by some perturbation such as that of an (unseen) stellar
cluster, and which were trapped on close orbits around {\SgrA} by
exchanges with less massive compact remnants~\citep{AL04}.

The presence of the stellar BHs around the MBH can in principle be
revealed through different kinds of observations. If one of the
objects acts as a secondary gravitational lens for a distant star
lensed by the central
MBH~\citep{CGME01,AlexanderLoeb01,Alexander03}. Unfortunately,
according to these studies, the rate of such double lensing occurring
at a detectable level is very low.

If the motions of the S-stars can be tracked with high enough a
precision, an extended distribution of non-luminous matter around the
Galactic MBH should signal itself through its effect on their
orbits. In a slightly non-Keplerian potential, the orbits are affected
by Newtonian retrograde precession \citep{RE01,WMG05}. Present-day
observations are insufficient to detect this effect
\citep{MouawadEtAl05}, but \citet{WMG05} have shown that future 
``extremely large telescopes'' (ELTs) with diameters of 30\,m or more
will likely be able to measure the mass and shape of a dark density
cusp of the form $\rho
\propto R^{-\gamma}$, if it tallies at least $\sim 2000\,\Msun$ within
$10^{-2}\,\pc$ of {\SgrA} and has $\gamma\le 2$. This effect is only
sensitive to the overall $\rho(R)$; it does not distinguish
between a population of stellar BHs and another type of non-luminous
component such as a cold dark matter, although the latter
probably contributes much less than 10\,\% of the density inside the 
innermost pc of the Galaxy
\citep[e.g.,][]{GP04,BM05b,BM05a}. However, if a concentration of 
$\sim 10\,\Msun$ BHs is indeed present, ELT observations should allow 
us to witness 2-body relaxation at work by the detection of $\sim 3$
gravitational encounters per year between any of $\sim 100$
monitored S-stars and a stellar BH \citep{WMG05}.

Radio pulsars on similar short-period orbits would allow the same kind
of measurements with a very high accuracy as well as precise tests of
the theory of general relativity
\citep{CordesEtAl04,KramerEtAl04,PL04}. Based on the semi-analytical
work of \citep{MEG00}, \citet{CG02} have suggested that stellar BHs,
by concentrating around {\SgrA} will push out lighter objects,
possibly creating a central dip in their density profile and have
pointed out that pulsars would be the ideal probes to detect this
effect if the sky position of some 50 of them within a few arcsec of
{\SgrA} can be obtained. Unfortunately, because of extreme dispersion
suffered by radio signals traveling from the Galactic center, the
detection of pulsars in this region will probably require future radio
telescopes with high sensitivity at frequencies $\gtrsim 10\,$GHz,
such as the SKA\footnote{Square Kilometer Array,
\texttt{http://www.skatelescope.org}} \citep{CL97,CordesEtAl04,KramerEtAl04}.

Nevertheless, relatively direct evidence for the presence of an
abundant population of stellar BHs around {\SgrA} may not need to
await next-generation telescopes. Recently, observations with the
Chandra X-ray satellite have revealed 7 transient sources within
23\,pc of projected distance of {\SgrA} \citep{MunoEtAl05}; 4 of them
have projected distances smaller than 1\,pc, indicative of an
overabundance in this central region by a factor $\sim 20$ when
normalized to the total enclosed mass at 1 and
23\,pc~\citep{LZM02}. These sources are believed to be X-ray binaries,
i.e., compact objects accreting from a binary companion; however
current observations do not shed light on whether these are neutron
stars or black holes with low or high mass companions. However, for one
case, there is strong evidence for a low-mass ($< 1\,\Msun$) donor
and some preference for a BH accretor
\citep{BowerEtAl05,MunoEtAl05b,PorquetEtAl05}.

In the present study we undertake a careful numerical model
exploration of the distribution of compact objects in galactic
centers. Our goal is to assess the importance and detectability of
these various effects of mass segregation in the context of existing
observations. These models are obtained by explicit integration of the
long-term stellar dynamical evolution of spherical nucleus models with
account for self-gravity, 2--body relaxation, interactions between
stars and the MBH, and, in some cases, additional physics such as
large-angle scatterings, collisions or stellar evolution. The models
presented here constitute a noticeable improvement over the very few
simple estimates of mass segregation available in the
literature~\citep{Morris93,MEG00}.

``Extreme-mass-ratio events'' (EMREs) in galactic nuclei is our other
key motivation. EMREs are events where a stellar object
interacts strongly with an MBH. The best studied case so far (first
considered by \citet{Hills75}) is that of an extended star (main
sequence or giant) coming so close to the MBH that it is partially or
totally disrupted by the intense tidal forces. The hydrodynamical and
stellar dynamical aspects of such tidal disruptions have been the
object of scores of articles (see references at\\
\texttt{http://obswww.unige.ch/$\sim$freitag/MODEST\_WG4/TidalDisrupt.html} for
the former aspect and \citealt{FR76,Rees88,MT99,SU99,FB02b,WM04},
amongst others, for the latter). Although our models also include a
simple treatment of tidal disruptions and yield rates for these
events, we are more specifically interested in another class of EMRE,
namely the coalescence between a compact star and the
MBH. ``Coalescences'' as defined here include ``plunges'' when a star
suddenly finds itself on a radial, relativistically unstable orbit and
disappears through the MBH horizon at its next periapse passage, and
``inspirals'' (EMRIs) during which the orbit of the stellar object
progressively shrinks by emission of gravitational waves (GWs) until
it plunges.

EMRIs will be of prime interest for LISA (``Laser Interferometer Space
Antenna'', see \citealt{Danzmann96,Danzmann00} and
\texttt{http://lisa.jpl.nasa.gov/}), the future space borne mission to
detect GWs with frequencies in the range $\sim 10^{-4}-0.1$\,Hz. The
waves emitted during the last year of inspiral, as the stellar object
orbits in the deep gravitational field of the MBH, if detected and
analyzed successfully, will inform us about the geometry of the space
time in the immediate vicinity of the massive object, thus allowing to
probe general relativity in the strong field regime, to establish the
existence of MBHs and measure with high accuracy their masses and
spins \citep{Ryan95,Ryan97,Thorne98,Hughes03}.

Predictions of EMRE rates and properties (especially the mass of
stellar object and the orbital eccentricity when the signal starts
contributing to the LISA stream) are important for the design and of
LISA and the development of data analysis tools required to extract
weak EMRE signals from a data stream containing noise and a large
number of other astrophysical sources
\citep{Phinney02,Prince02,GairEtAl04}. For the GW signal to be in the
frequency range of optimal LISA sensitivity during the last year of
inspiral (when wave amplitude and the interesting strong-field effects
are the strongest), the MBH mass must be in the range
$\MBH\simeq10^5-10^7\,\Msun$. In what follows we argue that such MBHs
are likely to inhabit stellar spheroids in which relaxation time is
relatively short causing mass segregation close to the MBH. This is of
great importance for EMREs as the inspiral a stellar BH, with a mass
$\simeq 10\,\Msun$ can be detected in galaxies $\sim 10$ times more
distant (and therefore $\sim 10^3$ more numerous) than that of a $\sim
1\,\Msun$ object.

Determining rates and characteristics of EMREs for LISA is beyond the
scope of this paper. A few estimates for those exist in the
literature, based on the same stellar dynamical code as used here
\citep{Freitag01,Freitag03} or on other semi-analytical or numerical methods 
\citep{HB95,SR97,MEG00,Ivanov02,HA05}. The results from 
these studies are scattered over a disquieting large range,
approximately from $5\times 10^{-9}\peryr$ \citep{HA05} to
$10^{-6}\peryr$ \citep{Freitag01}, for EMRIs of stellar BHs in a
MW-like nucleus (see \citealt{Sigurdsson03} for a brief discussion of
these various studies and EMRIs in general). This is witness, in part,
to the lack of realistic agreed-on models for the structure of
galactic nuclei, causing different authors to adopt different
approximations and values to describe the stellar distribution around
a MBH. With this study, we strive to improve this situation.

Another cause for the disagreement found among EMRI studies is the
poor understanding of the mechanisms responsible for EMRIs. All
studies have assumed unperturbed spherical galactic nuclei in
dynamical equilibrium, in which case 2-body relaxation is certainly
the main agent for bringing stars onto very elongated orbits which
may result in EMRIs. At the same time, as already pointed out in the
pioneering work of \citet{HB95} and analyzed in detail by
\citet{HA05}, encounters with other stars may cause an inspiralling
star to plunge prematurely, before its orbital frequency has entered
the LISA band. Although GWs emitted during the plunge itself will
contain some high-frequency components, it is unlikely to be
detectable by LISA as a resolved source because, typically, tens to
hundreds of thousands of cycles are required to accumulate enough
signal-to-noise ratio \citep{BC04b,BC04,GW05,WG05}. Hence, for LISA, the
problem is not limited to the determination of the rate of
coalescences, $\dot{N}_{\rm coal}$; one also needs to compute the
fraction of those, $f_{\rm EMRI}
\equiv \dot{N_{\rm EMRI}}/\dot{N}_{\rm coal}$, that are ``clean''
inspirals instead of plunges. However, here, we limit ourselves to
$\dot{N}_{\rm coal}$, the quantity for which the type of simulations
we carry out yield robust predictions. We think that a real
trustworthy estimate of $f_{\rm EMRI}$ can only be arrived at through
the use of novel methods, to be developed in the future (see
\S~\ref{sec:concl}).

The rest of this paper is organized as follows. In \S~\ref{sec:review},
a quick review of the relevant aspects of stellar dynamics in
galactic nuclei and the previous relevant work is presented. In
\S~\ref{sec:method}, we describe the numerical method used in our simulations, 
as well as the physics and initial conditions implemented. Our main
results from our $\simeq 80$ simulations are described in
\S~\ref{sec:results} and we conclude in \S~\ref{sec:concl} with a 
discussion of the astrophysical implications of our results and an
outlook for future work.

\section{Review of the Theory and Previous Studies}
\label{sec:review}

\subsection{Collisional Dynamics in Galactic Nuclei}
\label{subsec:colldyn}

In stellar dynamics, the term ``collisional'' refers to all situations
in which the discrete nature of stars, i.e.\ the fact that a stellar
system is not composed of a continuous fluid but of individual
objects, plays a role. In the context of galactic nuclei, these
effects include 2-body relaxation, direct (hydrodynamical) collisions
between stars and close, dissipative interactions between stars and a
central MBH. 

In this work, we restrict ourselves to the situation of isolated,
spherical systems in dynamical equilibrium. These assumptions are
made necessary by the numerical method we use (the Monte-Carlo code,
see \S~\ref{sec:method}) which is still, at the present day, the only
stellar dynamical scheme able to treat the collisional evolution of
systems consisting of more than one million stars with acceptable
realism. They are also adopted in almost all other works on the
subject because they introduce a well-defined theoretical framework,
allowing in particular to make use of methods developed for the study
of globular clusters. Here we use the term ``cluster'' for any
collisional stellar system, including galactic nuclei. Collisional
dynamics, generally with a focus on globular or smaller clusters is
covered by several textbooks \citep{BT87,Spitzer87,HH03}; therefore we
only recall here the few concepts needed to understand the rest of the
paper. More detailed explanations about collisional dynamics in the
context of Monte-Carlo simulations can be found in our previous papers
\citep{FB01a,FB02b,FRB05}.

Barring the effects of mass loss due to stellar evolution, in a
stationary smooth, spherical potential, stellar orbits would be
energy- and angular-momentum-conserving rosettes of fixed shape and
the cluster structure would show no secular evolution. But the
potential is the sum of the contribution of a finite number of stars
(and a MBH) and is affected by short-scale and short-time fluctuations
which causes the orbital parameters to slowly change. In effect, stars
are exchanging energy and angular momentum with one another and, to a
very good approximation this ``relaxation'' can be idealized as due to
the sum of a large number of uncorrelated 2-body encounters leading to
small deflection angles. This is the base for the Chandrasekhar theory
of relaxation \citep{Chandrasekhar60} on which the Fokker-Planck
equation and other approximate treatment of collisional dynamics are
based \citep{BT87,Henon73}.

In this picture, one can define a local relaxation time
\begin{eqnarray}
\trlx =& \frac{\sqrt{2}\pi}{64}
\frac{\sigma^3}{\ln\Lambda \,n G^2 \Mstavrg^2} = \tento{3.67}{8}\,{\rm yr}\,\times\\
& \left(\frac{\ln\Lambda}{10}\right)^{-1} 
\left(\frac{\sigma}{100\,\kms}\right)^3 
\left(\frac{n}{10^6\,\pccube}\right)^{-1}
\left(\frac{\Mstavrg}{\Msun}\right)^{-2} \nonumber
\label{eq:trlx}
\end{eqnarray}
where $\ln\Lambda$ is the Coulomb logarithm, $\sigma$ is the 1D
velocity dispersion, $n$ the number density of stars and $\Mstavrg$
the average stellar mass. The slightly unusual numerical coefficient
is devised such that a particle of mass $\Mstavrg$ traveling for a
time $\delta t$ through a field of particles of same mass at a
relative velocity $v_{\rm rel} =
\sqrt{2}\sigma$ would have its trajectory deflected by an angle
$\delta \theta$, with $\langle \delta \theta^2 \rangle = (\pi/2)^2
\delta t/\trlx$.

The argument of the Coulomb logarithm is $\Lambda = b_{\rm max}/b_0$
where $b_0=G\Mstavrg/\sigma^2$ is the typical impact parameter leading
to a deflection angle of $\pi/2$ in gravitational encounters between
stars. In a virialized, self-gravitating system, $b_{\rm max}$ is of
order the half-mass radius $\Rh$ and $\Lambda = \GCoulomb\Nstar$ with
$\GCoulomb\approx 0.01$ if stars have a mass spectrum \citep[][ and
references therein]{FRB05}. In the region where the gravitational
force is dominated by the central object,
one finds $\Lambda \approx
\MBH/\Mstavrg$ \citep[e.g.,][]{BW76}. In practice, this does not 
lead to an important difference, thanks to the damping effect of the
logarithm. For instance, one finds $\ln (\GCoulomb\Nstar) \simeq 15$
for $\Nstar = \tento{3}{8}$ and $\ln (\MBH/\Mstavrg) \simeq 11.5$ for
$\MBH=10^5\,\Msun$ and $\Mstavrg=1\,\Msun$.  Therefore, in most
studies, including the present one, a fixed value of $\Lambda$
($=\GCoulomb\Nstar$) is adopted. Comparisons with direct $N-$body
integrations, presented in \S~\ref{sec:tests}, as well as with a
version of the Monte-Carlo code in which $\Lambda$ varies with the
distance to the center, from $\MBH/\Mstavrg$ to $\GCoulomb\Nstar$
\citep{TheseFreitag}, confirm the validity of this approximation. We
set $\GCoulomb=0.01$.

The MBH dominates the gravitational force acting on stars within an
``influence sphere'' with a radius of order $G\MBH\sigma_0^{-2}$ where
$\sigma_0$ is the stellar velocity dispersion at larger distances (a
more practical definition is given in \S~\ref{sec:ini_nucl_models} for
the category of galactic nucleus models considered in our
simulations). In this central region, the velocity dispersion is
Keplerian, $\sigma(R) \simeq G\MBH/R$. If the stars are distributed
according to a a power-law density profile, $n\propto R^{-\gamma}$,
the relaxation time gets shorter closer to the MBH when $\gamma>1.5$
and longer if $\gamma<1.5$.

In what follows, we call ``collision'' the event in which two stars
actually come so close to each other as to touch. Neglecting
deformations due to mutual tidal interactions, a collision between
stars of radii $r_1$ and $r_2$ corresponds to their centers coming
within a distance $r_1+r_2$ of each other. The cross section for this
process is
\begin{equation}
S_{\rm coll}^{(1,2)} = \pi (r_1+r_2)^2\left[1+\frac{2G(m_1+m_2)}{(r_1+r_2)\Vrel^2}\right]
\end{equation}
where $m_{1,2}$ are the stellar masses and $\Vrel$ their relative
velocity at a large separation. If field-stars of type ``2'' have
number density $n_2$, and all stars of this type have the same velocity, the average
time for test-star ``1'' to collide with one of type ``2'' is
\begin{equation}
\tcoll^{(1,2)} = \left(n_2 S_{\rm coll}^{(1,2)}\Vrel\right)^{-1}
\label{eq:tcoll}
\end{equation}
To estimate the importance of collisions in the dynamics, we assume
all stars have the same mass and radius, $\Mstar$, $\Rstar$ and their
velocity distribution is Maxwellian with dispersion $\sigma$. The
collision time is then \citep{BT87}
\begin{eqnarray}
\tcoll =& \left[16\sqrt{\pi}n\sigma\Rstar^2 
\left(1+\frac{G\Mstar}{2\sigma^2\Rstar}\right)\right]^{-1} 
\simeq \tento{8.9}{10}\,{\rm yr}\,\times \nonumber \\
 & \left(\frac{n}{10^6\,\pccube}\right) ^{-1} 
\left(\frac{\sigma}{100\,\kms}\right)
\left(\frac{\Rstar}{\Rsun}\right)^{-1}
\left(\frac{\Mstar}{\Msun}\right).
\end{eqnarray}
The numerical relation is valid when the velocities are much smaller
than the stellar escape velocity, $\sigma \ll \Vstar =
(2G\Mstar/\Rstar)^{1/2} = 617.5\,\kms\,
(\Mstar/\Msun)^{1/2}(\Rstar/\Rsun)^{-1/2}$ so that the cross section
is dominated by gravitational focusing. This ceases to be
applicable at distances from the MBH smaller than
\begin{equation}
\Rcoll = \Rstar\frac{\MBH}{\Mstar} \simeq 
\tento{2.3}{-2}\,\pc \left(\frac{\MBH}{10^6\Msun}\right) 
\left(\frac{\Mstar}{\Msun}\right)^{-1}.
\end{equation}
For $R < \Rcoll$, $\sigma > \Vstar$, so the collision time reduces
to (note the different normalization for $n$ and $\sigma$)
\begin{eqnarray}
\tcoll \simeq & \tento{6.7}{10}\,{\rm yr}\,\times \\
 & \left(\frac{n}{10^7\,\pccube}\right)^{-1} 
\left(\frac{\sigma}{10^3\,\kms}\right)^{-1}
\left(\frac{\Rstar}{\Rsun}\right)^{-2}. \nonumber
\end{eqnarray}
The condition for the collision time to become shorter than the
relaxation time is also $\sigma>\Vstar$, for $\ln\Lambda\approx
10-20$. Collisions at such velocities are unlikely to lead to mergers;
a fly-by with partial mass loss is the most likely outcome
\citep{FB05}. Only within $\Rcoll$ can collisions noticeably affect
the density profile
\citep{FR76,SR97}. However, hydrodynamical simulations of collisions 
between MS stars show that complete stellar disruptions require
$\sigma \gtrsim 5\Vstar$ and nearly head-on geometry
\citep{BH87,BH92,LRS93,FB00c,FB05}. Disruptions are therefore rare and
the effect of collisions on the stellar distribution is weak, even for
$R<\Rcoll$~\citep{FB02b}.

Gravitational encounters with an impact parameter smaller than a few
$b_0$ lead to deflection angles that are relatively large and cannot
be accounted for in the standard, ``diffusive'' theory of
relaxation. Therefore in most approaches, both analytical and
numerical, these large-angle scatterings have to be considered as a
separate process. We call them ``large-angle scatterings'' and reserve
the word ``relaxation'' to the effect of 2-body encounters with larger
impact parameters. On average, a star will experience an encounter
with impact parameter (with $f_{\rm LA}$ of order a few) over a
timescale
\begin{equation}
  \tLA \simeq \left[\pi(f_{\rm LA}b_0)^2n\sigma\right]^{-1} 
   \approx \frac{\ln\Lambda}{f_{\rm LA}^2}\trlx.
\end{equation}
The effects of large-angle scatterings on the overall evolution of a
cluster are negligible in comparison with ``diffusive'' relaxation
\citep{Henon75,Goodman83}. However, unlike the latter process,
they can produce velocity changes strong enough to eject stars from an
isolated cluster \citep{Henon60b,Henon69,Goodman83} or, more
importantly, from the ``cusp'' around the central
BH~\citep{LT80,BME04a}. Therefore they may be important for the
dynamics of the innermost regions just where mass segregation is
relevant too.

A central MBH represents a sink for the stellar system as it destroys,
captures or --if it forms a very compact binary with another object--
ejects stars that venture very close to it, i.e., within some distance
$\Rloss$. In particular, tidal disruption, for a star of
radius $\Rstar$ and mass $\Mstar$, occurs at $\Rloss=\Rtd
\simeq 1.25\,\Rstar (\MBH/\Mstar)^{1/3}$ \citep[][ and
references therein]{FB02b}. A quasi parabolic orbit whose Newtonian
periapse distance would be smaller than $\Rloss=\Rplunge = 8G\MBH
c^{-2}$ actually plunges directly through the horizon \citep{ZN99}. Orbits with
periapse distance $<\Rloss$, corresponding to angular momentum (per
unit mass) $J< J_{\rm LC}\simeq \sqrt{2G\MBH\Rloss}$, form the ``loss
cone''. For a star with velocity $v$ at distance $R$ from the center,
the loss cone has an aperture angle $\ThLC=J_{\rm LC}/(Rv)$.

If the star is removed from the cluster in a single close
encounter with the MBH, a mature loss cone theory has been developed
which predicts rates and orbital characteristics of such events
\citep{FR76,BW77,LS77,CK78,ASFS04}. The notion of critical radius ($\Rcr$) is central 
in these cases; it is basically the semi-major axis of an orbit for
which the relaxation processes cause a change of angular momentum per
orbital time of order $J_{\rm LC}$. Inside $\Rcr$ loss cone orbits are
nearly completely depleted (``empty loss cone'' regime). On the scale
of the loss cone, the change of orbital parameters due to relaxation
can be treated as a diffusion process and a direct analogy with the
heat equation can be used to obtain the average time for a star to be
destroyed, $t_{\rm destr,e}\simeq \ln(\ThLC^{-2})\trlx$. At distances
larger than $\Rcr$, relaxation is efficient enough to bring stars into
and out of the loss cone over an orbital time $\Porb$. The loss cone
is therefore full and $t_{\rm destr,f}\simeq
\ThLC^{-2}\Porb$. The total rate of interactions with the MBH is given 
by $\Gamma=4\pi\int R^2 n t_{\rm destr}^{-1} dR$. It peaks
around $\Rcr$ for many density profiles $n(R)$.

In cases, such as non-destructive tidal interactions and GW emission,
in which the star looses energy gradually and is only destroyed after
a large number of periapse passages, the interplay between
relaxation and dissipative processes is not directly amenable
to the relatively simple loss cone formalism. The detailed analysis of
such situations has only recently been pioneered
\citep{AH03,HA05}.

In the sphere of influence of the MBH, orbits of bound stars are
essentially ellipses precessing on a time scale of order
$(\MBH/M_{\ast,orb})\Porb\gg \Porb$ where $\MBH$ is the mass of the
central object, $M_{\ast,\rm orb}$ the mass in stars within the
apocenter distance of the orbit and $\Porb$ the orbital period. On
shorter timescales, orbits exert torques on each other, thus
introducing so-called ``resonant relaxation'' which affects the
angular momentum on a time scale $t_{\rm res}\approx
(\MBH/\Mstar)\Porb
\approx \ln\Lambda^{-1}(\MBH/M_{\ast,orb})\trlx$ \citep{RT96}. Resonant relaxation 
is suppressed by relativistic precession for very close-by orbits satisfying $\Rperi/\RS <
\MBH/M_{\ast,orb}$ with $\Rperi$ the periapse distance and 
$\RS=2G\MBH/c^2$ the Schwarzschild radius of the central BH. Although
resonant relaxation may be much faster than ``normal'' relaxation in
the sphere of influence of the MBH, it was shown to have only a
moderate impact on the rate of tidal disruptions in galactic nuclei,
because these events are dominated by stars with semi-major axis of
order the critical radius (see \S~\ref{sec:ini_nucl_models}) and
$\MBH/M_{\ast,orb}\ll 10$ \citep{RT96,RI98}. On the other hand, the
effects on the coalescence of compact objects is likely to be weak due
to relativistic precession. In any case, the study of this question
requires a method that can account for non-local gravitational
interactions between orbits (``2-orbit'' effects) and is not
undertaken here.

\subsection{Single-Mass Clusters with a Central Object}

The question of how relaxation will shape the distribution of a large
number of point-like objects of the same mass orbiting a massive
object has been addressed in the 70's, shortly after the detection of
X-ray sources in globular clusters triggered the hypothesis that there
may be IMBHs at their center
\citep{Peebles72,BW76,SL76,LS77,CK78}. The 
approximate solution, first found by \citet{BW76} is this context
through a Fokker-Planck-type treatment of the stellar dynamics, is the
formation of a power-law density, $n(R) \propto R^{-7/4}$. In this
simplified treatment, stars are only destroyed if they reach a very high
binding energy (typically $E_{\rm loss}\approx G\MBH/\Rtd$ for tidal
disruptions).
As it neglects the disruption of stars on (very) elongated orbits 
($J<J_{\rm LC}$), this idealized
configuration corresponds to an isotropic distribution with a zero net
diffusive flux of stars in $E$ space and a constant outward energy
flux\footnote{Treating the cluster as a conducting gas, the
Bahcall-Wolf solution can be found by imposing $dF/dR=0$ with
$F=-4\pi R^2\cdot\kappa(d\sigma^2/dR)$ the rate of ``thermal'' energy 
conducted across a sphere of radius R. $\kappa$ is the thermal
conductivity, $\kappa=\rho \lambda^2/\tau$ with $\rho$ the mass density, 
$\lambda\approx R$ the
effective mean free path and $\tau \approx \trlx$ the timescale for energy
exchange.}. More detailed Fokker-Planck treatment accounting for
loss-cone effects and other realistic bounding conditions confirmed
the Bahcall-Wolf cusp as a very good approximation
\citep{BW77,LS77,CK78}. It has since be found with other methods also 
based on the diffusive, local theory of relaxation: two types of Monte
Carlo codes (\citealt{Shapiro85} and references therein; \citealt{FB02b}) 
and a gas-dynamical
approach \citep{ASFS04}. Very recently, the approximations involved
in these computations have been vindicated by direct $N-$body
simulations in which the formation of the $R^{-7/4}$ profile over a
relaxation time was indeed witnessed \citep{BME04a,PMS04}.

Using a homological model for the evolution of a cluster,
\citet{Shapiro77} showed how a central BH can power the expansion of 
the stellar system, by destroying stars that have diffused deep into
the cusp. A central BH is therefore able to drive gravothermal
expansion in a way similar to hardening binaries, but without leading
to core oscillations \citep{HH03}. The central BH acts as a heat
source for the whole cluster only if, on average, destroyed stars have
negative orbital energies relative to the BH, a condition roughly
equivalent to $\Rcr<\Rinfl$ \citep{DS83}.

\subsection{Multi-Mass Clusters with a Central Object}

Surprisingly, the effects of relaxation in a multi-mass cluster
containing a central massive object have been little studied. To our
knowledge, the only in-depth theoretical study of mass segregation in
the Keplerian potential of a MBH is the work of
\citet{BW77}. The long-term evolution of a few models of MBH-hosting 
galactic nuclei with a mass spectrum was followed numerically using a 
Fokker-Planck code by
\citet{MCD91} and with the same Monte-Carlo code as the present study 
by \citet{FB02b} and \citet{Freitag03}. However, in those studies, 
rich physics was included which complicates the interpretation of the
results (collisions, stellar evolution,\ldots) and their authors did
not present detailed results concerning mass segregation. Also, with
the exception of \citet{Freitag03}, the initial conditions used were
not tailored to represent any specific galactic nucleus. Recently,
\citep{BME04b} carried out direct $N-$body simulations of multi-mass
clusters with some $\tento{1.6}{4}$ to $\tento{1.3}{5}$ stars hosting
a central IMBH and discussed how stars of different masses distribute
themselves around the central object. This study offers the only direct
characterization of mass segregation around a massive object. One
should be cautious, however when trying to apply these $N-$body
results to larger systems such as galactic nuclei because small$-N$
effects (large-angle scatterings, binary interactions, IMBH
wandering,\ldots) may play a significant role there \citep{LT80}.

A first step towards the understanding of mass segregation in
galactic nuclei is to consider the simpler problem of the evolution of
one or a few massive ``tracers'' in a non-evolving stellar
background. We undertake this step here for illustrative purposes.
This is a useful idealization for the early dynamical
evolution of the population of stellar BHs. Those are very rare
objects so, until they have concentrated in the innermost regions,
they will mostly interact with other stars and not with one another. We
assume all stellar BHs have mass $m_{\rm BH}$ and all other stars have
mass $m$ with $q=m_{\rm BH}/m\gg 1$; $q=30$ is a realistic value. The
effects of 2-body relaxation on the orbit of a massive particle
(``test particle'') in a field of much lighter field particles is
embodied in the classical {\it dynamical friction} (DF) formula
\citep[see][ \S~7.1]{BT87}
\begin{equation}
\vec{a}_{\rm DF} = -t_{\rm DF}^{-1}\,\vec{v}
=-\frac{4\pi\ln\Lambda G^2nm(m+m_{\rm BH})}{v^3}\kappa(X)\,\vec{v}
\label{eq:dynfric1}
\end{equation}
with $\kappa(X)={\rm erf}(X)-2\pi^{-1/2}X e^{-X^2}$ and
$X=v/(\sqrt{2}\sigma)$. In this formula, $\vec{a}_{\rm DF}$ is the
force per unit mass on the test particle due to DF, $\vec{v}$ its
velocity, $n$ the density of field particles, and $\sigma$ the (1D)
dispersion of their velocities, assumed to have a Maxwellian
distribution. $t_{\rm DF}$ is of order the local relaxation time
divided by $1+q$ so the massive particles should already experience
significant mass segregation after a small fraction of the relaxation time.

For an object on a circular orbit of radius $R$,
$v=v_{\rm c}\equiv\sqrt{GM_{\rm encl}(R)/R}$ where $M_{\rm encl}(R)$ is
the total mass within $R$, and a differential equation for the
evolution of $R$ is easily derived from ${a}_{\rm DF}R=d(Rv)/dt$,
\begin{equation}
\frac{dR}{dt} = -\left(\frac{2\pi G R^2n(R)m}{v_{\rm c}(R)^2}
+0.5\right)^{-1}\frac{R}{t_{\rm DF}(R)}.
\label{eq:dynfric2}
\end{equation}

Although it can yield a qualitative understanding and a first
approximation to the development of mass segregation, a treatment
based on the use of Eq.~\ref{eq:dynfric2} falls short of physical
realism. First, relaxation reduces to dynamical friction only in the
limit of very large mass ratio. In general, the {\it direction} of
$\vec{v}$ (and not only its modulus) is also affected by 2-body
encounters, causing the eccentricity of a circular orbit to drift away
from zero. Second, if massive objects are numerous enough, they will
eventually come to dominate the central region. There, they will push
the lighter objects away by heating them and start interacting with
each other in a way more similar to the single-mass situation. The
dynamical friction picture does not provide a way to determine the
quasi-stationary distribution the particles of different masses will
adopt on the long term.

In another seminal paper, \citet{BW77} studied the possibility for a
multimass system dominated by the potential of a central MBH to settle
into a relaxational steady-state configuration (a cusp), provided
stars lost to interactions with the MBH are replaced by stars coming
from more distant regions. By solving the coupled Boltzmann equations
for stars of various masses, they found that the stars of different
masses, $m_i$, should approximately follow one-particle distribution
functions that are power law of the binding energy,
$f_i(\vec{x},\vec{v}) = F_i(E)
\propto E^{p_i}$, with indices scaling like
\begin{equation}
\frac{p_i}{m_i} = \frac{p_j}{m_j}.
\label{eq:BW77}
\end{equation}
These correspond to density profiles $n_i\propto R^{-\gamma}$ with
$\gamma=\frac{3}{2}+p$. They found $p\simeq0.30$ for the most massive
objects, who dominate the central density, close to the value for a
single-mass distribution, $p\simeq 0.25$. For much lighter objects in
the innermost regions, $\gamma\simeq 1.5$ is expected (see also
\citealt{Merritt04b}).

It is interesting to note that the massive stars concentrate to the
center because they lose energy to lighter ones during 2-body
encounters. This tendency would yield statistical equipartition of
kinetic energy if it wasn't for the overall gravitational potential in
which the heavy objects sink, thus increasing their velocities. In a
cluster without a central black hole, equipartition can only be
reached at the center and only if the massive particles are in small
number or have a mass not much exceeding that of the lighter ones, so
that they cannot form a self-gravitating system, with negative heat
capacity, on their own
\citep{Spitzer69,Vishniac78,IS85,WJR00,GFR04,KASS05}. For all
realistic mass spectrum, mass segregation will trigger the core
collapse of the sub-system of massive bodies, a process known as
``Spitzer instability''. 

Clearly, in a fixed Keplerian potential, massive stars can never reach
equipartition with lighter ones; as they concentrate to the center,
their velocity dispersion must increase and the thermal imbalance with
the lighter objects is maintained. An accelerated, catastrophic
collapse of the population of massive objects is prevented, however,
by the heating effect of the central MBH which eventually compensates
for the energy lost to the light stars. Hence, a cusp of massive
objects is expected to form and maintain itself in thermal quasi
equilibrium while it drives the expansion of the distribution of
lighter objects.


Published simulations of multi-mass clusters with a central (I)MBH are
few and far apart. The work of \citet{MCD91} stands out as a
pioneering effort to follow the evolution of galactic nuclei taking into
account relaxation, stellar evolution and collisions. These authors
have published limited data from one run without stellar evolution or
collisions. They report a good agreement with the prediction of
\citet{BW77} relative to the cusp exponents for stars of different 
masses (Eq.~\ref{eq:BW77}). From their Figure~9, however, it seems
that the region for which this applies encompasses of order
$10^4\,\Msun$ only, at a time when, judging from their case 4C, the
MBH has certainly grown past $10^6\,\Msun$.

To our knowledge, \citet{BME04b} have presented the only direct
$N-$body simulations of a multi-mass system with a central massive
object. Although they observe that the most massive objects form a
power-law cusp of exponent compatible with $\gamma=1.75$, the central
profiles of the lighter species are found to be much shallower than
predicted by Eq.~\ref{eq:BW77}, with $\gamma \simeq 0.75 +
\Mstar/(1.1\,\Msun)$. However, in the light of our comment on
\citet{MCD91} and of our simulations, this cannot be interpreted as a
rebuttal of
\citet{BW77} but more likely is an indication that the appropriate 
regime is only reached deep 
in the influence region, a region not probed by $N-$body simulations
with $\Npart \lesssim 131\,000$.

\section{Simulations: Method and Initial Conditions}
\label{sec:method}

\subsection{The Monte-Carlo Code for Nucleus Dynamics}
\label{sec:MCcode}

This work is based on simulations of the long-term stellar dynamical
evolution of galactic nuclei performed with {\MESSY}. This code is
based on the Monte-Carlo algorithm first described and implemented by
H\'enon \citep{Henon71a,Henon71b,Henon73,Henon75}. It has been
described in detail by \citet{FB01a,FB02b}. Here, we succinctly remind
the basics of the method and the included physics.

The Monte-Carlo method is based on the assumptions of spherical
symmetry and dynamical equilibrium. The cluster is represented by a
number (typically $\Npart=10^5-10^7$) of particles, each of which is a
spherical shell. These shells constitute a sampling of 1-particle the
distribution function in the phase and stellar-parameters spaces. In
other words, a shell corresponds to stars with a given orbital energy
$E$ and angular momentum (in modulus) $J$ and given stellar properties
(mass, age, etc.). At any time a shell also has a given radius
$R$. Each shell represents the same number of stars, $\Nstar/\Npart>1$
(for small systems, one may set $\Npart=\Nstar$ and $\Npart>\Nstar$ is
formally possible).

Orbital motion is not followed as dynamical equilibrium is assumed
(the system is phase-mixed); instead, the position of a particle on
its orbit, i.e.\ its radius $R$, is selected with probability
reflecting the time spent at each $R$ on the orbit specified by $E$
and $J$ in the potential of the other shells and the central object.

Gravitational relaxation is treated in the Chandrasekhar picture,
similarly to what is done to derive the orbit-averaged Fokker-Planck
equation
\citep{BT87}. It is assumed to reduce to the effect of a large number 
of uncorrelated, small-angle 2-body scatterings dominated by impact
parameters $b_0\ll b \ll b_{\rm max}$ (the value of $b_{\rm max}$ is
discussed in \S~\ref{subsec:colldyn}). Consequently, relaxation is
implemented as a series of velocity perturbations between neighboring
particles. In {\MESSY}, time steps $\delta t$ are a function of the
radius $R$ and are set to be smaller or equal to a fraction $f_{\delta
t}$ of the local $\trlx$. For the present work, we set $f_{\delta
t}=0.04$ and checked that $f_{\delta t}=0.01$ does not lead to
significantly different results.

Stellar collisions can also be treated by computing the collision time
for a pair (Eq.~\ref{eq:tcoll}) and comparing $\delta t/\tcoll$ to a
uniform $[0,1[$ variate. For the simulations of the present work where
collisions were included, interpolation from a large database of
Smoothed Particle Hydrodynamics (SPH) simulations \citep{FB05} was used to determine the
outcome, as described in \citet{FRB05}. The simulations of
\citet{FB05} specifically probe the high-velocity regime found in the
vicinity of MBHs.

An accurate treatment of the loss-cone process is not possible in the
framework of the present version of {\MESSY} because it would require
to endow particles in on near the loss cone (or on orbits eccentric
enough to possibly lead to EMRIs) with time steps shorter than the
timescale taken by relaxation to modify significantly the pericenter
distance $\trperi\simeq(1-e)\trlx\ll \trlx$. $\delta t(R)$ must be an
increasing function so that setting a short time step for some
particle would, in practice, reduce the time steps of all particles
with positions lying inside its apocenter. This difficulty is
circumvented by an approximate treatment of the relaxation-induced
random-walk of the direction of a particle's velocity vector during a
time step \citep{FB02b}.

A novelty introduced in a few runs presented here is the treatment of
large-angle scatterings. They are treated in a way similar to 
collisions but with a cross-section 
\begin{equation}
S_{\rm LA}^{(1,2)}=\pi(f_{\rm LA}b_0^{(1,2)})^2 \mbox{\ \ with\ \ }
b_0^{(1,2)} = \frac{G(m_1+m_2)}{\Vrel^2}.
\end{equation}
When a large-angle scattering is deemed to occur, the impact parameter
$b$ is selected at random between 0 and $f_{\rm LA}b_0$ with
probability density $dP/db \propto b$. The outcome, in the
center-of-mass frame of the pair, is a deflection of the velocity
vectors by an angle $2\arctan(b_0^{(1,2)}/b)$. When large-angle scatterings
are included, the Coulomb logarithm is reduced to
$\ln(\GCoulomb\Nstar/f_{\rm LA})$ to account for the fact that
gravitational encounters with $b\lesssim f_{\rm LA}b_0$ are now treated
separately.
 
\subsection{Initial Nucleus Models}
\label{sec:ini_nucl_models}



As is customary in cluster simulations, we use the ``$N-$body'' system
of units \citep{Henon71b,HM86}. Unlike the situations for which this
system was first introduced, we deal here with stellar systems that
are not strictly self-gravitating; instead their central regions are
dominated by the potential of a massive, fixed object. Hence we define
the unit system such that the constant of gravity is $G=1$, the total
{\em stellar} mass is initially $\Mcl(0)=1$, and the total initial
{\it stellar} gravitational energy (not accounting for the
contribution of the MBH to the potential) is $-1/2$. We denote by
$\Rnb$ the $N-$body length unit.
 
As a time unit, we use the ``Fokker-Planck time'' $\Tfp$ which is
connected to the $N-$body time unit $\Tnb$ through $\Tfp =
(\Nstar(0)/\ln\Lambda) \Tnb$ were $\Nstar(0)$ is the initial number of
stars. We prefer to use $\Tfp$ rather than $\Tnb$ because the former
is a relaxation time while the latter is a dynamical time. We consider
systems in dynamical equilibrium whose evolution is secular, in most
cases driven by $2-$body relaxation. For a large variety of cluster
structures, $\Tfp
\approx 10\,\trh$ where $\trh$ is the half-mass relaxation time 
\citep{Spitzer87},
\begin{equation}
\label{eq:trh}	
\trh = \frac{0.138 \Nstar}{\ln\Lambda} \left(\frac{\Rh^3}{G\Mcl}\right)^{1/2}
\end{equation}
with $\Rh$ the radius enclosing half of the stellar mass.

\begin{figure}
  \resizebox{\hsize}{!}{%
    \includegraphics[bb=19 154 565 691, clip]{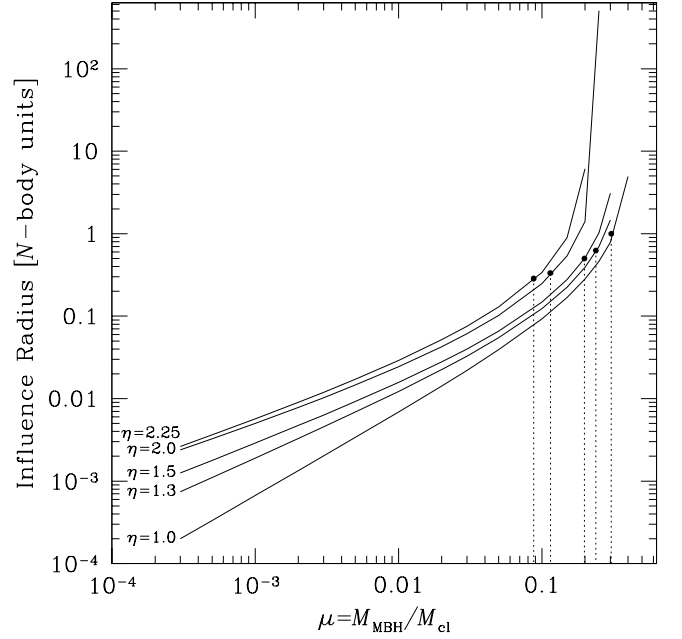}
    }
  \caption{Influence radius $\Rinfl$ as a function of the parameters
  $\mu$ in eta-models. Each curve is for a value of $\eta$. The dots
  indicate the value of the break radius $R_{\rm b}$. This diagram
  allows to find the maximum $\mu$ value for which $\Rinfl<R_{\rm b}$.}
  \label{fig:Rinfl}
\end{figure}

\begin{figure}
  \resizebox{\hsize}{!}{%
    \includegraphics[bb=17 147 585 710, clip]{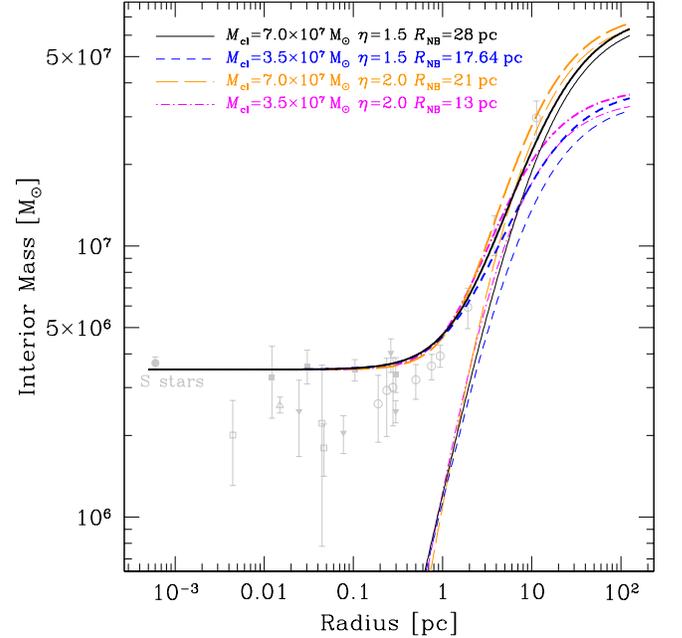}
    }
  \caption{Enclosed mass as function of radius for some of our
  models. The points in gray are observational constrains from the
  kinematics of stars and gas at the center of the MW. The point at
  the smallest distance corresponds to the simultaneous fit of the
  orbits of S-stars by \citet{GhezEtAl05}. Other data points have
  been compiled by \citet{SchoedelEtAl03}. The thin lines show the
  stellar contribution and thick lines include the central MBH with
  $\MBH=\tento{3.5}{6}\,\Msun$. The solid line is our reference
  model. The short-dashed line represents a model with a total stellar
  mass two time smaller but same stellar density at small distances
  ($R\ll \Rnb$).}
  \label{fig:MenclMW}
\end{figure}

\begin{figure}
  \resizebox{\hsize}{!}{%
    \includegraphics[bb=32 154 568 712, clip]{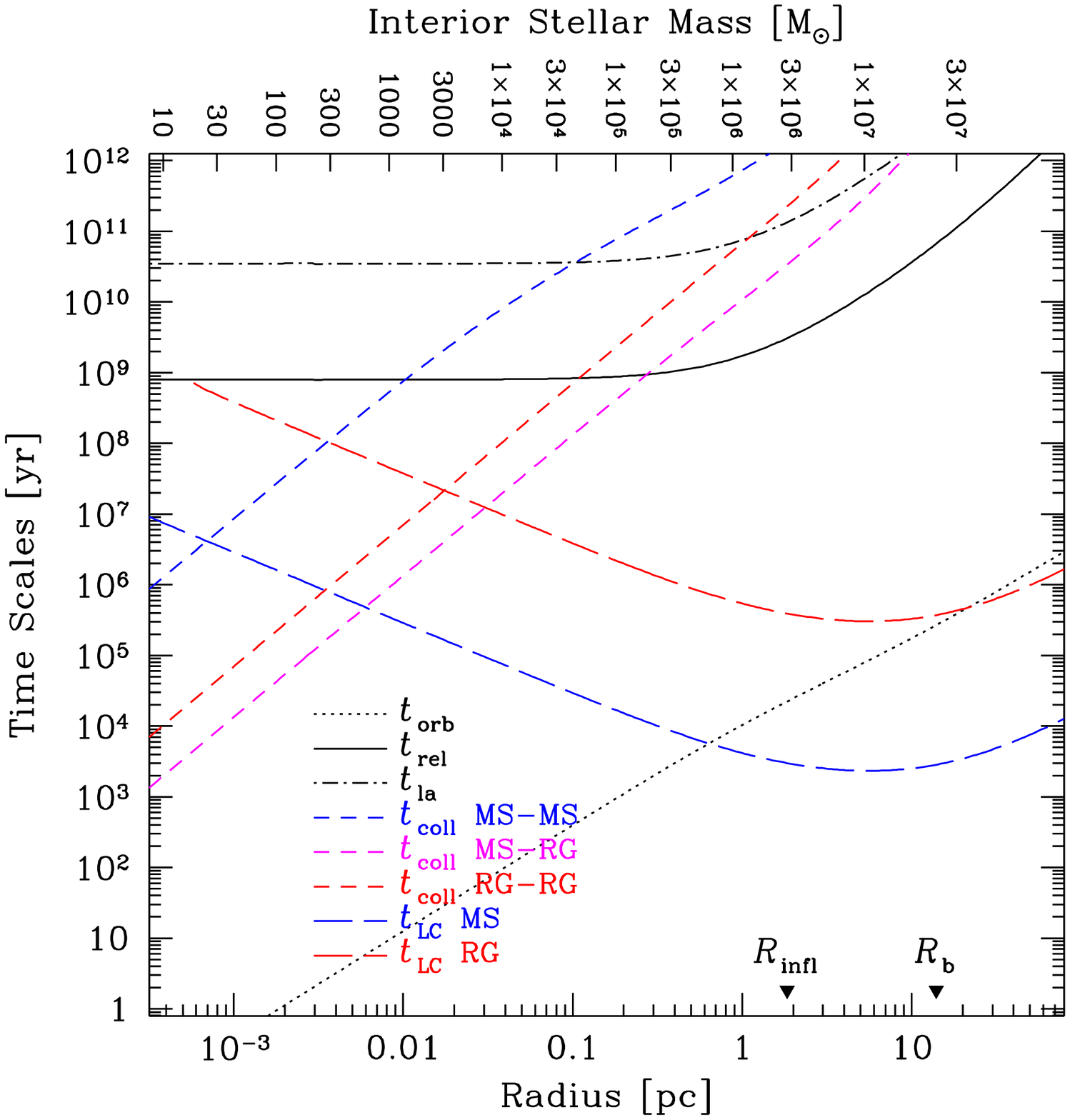}
    }
  \caption{Time scales in our reference model of the MW nucleus
  ($\Mcl=\tento{7}{7}\,\Msun$, $\eta=1.5$, $\Rnb=28\,$pc). We plot
  the orbital time $t_{\rm orb}$ (assuming a circular orbit), the
  relaxation time $\trlx$, the time scale for large-angle deflections
  $t_{\rm la}$, collision times $\tcoll$ and the timescale for diffusion
  by 2-body relaxation over the loss cone for tidal disruption,
  $t_{\rm LC}=\ThLC^2\trlx$. For collisions we indicate the average time for a MS
  star to collide with another MS star or with a red giant (RG) and
  the average time for a RG to collide with another RG. We assume
  $M_{\rm MS}=M_{\rm RG}=1\,\Msun$, $R_{\rm MS}=1\,\Rsun$, $R_{\rm
  RG}=50\,\Msun$, and 5\,\% of RGs. The radius where $t_{\rm
  orb}=t_{\rm LC}$ is the critical radius.}
  \label{fig:TscalesMW}
\end{figure}
There are only few published models for (spherical) clusters in
dynamical equilibrium and containing a massive central object. The
best described and most convenient ones are the ``eta-models''
introduced by \citet{Dehnen93} and extended to systems with a
central object by \citet{Tremaine94}. The density profile is
\begin{equation}
\rho(R) = \frac{\eta \Mcl}{4\pi R_{\rm b}^3} \left(\frac{R}{R_{\rm b}}\right)^{\eta-3} 
\left(1+\frac{R}{R_{\rm b}}\right)^{-\eta-1}.
\end{equation}
The exponent $\eta$ can take any value between 0 and 5/2. At small
radii, $\rho \propto R^{-\gamma}$ with $\gamma \equiv 3-\eta$ while at
large distances, density falls off like $R^{-4}$. The break radius can
easily be expressed in terms of other important length scales,
\begin{equation}
R_{\rm b} = (2\eta-1)^{-1}\Rnb = (2^{1/\eta}-1)\Rh.
\end{equation}
The fraction of the stellar mass enclosed by $R_{\rm b}$ is $2^{-\eta}$.
The central MBH defines a second dimensionless parameter $\mu\equiv
\MBH/\Mcl$.

At short distances from it, the MBH dominates the dynamics and therefore
$\sigma^2(R) \approx \sigma^2_{\rm MBH}(R) = (4-\eta)^{-1}G\MBH/R$. We
define the influence radius $\Rinfl$ implicitly through
$\sigma^2(\Rinfl) = 2\sigma^2_{\rm
MBH}(\Rinfl)$. Figure~\ref{fig:Rinfl} shows how $\Rinfl$ depends on
$\mu$ for various values of $\eta$. In the present study, we use
eta-models as a way to carry out simulations with a power-law density
cusp of controlled exponent $\gamma$ as initial conditions. We view
the steeper density decrease at large radii, $R>R_{\rm b}$, as a
cut-off to avoid wasting computer memory and CPU time by putting a
large number of particles at distances that should not be influenced
by the presence of the MBH through relaxation effects. In other
words, the value of $R_{\rm b}$ should be irrelevant as long as it is
large enough to encompass the region within which the collisional physics
takes place. It is therefore important to have $R_{\rm b}>\Rinfl$ and,
from Figure~\ref{fig:Rinfl}, we see that this will be the case for
$\eta=1-2.25$ provided that $\mu\le 0.05$. For $\eta\le 1.5$, $\mu \le 0.1$
should be sufficient.

Another important radius is the critical radius for
tidal disruptions, $R_{\rm cr,td}$ 
\citep{FR76,LS77,MT99,SU99,ASFS04}.
It is defined as the position in the cluster where the diffusion angle
caused by relaxation per orbital time equals the opening angle of the
loss cone, for a typical MS star,
\begin{equation}
\left(\frac{\pi}{2}\right)^2 \frac{\Porb}{\trlx} = \ThLC^2.
\end{equation}
A local, typical value of $\ThLC$ can be obtained by computing it for
a star whose velocity would be equal to the (3D) velocity dispersion,
i.e., solving Eq.~23 of \citet{FB02b} with $v^2=3\sigma^2(R)$.

The rate of tidal disruptions is dominated by the contribution of
stars with apocenter distances of order of the minimum between
$\Rinfl$ and $R_{\rm cr,td}$. For all models considered in this study,
$R_{\rm cr,td}<\Rinfl<R_{\rm b}$ (see Figure~\ref{fig:TscalesMW}) so
the loss-cone effects should be little affected by the existence of a
steeper density decrease beyond $R_ {\rm b}$.

For the present study we construct most models so that they best
approximate the conditions in the MW nucleus. In
Figure~\ref{fig:MenclMW}, we plot the enclosed mass as a function of
radius for some of our initial models and compare with observational
data \citep{SchoedelEtAl03,GhezEtAl05}. Our reference cluster model is
described by $\MBH=\tento{3.5}{6}\,\Msun$, $\Mcl=\tento{7}{7}\,\Msun$
(hence $\mu=0.05$), $\eta=1.5$ and $\Rnb=28\,$pc. This model has a
central density cusp with $\gamma=1.5$, a value consistent with the
stellar counts at the galactic center
\citep{Alexander99,GenzelEtAl03}. However, a detailed modeling of the
Galactic center is not our goal. This would in particular require ad
hoc assumptions regarding the history and locations of star formation
to account for a population with a variety of ages
\citep{FRKMS04}. This variety could be the result of the intermittent
formation, at a few pc from the center, of small clusters that then
spiral in and deposit their stars in the nucleus (see, e.g.,
\citealt{MPSR04,PaumardEtAl04,LuEtAl05} for observations and
\citealt{KM03,PZMcMG03,KFM04,GR05} for simulations). 

Separate from the MW-like models, we explore the effects of mass
segregation in idealized galactic nucleus models with a variety of
structural parameters. We consider nuclei with $\MBH$ in the range
$10^5-10^7\,\Msun$. To decrease the dimensionality of the parameter
space, we assume the $M-\sigma$ relation
\citep{MF01,TremaineEtAl02,BGH05} to hold perfectly,
\begin{equation}
\MBH \simeq M_{100}\left(\frac{\sigma}{100\,\kms}\right)^\beta.
\label{eq:MSigma}
\end{equation}

\citet{TremaineEtAl02} find $\beta=4.0\pm 0.3$ and 
$M_{100}=\tento{8.3^{+5.5}_{-3.3}}{6}\,\Msun$. With a velocity
dispersion of $\sigma\simeq 100\,\kms$, the MW harbors an
under-massive MBH. We note, however, that the velocity dispersion of
the MW central region, as defined for use in Eq.~\ref{eq:MSigma}, is
dominated by stars located at a few hundreds pc from the center
\citep[][ and references therein]{TremaineEtAl02}, a region we do 
not attempt to model. The MW nucleus is the only one whose structure 
is relatively well constrained by observations at the scales of interest 
here ($R<10\,$pc). Hence, for simplicity we adopt the MW nucleus as typical. 
A model for a nucleus with an MBH of mass $\MBH$ is
obtained from a MW model of same $\eta$ and $\mu$ by simple length
(and mass) rescaling. As $R\propto M\sigma^{-2}$ and using $\beta=4$ we obtain:
\begin{equation}
\Rnb = \left.\Rnb\right|_{\rm MW} 
\left(\frac{\MBH}{\tento{3.5}{6}\,\Msun}\right)^{1/2}
\label{eq:size_scaling}
\end{equation}
where $\left.\Rnb\right|_{\rm MW}$ is the $N-$body radius of the MW
model. Neglecting the dependence of the Coulomb logarithm on $\Nstar$,
the relaxation time of the model scales like $\trlx =
\left.\trlx\right|_{\rm MW} ({\MBH}/{\tento{3.5}{6}\,\Msun})^{5/4}$. 
It follows that $\trlx(\Rinfl)$ exceeds $\sim 10$\,Gyr for
$\MBH\gtrsim 10^7\,\Msun$ and we expect only minor relaxation
effects in such massive nuclei. The lowest MBH mass we consider,
$10^5\,\Msun$, corresponds to that of the smallest MBH detected with
some confidence in a galactic center so far \citep{BHS04,BGH05}. Even
smaller systems, such as the nuclei of dwarf galaxies or globular
clusters may host intermediate mass black holes (IMBHs) with
$\MBH<10^5\,\Msun$. We do not address the evolution of low-mass
objects here because their central dynamics may be significantly
influenced by small-$\Nstar$ effects not included in
{\MESSY}. Recently, the very significant increase of computational power offered
by special-purpose GRAPE hardware \citep{MFKN03}, combined with a
variety of mathematical and numerical techniques to speed up
computations \citep{Aarseth03b} have made it possible to follow the
relaxation evolution of clusters with a central massive object and up
to $\tento{2.5}{5}$ stars by ``direct'' $N-$body integrations
\citep{BME04a,BME04b,BMH05,PMS04}. However, because of the steep dependence 
of the CPU time on the number of particles imposed by direct force
computations in the $N-$body algorithm (approximately $T_{\rm
cpu}\propto
\Npart^3$ per relaxation time), systems containing $\sim 10^6$ stars or 
more can only be studied with more approximate methods such as MC
codes.

The range $\MBH=10^5-10^7\,\Msun$ also corresponds to the MBH around
which an EMRI has the best chance to be detected by LISA. The orbital
period of a test particle on the innermost stable circular orbit
around a non-rotating BH of mass $\MBH$ is
\begin{equation}
f_{\rm ISCO} \simeq \frac{c^3}{2\pi6^{3/2}G\MBH} = 
\tento{2.2}{-3}\,{\rm Hz}\left(\frac{\MBH}{10^6\,\Msun}\right)^{-1}.
\end{equation}
Consequently, inspirals into MBH more massive than $\sim 10^7\,\Msun$
produce signals with frequency too low for LISA to detect, while the
final inspiral into an MBH with $\MBH\lesssim 10^5\,\Msun$ occurs at
periods higher than the time taken by light to travel along LISA arms,
which strongly reduces sensitivity at those frequencies
\citep{LHH00}. In principle EMRIs into such lower-mass MBH could be
caught at an earlier phase in their orbital evolution but the emitted
waves have much lower amplitude then, thus severely limiting the
detection range \citep{Will04}. Furthermore, the analysis of
\citet{HA05} indicates that most stellar objects closely bound to an 
IMBH will be scattered on to a direct plunge orbit before they 
enter the LISA band. These authors predict that successful (i.e.\
gradual) LISA inspirals around IMBH have to start at very high
eccentricities and small semi-major axis and should last only $\sim 1$
year before coalescence.

In this study, we concern ourselves with the idealized situation of an
isolated, gas-free galactic nucleus. In particular, we do not consider
the effects of interactions with other galaxies, such as mergers with
other nuclei or gas inflow. Similarly, we neglect the possibility of
smaller stellar clusters spiraling down to the galactic center or
non-spherical mass distributions. Finally, we assume that all stars
have formed in a single burst with no further star formation. 
For the models in which stellar evolution or collisions are included,
the gas lost by stars is considered instantaneously lost from the
system, with, in some cases, a fraction being accreted by the central
MBH. 

Some of these simplifications, most noticeably that of spherical
symmetry and absence of gas, are required by the numerical methods
used. Others are made in order to reduce the complexity of the problem
and the dimensionality of the parameter-space, hence allowing a better
understanding of the systems under study.

Most of our simplifying assumptions favor mass segregation of stellar
BHs, our primary object of study. For instance, it seems likely that a
merger between nuclei induces violent relaxation, thus erasing --at
least partially-- any previous mass segregation. If both nuclei
contain a MBH, the binary MBH will eject stars from the central
regions and strongly decrease the density there, thus lengthening
relaxation time
\citep[e.g.,][]{MM01,MF04}. Also, if stars are formed over an extended
period of time instead of all being born at some ``initial'' time,
stellar BHs will, on average, have less time to experience mass
segregation.

Cosmological simulations indicate that most normal galaxies have not
suffered a major merger for several Gyrs. In particular, some 5-7 Gyr
are probably required for a disk to (re)form after a merger
\citep{GCM94,ANSE03I}. Therefore our simulations can be considered to 
cover the evolution of a galactic nucleus since it experienced its
last major merger. We will focus our analysis on the structure of the
nucleus after 5 and 10 Gyr of simulated evolution; 5 Gyr is a
reasonable value for the period of time during which a nucleus in the
present-day universe may have evolved without strong interactions;
10 Gyr is an upper limit that enables us to see what the maximum
effects of relaxation are likely to be. Mergers probably lead to
important gas inflow into the central regions, triggering stellar
formation and accretion onto the MBH, in a complex interplay
\citep[e.g.,][]{SDMH05}. In such episodes the MBH may grow substantially on 
time scales shorter than the relaxation time, but still significantly
longer than stellar orbital periods. The stellar nucleus then
contracts adiabatically in response to the deepening of the MBH
potential \citep{Young80,QHS95,FB02b}. To investigate the impact of
such episodes on the structure of the nucleus several Gyrs later, and
contrast it with our standard models where the mass of the MBH
increases only little during the course of the simulation (by tidally
disrupting and capturing stars), we computed a few models in which a
central BH of small mass ($\mu(t=0)=10^{-5}$) grows rapidly by
accreting some fraction of the gas released due to stellar evolution.

\subsection{Stellar Population and Evolution}

Except for a few test-case models presented in \S~\ref{sec:tests}, we
use the ``Kroupa'' initial mass function (IMF) for all our models
\citep{KTG93,Kroupa00b,Kroupa01}. It is a broken power-law, 
$d\Nstar/d\Mstar \propto \Mstar^{-\alpha}$, with $\alpha=0.3$, $1.3$
and $2.3$ in the ranges $\Mstar/\Msun \in [0.01,0.08[$, $[0.08,0.5[$
and $[0.5,120]$, respectively. We generally consider the range
$0.2-120\,\Msun$ for stellar masses on the MS. 

In most simulations, we do not include stellar evolution but start
with a stellar population in which all stars already have an age of 10
Gyr. This is of course not a physically consistent treatment but we
choose it for the sake of simplicity. For comparison purposes, in a
few simulations, stellar evolution is included and those simulations
are started with zero-age MS stars. The main impact of stellar
evolution is to induce significant mass loss in the first $\sim 10^8$
years. As we will see, the nucleus experiences strong expansion if
this gas is expelled from it. To produce such a model for a nucleus
with specific current properties (as those of the MW), we have to
find, by trial-and-error, initial cluster structural parameters
leading, after $5-10$ Gyr, to a nucleus model fitting the observations
(in our case, the enclosed mass as function of radius).

We use a simple stellar evolution prescription according to which
stars keep a fixed mass and radius while on the MS and instantaneously
turn into compact remnants (CRs) at the end of their MS lifetime,
$t_{\rm MS}(\Mstar)$. Data for $t_{\rm MS}(\Mstar)$ were provided by
K.~Belczynski \citep{HPT00,BelczynskiEtAl02}. As for the relation
between the stellar mass on the MS and the nature and mass of the CR,
we consider three models, presented in Table~\ref{table:remnants}. In
the first one, dubbed ``fiducial'' (F), we assume all white dwarfs
(WDs), neutron stars (NSs) and stellar BHs have a mass of $0.6$, $1.4$
and $10\,\Msun$ respectively. The two other models make use of the
prescriptions developed by \citet{BelczynskiEtAl02}, assuming either
solar ($Z=0.02$, model ``BS'') or metal-poor ($Z=10^{-4}$, model BP)
chemical composition. These prescriptions represent our current
understanding (although incomplete) of massive-star core collapse and
possible fallback onto the nascent compact remnant. The quantitative
aspects are consistent with the hydrodynamic calculations presented in
\citet{Fryer99} and the resulting relations between MS and CR masses are
shown in Figure 1 of \citet{BelczynskiEtAl02}. When stellar evolution
is included, we impose that the time step is smaller than a factor $f_{\rm
SE}$ times the MS life time $t_{\rm MS}$ for all stars still on the MS. We have set $f_{\rm SE}=0.1$
after have checked that results are essentially the same
as with $f_{\rm SE}=0.025$.

To explore the effect of supernova kicks in some simulations with
stellar evolution, we give NSs and BHs a velocity kick at
birth. Although the mechanism responsible for such ``natal kicks'' is
still not understood, they are required to explain the high spacial
velocities of observed field pulsars \citep[][ and references
therein]{HLLK05} as well as other observed characteristics of neutron
star binaries \citep[e.g.,][and references therein]{WKH04,TDS05}.

 There are also observations and interpretation analyses suggesting
 that some BHs receive a kick at birth
\citep{MirabelEtAl01,MirabelEtAl02,GCPZP05,WillemsEtAl05}. It is 
generally accepted that a supernova explosion is required to provide
the natal kick. Consequently, it is likely that only BHs formed
through the fallback mechanism, with a progenitor less massive than
$m_{\rm FB} \approx 42\,\Msun$ receive kicks \citep{FK01,HFWLH03}. In
the MC simulations with natal kicks, we base our prescription on the
results of
\citet{HLLK05}. The kick velocity is picked from a single Maxwellian 
distribution with a one-dimensional dispersion of
$\sigma_{\rm NK}=265(1.4\,\Msun/m)\,\kms$ where $m$ is the mass of the
NS or BH. BHs resulting from the evolution of a MS star more massive
than $42\,\Msun$ are not given any kick.

\section{Results of Simulations}
\label{sec:results}

Our simulations fall into two categories.  First are a few cases with
a single-mass or a two-component stellar population. They are used to
test the MC algorithm by comparison with analytical or $N-$body
results. The second category consists of more than 80 galactic nucleus
models with more realistic choices of parameters and stellar
populations. In what follows we describe the results of some
representative runs and explain how the important outcomes are
affected by the initial conditions and physics.

\subsection{Test Models}
\label{sec:tests}
\begin{figure}
  \resizebox{\hsize}{!}{%
\includegraphics[bb=37 154 583 697,clip]{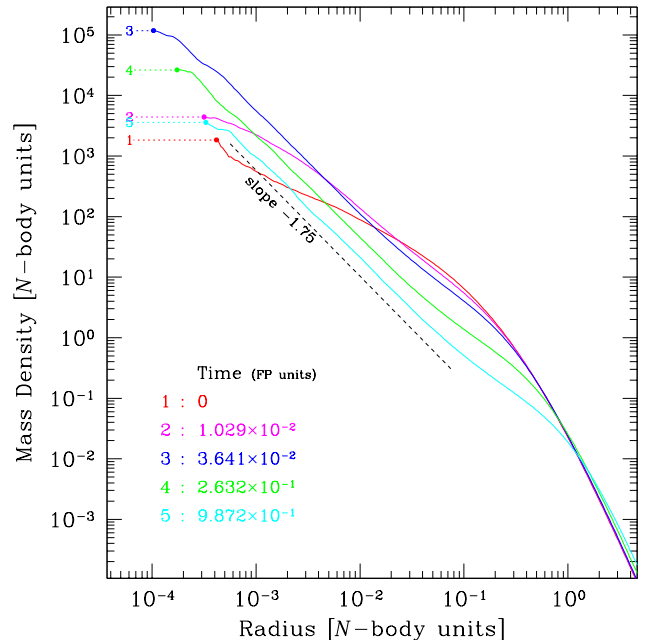}
}
  \caption{Evolution of the density profile for a single-mass cluster
  with $\eta=2.25$ and $\mu=\MBH/\Mcl=0.05$ simulated with 4 million
  particles. Note the establishment of a $\rho \propto R^{-1.75}$ cusp
  and the expansion of the cluster, driven by diffusion of stars
  towards the MBH.}
  \label{fig:ProfDensSingleMass}
\end{figure}

\begin{figure}
  \resizebox{\hsize}{!}{%
    \includegraphics[bb=18 148 579 692, clip]{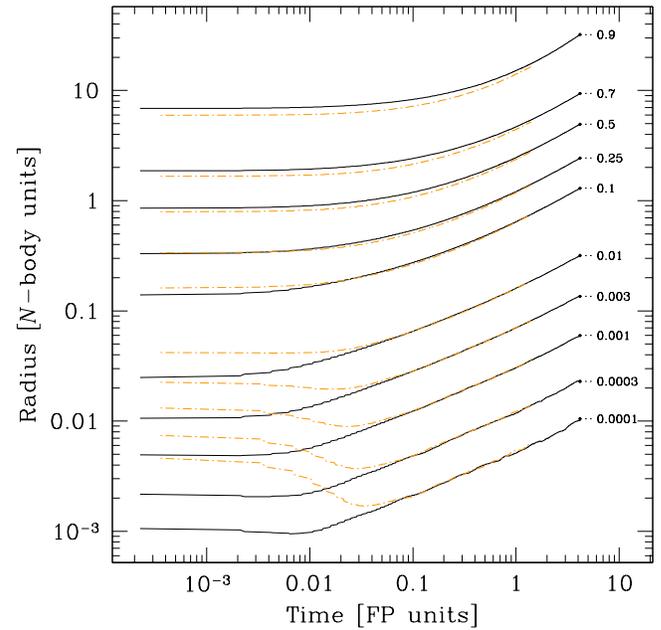}
    }
  \caption{Comparative evolution of single-mass models with different
  initial central density profiles. We plot the Lagrange radii, i.e.\
  the radius of spheres enclosing the indicated fraction of the total
  stellar mass. Models with $\eta=1.5$ (solid lines) and $\eta=2.25$ (dash-dot)
  are compared. Both runs have $\mu=0.05$ and
  $\Npart=\tento{4}{6}$. At late times, the two cases have converged
  to the same structure and evolution.}
  \label{fig:LagRadSingleMass}
\end{figure}

\begin{figure}
\begin{center}
  \resizebox{0.75\hsize}{!}{%
    \includegraphics{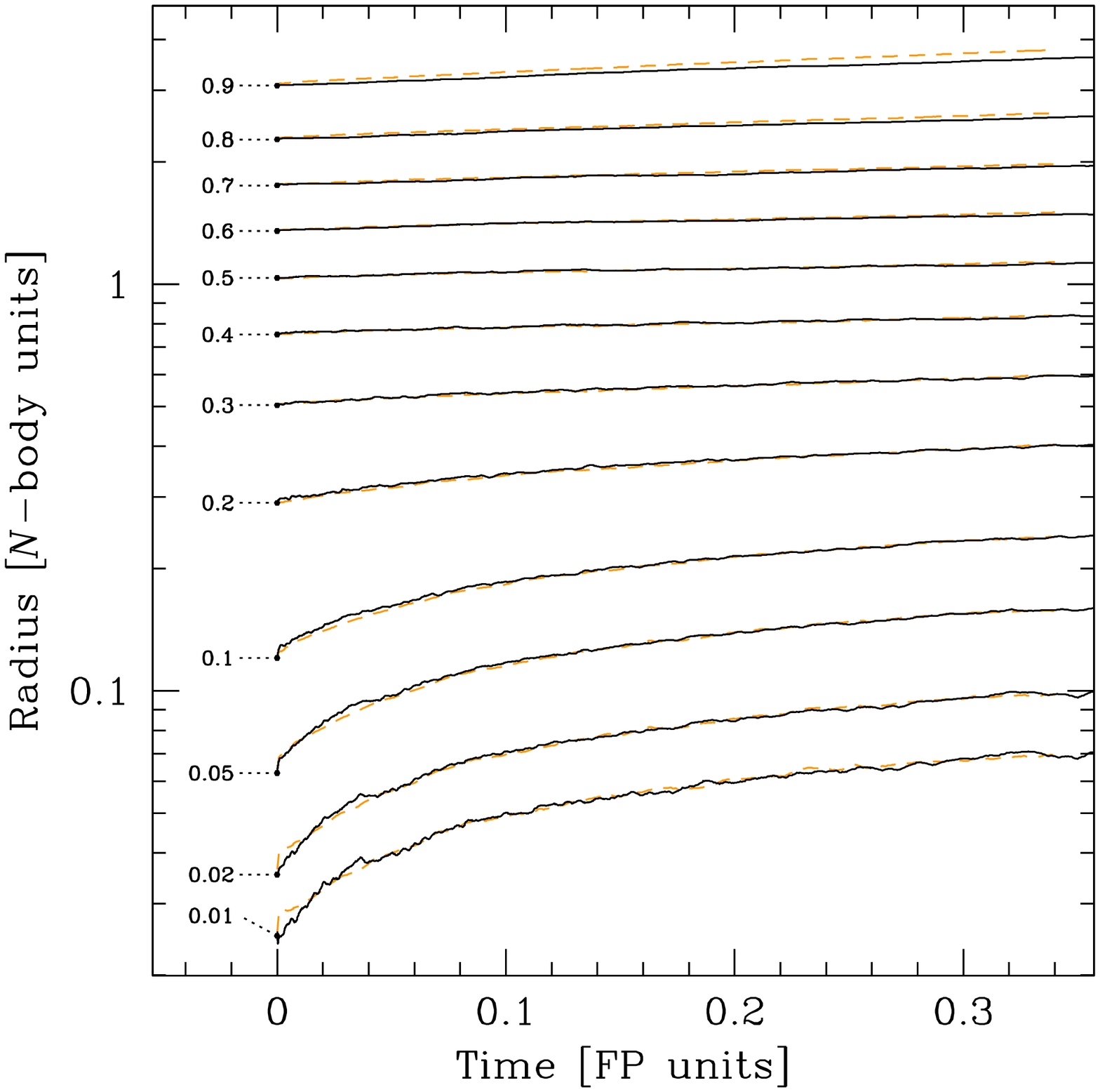}
    }\\
  \resizebox{0.75\hsize}{!}{%
    \includegraphics{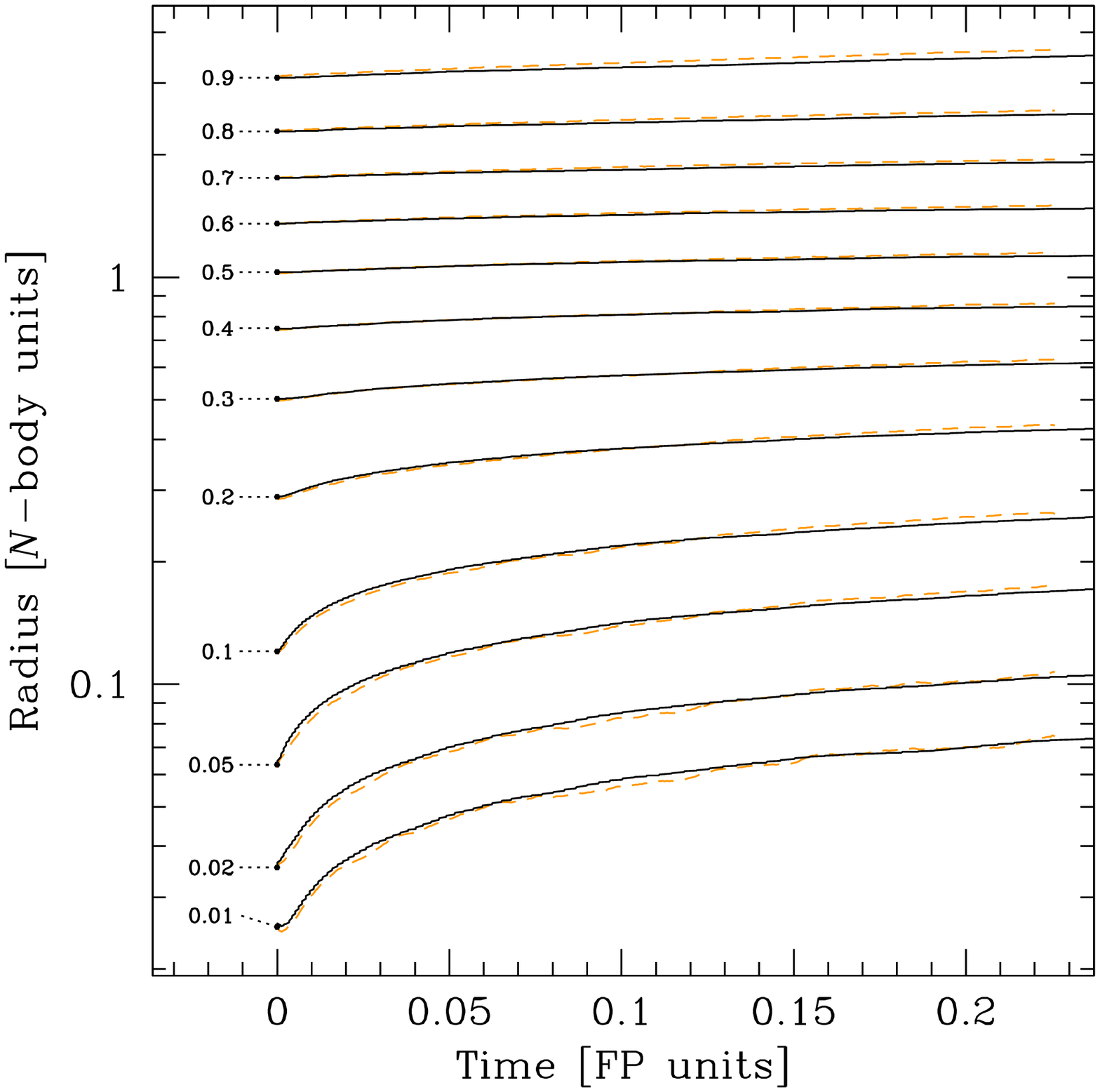}
    }\\
  \resizebox{0.75\hsize}{!}{%
    \includegraphics{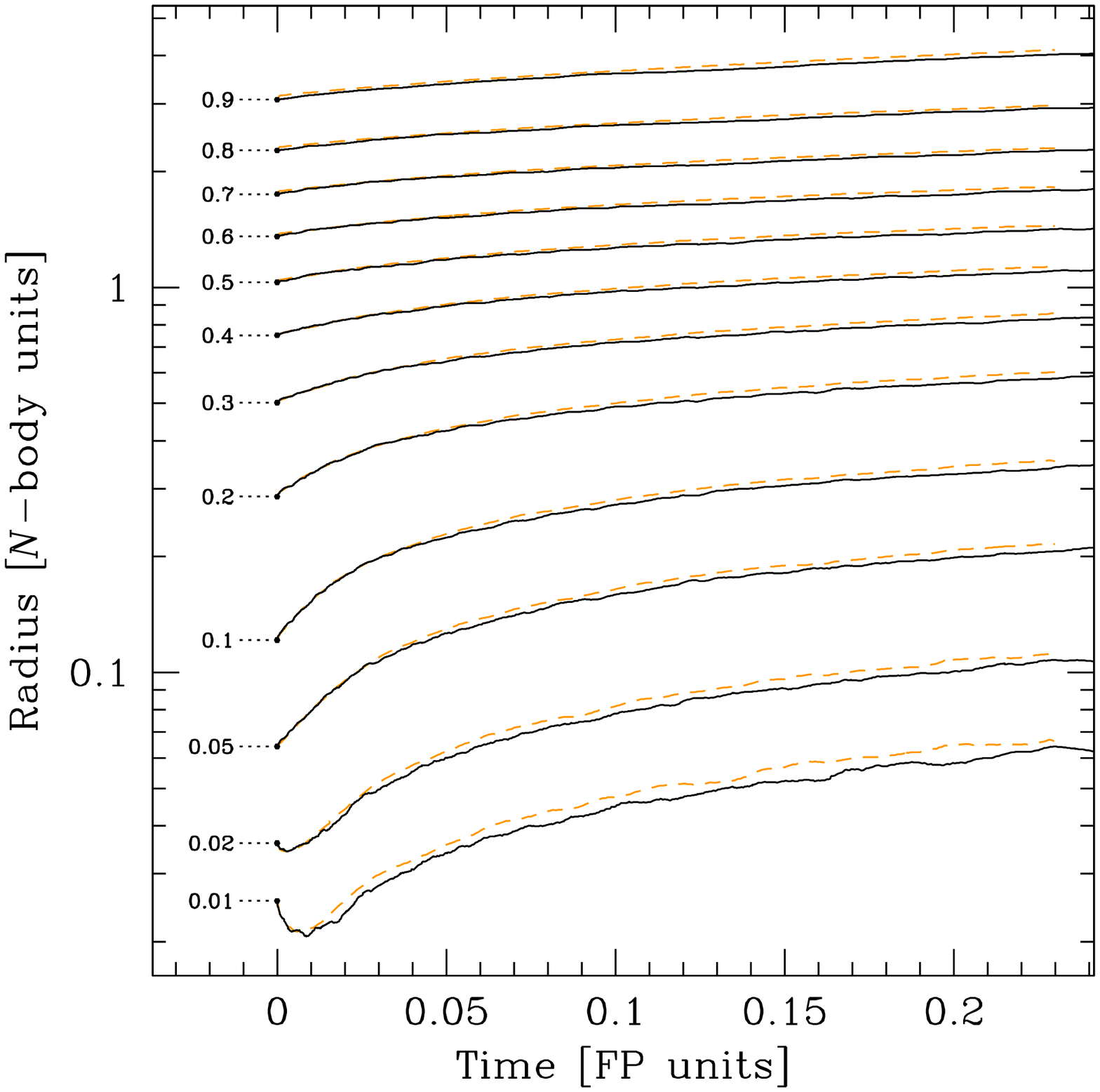}
    }
\end{center}
  \caption{ Comparison of the Lagrange radii evolution between
  $N-$body results from \citet{BME04a} and our {\MESSY} simulations of
  single-mass cluster models. From top to bottom, the panels
  correspond to models 1, 2 and 4 in Table~1 of \citet{BME04a}. The MC
  results are plotted with solid lines, the $N-$body results with
  dashed lines. We present MC results obtained with
  $\Npart=\Nstar=80\,000$ for case 1 and 4 (top and bottom) and
  $\Npart=5\,\Nstar=320\,000$ for case 2. A run with
  $\Npart=\Nstar=80\,000$ gives very similar (but noisier) results. The
  $N-$body time unit is converted into FP unit assuming
  $\GCoulomb=0.11$. The curves have been smoothed with a running
  averaging procedure using a Gaussian Kernel.}
  \label{fig:testBME04_LagRad}
\end{figure}
Since {\MESSY} was originally developed and tested
\citep{FB01a,FB02b}, the code has gone through many small revisions. 
Furthermore, at that time, only few direct $N-$body simulations had
been published with high enough resolution to yield test cases to
which the results of the more approximate MC code could be usefully
compared. The advent and spectacular increase of computing speed of
GRAPE boards now permits more comparisons, although restrictions in
the applicability of comparisons still exist. We have recently carried
out new tests for the core-collapse evolution of clusters with a
variety of stellar mass functions but no central object
\citep{FRB05}. Here, we investigate models with a central MBH. We
compare MC results with simple semi analytical predictions as well as
published and original $N-$body simulations, presented here for the
first time.

The development of a $\rho\propto R^{-1.75}$ density cusp in a
single-mass cluster hosting a central MBH has been a well-accepted
theoretical prediction for nearly 30 years \citep{BW76}, but has
only recently been verified by direct $N-$body simulations
\citep{PMS04,BME04a}. In Figure~\ref{fig:ProfDensSingleMass}
we show how such a profile forms in one of our single-mass MC
simulations of a cluster model with $\eta=2.25$, $\mu=0.05$ and 
$\Npart=\tento{4}{6}$. It is evident that at late times, the
evolution is an approximately self-similar expansion of the cluster,
driven by destruction of stars by the MBH (whose mass was kept
constant in this simulation). Models with different initial $\eta$
values converge to the same structure and evolution after $t\approx
(0.05-0.1)\,\Tfp$, as illustrated in
Figure~\ref{fig:LagRadSingleMass}. To measure the speed at which the
central regions evolve, the relaxation time at the influence radius,
$\trlx(\Rinfl)$ (using Eq.~\ref{eq:trlx}), is a more relevant
timescale than $\Tfp$; we find $\trlx(\Rinfl)=\tento{2.2}{-3}\,\Tfp$
for $\mu=0.05$, $\eta=1.5$ and $\trlx(\Rinfl)=\tento{4.9}{-3}\,\Tfp$
for $\eta=2.25$. Hence, the full development of a Bahcall-Wolf cusp
requires of order $10\,\trlx(\Rinfl)$ in a single-mass cluster.

With an $N-$body code, \citet{BME04a} have computed the evolution of
single-mass $W_0=10$ King models \citep{BT87,HH03} with a central BH
of $\mu$ in the range $0.0026-0.1$ (the stellar velocities were
modified to ensure approximate dynamical equilibrium). The central BH
was allowed to grow in mass by disrupting stars at
$\Rtd=(10^{-9}-10^{-7})\Rnb$ and fully accreting their mass. As
Figure~\ref{fig:testBME04_LagRad} clearly indicates, we can reproduce
the evolution of such systems in a satisfactory manner using
{\MESSY}. We have also checked that our results are insensitive to the
particle number (as long as it is large enough) by repeating a few
models with $\Npart=5\Nstar$ instead of $\Npart=\Nstar$. On the other
hand, we have found the MC results to be more sensitive on the time
step parameter than one might hope. For these single-mass models,
$f_{\rm
\delta t}=0.005-0.01$ gives the best results (see \citealt{FB01a} for
an explanation of how the time steps are determined in {\MESSY}; in
rough terms $f_{\rm \delta t}$ is a prescribed upper bound on $\delta
t(R)/\trlx(R)$). With larger values, the deflection angles in
``super-encounters'' become too large, leading to too little relaxation
(and hence evolution) per unit of simulated physical time.
\begin{figure}
  \resizebox{\hsize}{!}{%
    \includegraphics[bb=21 164 575 692, clip]{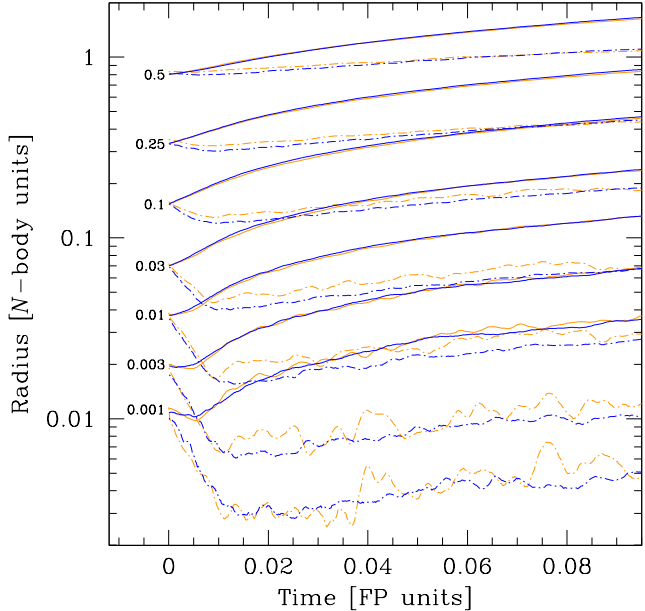}
    }
  \caption{ Comparison between our {\NBFOUR} and {\MESSY} simulations
  of a 2-component cluster model. The evolution of Lagrange radii for
  the indicated mass fractions of each component is plotted. The
  initial structure is an $\eta-$model with $\eta=2$ and
  $\mu=0.1$. The population consists of light stars with $m_{\rm
  light}=1\,\Msun$, shown with solid lines, and of 5\,\% (in number)
  of heavy objects with $m_{\rm heavy}=10\,\Msun$, shown with
  dash-dotted lines. For these runs both types are treated as MS
  stars, subject to tidal disruption by the central BH. The physical
  scales are set by $\Rnb=1\,\pc$ and $\Nstar=64\,000$. The gray
  (orange) curves show the results of the $N-$body simulation,
  realized with $\Npart=64\,000$. The black curves are for the MC run
  which used $\Npart=640\,000$. The $N-$body time unit is converted
  into FP unit assuming $\GCoulomb=0.02$. The Lagrange radii for the
  $N-$body run are determined, at each snapshot, through a procedure
  of ``orbital oversampling'' in which the position of each particle
  on its orbit is sampled many times, with probability density
  $dP/dR\propto v_r^{-1}(R)$, assuming a spherically symmetric
  potential centered on the IMBH. This way, one can follow a
  fractional mass as low as 0.001 which represents only 3.2 particles
  for the heavy stars.}
  \label{fig:compar2comp_LagRad}
\end{figure}

\begin{figure}
  \resizebox{\hsize}{!}{%
    \includegraphics[bb=34 152 569 692, clip]{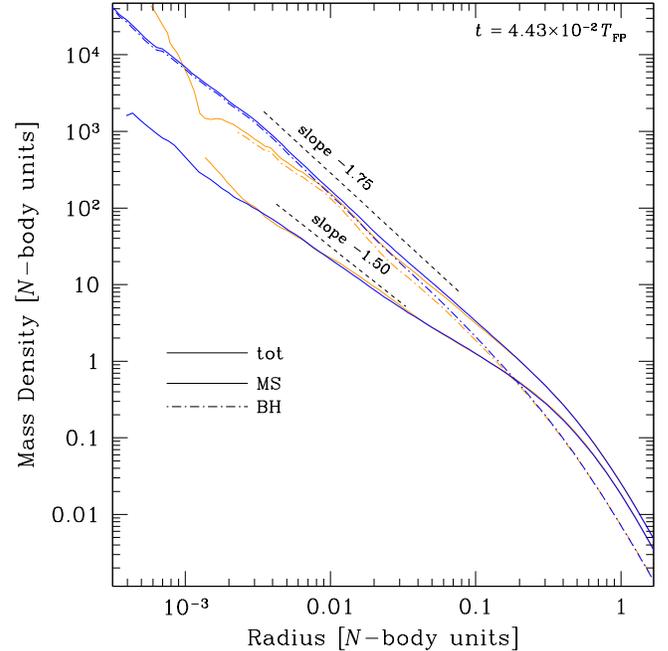}
    }
  \caption{ Comparison of the density profiles between our {\NBFOUR}
  and {\MESSY} simulations of a 2-component cluster model. Thick solid
  and dash-dotted lines show the mass density for light stars (``MS'')
  and massive ones (``BH''), respectively; the thin lines are the total
  densities. The $N-$body and MC results are shown in gray and black, 
  respectively. We also add straight lines
  representing power laws with $\gamma=1.75$ and $\gamma=1.5$.}
  \label{fig:compar2comp_Dens}
\end{figure}
The next step is to consider 2-component models in which a small
fraction $f_{\rm heavy}$ of the stars are significantly more massive
than the rest with $q=m_{\rm heavy}/m_{\rm light}\ge 10$. However,
there are no published results of this type using N-body simulations
that can provide a well-controlled test case. For this reason we have
undertaken our own $N-$body simulations using {\NBFOUR}, a code
developed and made freely available by Sverre
Aarseth\footnote{\texttt{http://www.ast.cam.ac.uk/$\sim$sverre/web/pages/nbody.htm}}.
Modifications were made to the code to include tidal disruptions and
BH mergers. Over the years, Aarseth's {\NBODY} family of codes
have become central workhorses in a great number of stellar dynamical
projects. They are described in detail in
\citet{Aarseth99,Aarseth03b}. {\NBFOUR} can exploit a 
GRAPE board to accelerate the computation by a very large factor
\citep{MFKN03}, which proved essential to obtain the results presented 
here.
 
To represent stellar BHs in a population of age 5--10\,Gyr 
$q\simeq 20-40$ and $f_{\rm heavy}\simeq\tento{(1-3)}{-3}$ would be adequate
but such small fractions cannot be adopted usefully in $N-$body
simulations with $\Npart\lesssim
\tento{1.3}{5}$, the highest number of particles that can be used 
on the micro-GRAPE hardware at our disposal. To have a reasonably
large number of heavy particles, we have chosen $f_{\rm heavy}=0.05$
and $q=10$ for a simulation with $\Npart=64\,000$. The initial
structure of the cluster is an $\eta-$model with $\eta=2$ and
$\mu=0.1$. For simplicity, we have assumed that
all stars have a MS size and are tidally disrupted if they come within
$\Rtd$ of the IMBH, itself treated as a massive particle (rather than an
external potential). When a star is tidally disrupted its whole mass
is given to the IMBH.
The size is set to
$\Rnb=1\,\pc$. MC models were run with $\Npart=\Nstar$ (``64k''),
$\Npart=5\Nstar$ (``320k'') and $\Npart=10\,\Nstar$ (``640k'') for
higher resolution and to permit a better determination of the local
density, particularly near the cluster center, as needed by the MC
algorithm for robust results. Actually, the results turn out to depend
very little on $\Npart$.\footnote{ The $N-$body model was run at the
Astronomisches Rechen-Institut in Heidelberg, on a PC equipped with a
micro-GRAPE board. It required approximately 2 weeks of
computation. In contrast, 64k and 320k MC runs took about 0.5 and 4
hours on a 1.7\,GHz laptop.}.

Figure~\ref{fig:compar2comp_LagRad} offers a global view on how the
spatial distributions of light and heavy particles evolve with time in
the $N-$body and MC simulations. For the $N-$body simulation, the
center of the system, from which distances are measured, was defined
to be the (instantaneous) position of the IMBH. As the natural time
scale is dynamical for the $N-$body code ($\Tnb$) but relaxational for
the MC algorithm ($\Tfp$), one needs to specify $\GCoulomb$ to compare
the results in the same time units. We find the best agreement with
$\GCoulomb=0.01-0.02$, as was the case for the (I)MBH-less multi-mass
systems simulated by \citet{FRB05}. For the light particles, the
concordance between the methods is excellent. The heavy particles, on
the other hand, show some discrepancy. The MC code produces mass
segregation at a rate almost equal to that seen in the N-body
runs. The heavy objects appear to concentrate slightly more at the
center before the whole cluster starts expanding
slowly.

The nature of the difference between the results from the two codes is
seen more clearly in Figure~\ref{fig:compar2comp_Dens} where a
snapshot of the central density profiles at nearly the same time is
shown. The MC run {shows a Bahcall-Wolf cusp of BHs that extends all
the way down to the resolution limit. In contrast, the $N-$body
profile appears to flatten slightly inside $R\approx
0.01\,\Rnb$. Given that the region with this flattened profile
involves only $5-10$ BH particles at a time in the $N-$body
simulation, this mismatch could be deemed of little significance, if
it were not consistently present in most snapshots. We have redone the
MC simulations with or without large-angle scatterings or tidal
disruptions of the MS stars and found that the results are not
altered: in all cases, the BHs develop a slightly more pronounced
innermost density peak than in the $N-$body run. The fact that in the
MC simulation the central BH is assumed to be fixed in position may be
the cause of the difference; this is supported by the amplitude of the
IMBH wandering in the $N-$body run, $\sim 0.01\,\Rnb$ (comparable to
the spatial extent of the flattened profile). If this is the case, the
effect should be less important in galactic nuclei, as far as the
distribution of stars around the MBH is concerned because the
wandering --essentially the manifestation of energy equipartition--
decreases with decreasing mass ratio $\Mstar/\MBH$ \citep[][ and
references therein]{LM04}. In our $N-$body simulation, this ratio
is is $\sim 10^{-4}$ and $\sim 10^{-3}$ for light and heavy stars
respectively. In a galactic nucleus with a $10^6\,\Msun$ MBH, the
ratio is $10^{-5}$ at most.

Last we examine the density profiles shown in
Figure~\ref{fig:compar2comp_Dens}. Specifically those obtained with
the MC code (less affected by small-number effects) clearly indicate
that the light objects follow a profile compatible with $\gamma=1.5$
only at distances smaller than $R_{1.5}\approx 0.01\,\Rnb$, whereas
the radius encompassing a mass of stars equivalent to $\MBH$ (an
approximation to $\Rinfl$) is of order $0.2-0.3\,\Rnb$. Only within
$R_{1.5}$ is the {\it number} space density of stars dominated by
BHs. Between $R_{0.5}$ and $\Rinfl$ is a transition region in which
$\gamma<1.5$ for the light objects even though $\gamma\simeq 1.75$ for
the heavy ones. Similar findings were obtained by
\citet{BME04b}. Although these results do not invalidate the prediction from 
the Fokker-Planck treatment that light objects should form a cusp with
$\gamma=1.5$ close to the central (I)MBH \citep{BW77,GP04}, they
indicate that, unless the fractional number of massive objects is
unrealistically high (as is the case in the test-computation presented
by \citealt{MCD91} in their Figure~9), this regime may only be
attained in a very small central volume and therefore will be of
relatively little
relevance to real systems.

\subsection{Realistic Models}

\subsubsection{{\SgrA}-type models}

\begin{figure}
  \resizebox{\hsize}{!}{%
\includegraphics[bb=23 154 583 697,clip]{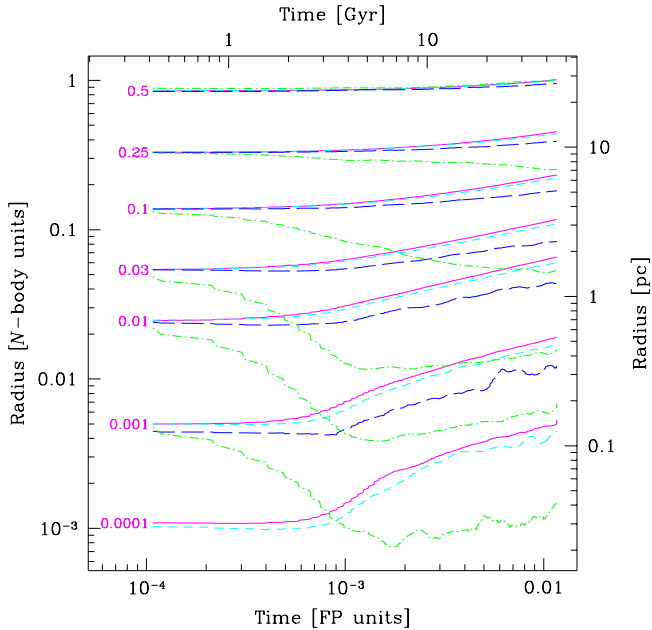} 
}
  \caption{Evolution of Lagrange radii for a ``standard'' MW nucleus
  model {\tt GN25}.  We plot the evolution of the radii of
  spheres that enclose the indicated fractions of the mass of various
  stellar species. Solid lines are for MS stars, short-dashed lines
  for white dwarfs, long-dashed lines for neutron stars and
  dash-dotted lines for stellar BHs.}
  \label{fig:LagRadStandModel}
\end{figure}

\begin{figure*}
  \resizebox{\hsize}{!}{%
    \includegraphics{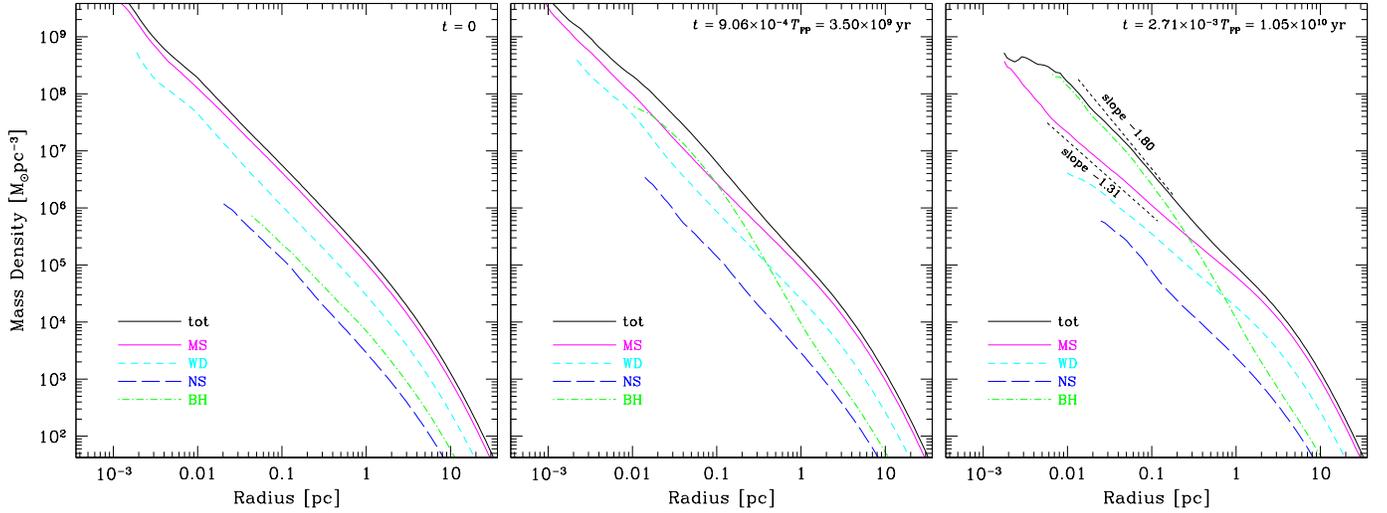}
    }
  \caption{Evolution of the density profiles for a ``standard'' MW
  nucleus model {\tt GN25} (same model as
  Figure~\ref{fig:LagRadStandModel}). On the last panel, we indicate
  the power-laws that appear consistent (visually) with the central 
  density  profiles of the MS stars and stellar BHs. The BH exponent,
  $\gamma\simeq 1.8$ is very close to $\gamma=1.75$
  predicted by \citet{BW76,BW77} but the lighter objects form a cusp
  shallower than $\gamma=1.5$ (see discussion in text).}
  \label{fig:DensStandModel}
\end{figure*}

\begin{figure*}
  \resizebox{\hsize}{!}{%
    \includegraphics{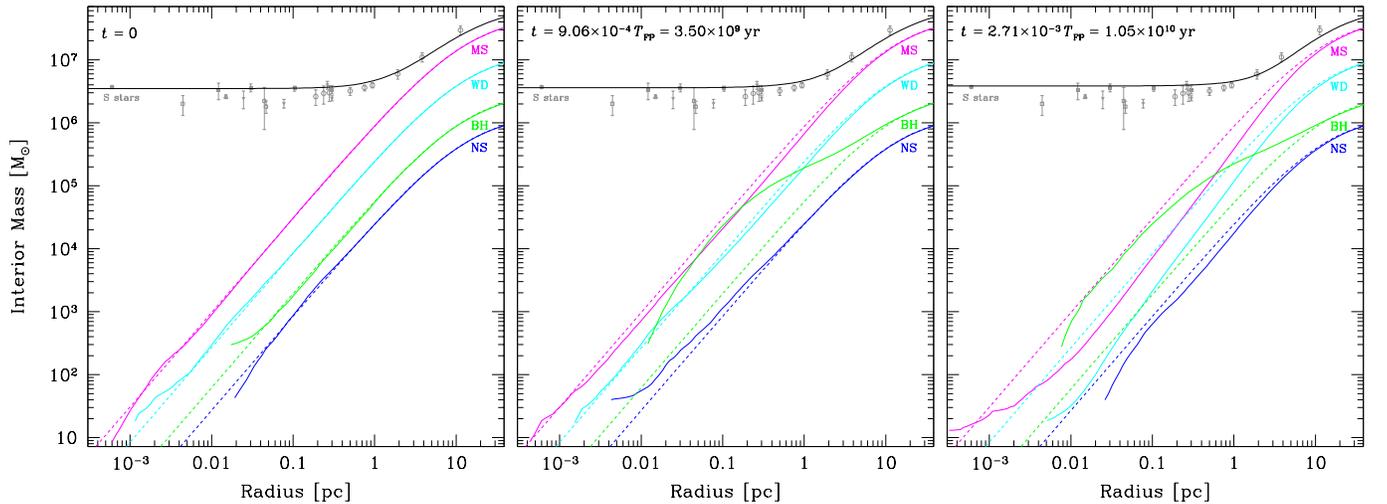}
    }
  \caption{Evolution of the profiles of enclosed mass for {\tt
  GN25}. The solid lines are the results of the MC simulation. For
  reference, he dashed lines show $\eta=1.5$ profiles adjusted on the
  total mass and half-mass radius of each component. The top thin line
  is the total mass, including the central MBH; it is compared to the
  observational constrains for the MW center (see
  Figure~\ref{fig:MenclMW}).}
  \label{fig:MenclStandModel}
\end{figure*}

\begin{figure}
  \resizebox{\hsize}{!}{%
    \includegraphics[bb=23 154 583 697, clip]{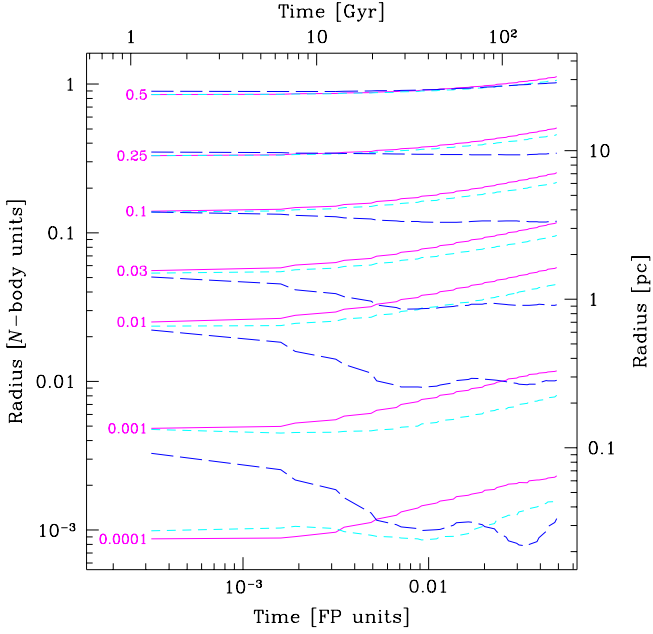}
    }
  \caption{Evolution of Lagrange radii for a nucleus model without stellar BHs ({\tt GN49}). }
  \label{fig:LagRadNoBHs}
\end{figure}

\begin{figure}
  \centerline{\resizebox{0.99\hsize}{!}{%
    \includegraphics[bb=30 147 590 694, clip]{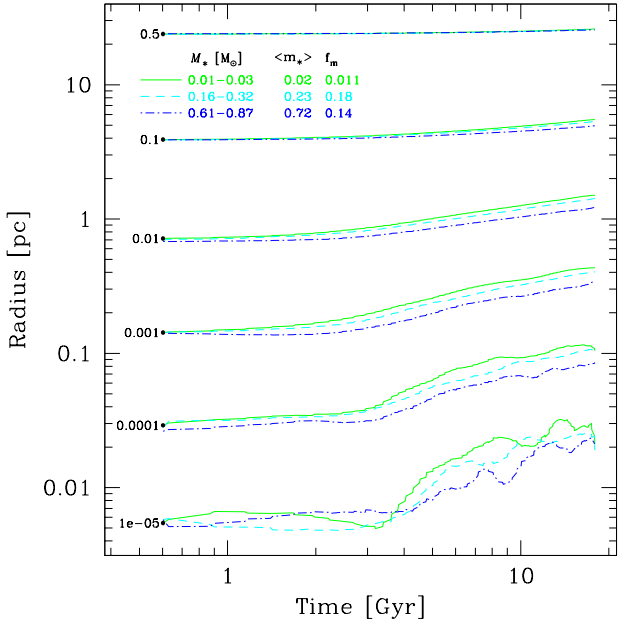}
    }}
  \caption{Differential mass segregation amongst MS stars. For model
  {\tt GN44}, in which the IMF extends down to $0.01\,\Msun$
  (instead of being truncated at $0.08\,\Msun$), we plot the evolution
  of Lagrange radii for MS stars in three different mass bins. The
  lightest objects expand only slightly more than the most massive
  ones. For each bin, we indicate the average mass
  $\langle\Mstar\rangle$ and mass fraction $\rm f_{\rm m}$.}
  \label{fig:LagRadMSStars}
\end{figure}

\begin{figure}
%
\begin{center}
\includegraphics[scale=.4]{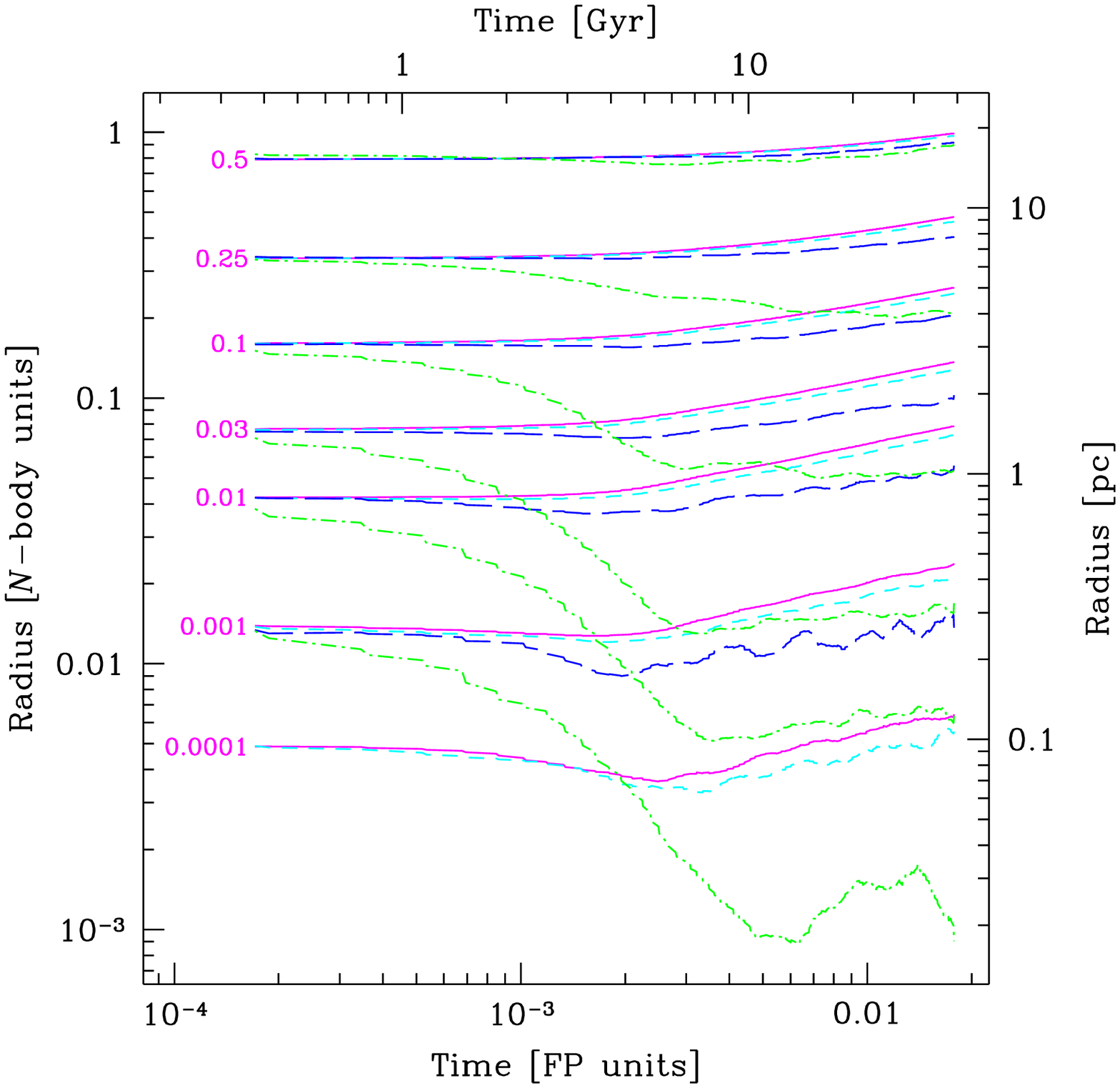}\\
\includegraphics[scale=.4]{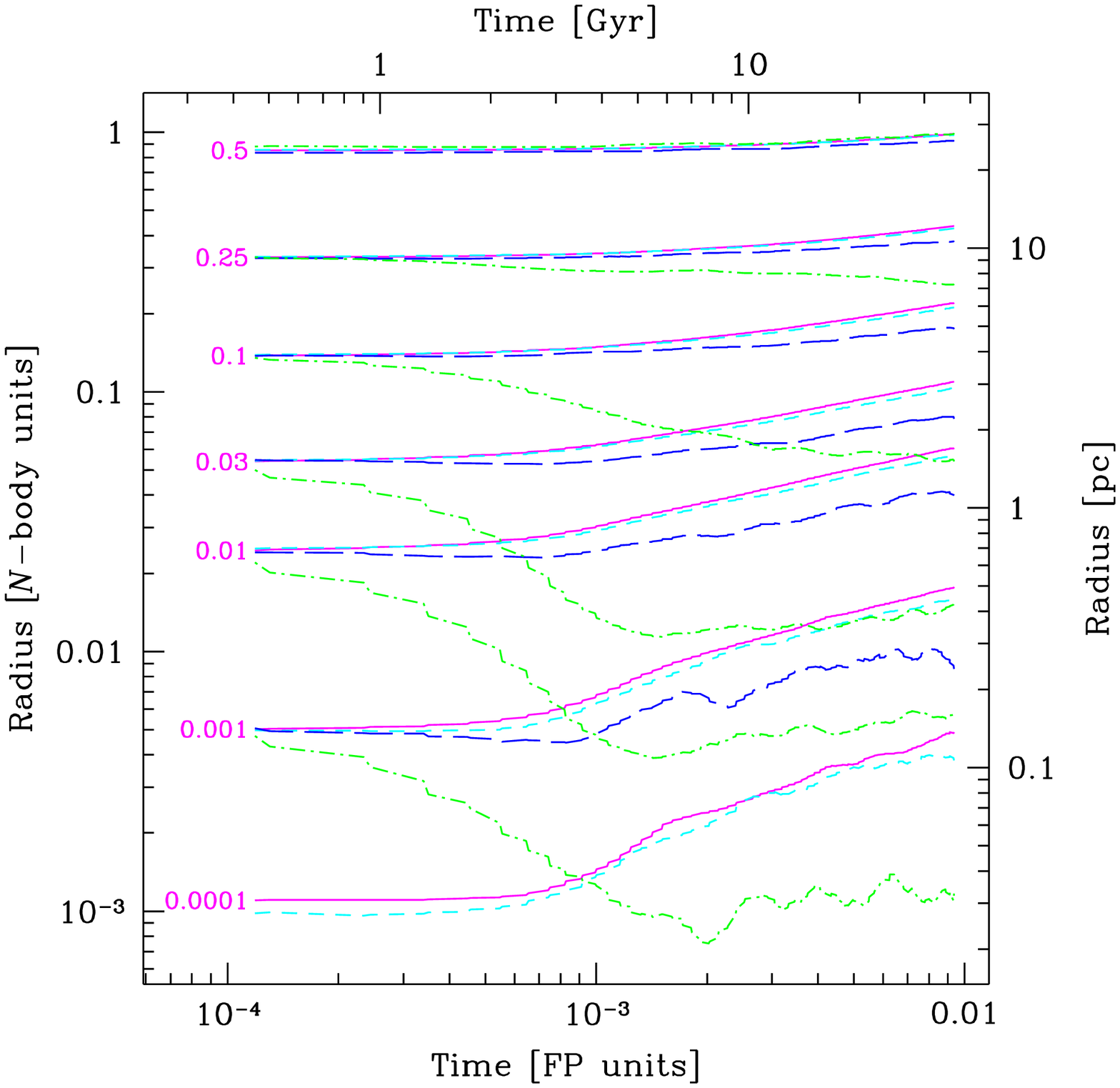}
\end{center}
  \caption{Comparison of the evolution of two nucleus models with
  similar initial conditions but different values of $\eta$:
  $\eta=2.25$ ({\tt GN20}; top panel) and $\eta=1.5$ ({\tt
  GN17}; bottom panel). Note that the conversion factors between $N-$body and FP
  units (for length and time, respectively) and physical ones are
  different in each case.}
  \label{fig:LagRad2Eta}
\end{figure}
Next we consider models specifically intended to represent galactic
nuclei. The parameters describing the initial conditions for these
simulations are listed in Table~\ref{table:ini_cond}. 

Let us first consider in some detail the evolution of our ``standard
model'', run {\tt GN25} with $\eta=1.5$ and $\mu=0.05$. These
parameters are adapted to fit the observed enclosed mass profile of
the MW center \citep{SchoedelEtAl03,GhezEtAl05}. The physics
implemented in this model is limited to 2-body relaxation, tidal
disruptions of stars by the central MBH (which accretes 50\,\% of the
stellar mass) and direct plunges through the horizon. We use stellar
population F. This is one of the highest resolution models with
$\Npart=\tento{8}{6}$ and each particle representing $26.65$ stars
(note that the MC code does not require a particle to stand for an {\it
integer} number of stars).

The overall evolution of the nucleus structure is depicted in three
different (but essentially equivalent) ways in Figures
\ref{fig:LagRadStandModel}, \ref{fig:DensStandModel} and 
\ref{fig:MenclStandModel}. In Figure \ref{fig:LagRadStandModel} we present
a general overview by showing how the Lagrange radii of the various stellar 
types evolve with time. The development of mass segregation is clearly 
apparent. Qualitatively, the region of influence of the MBH
corresponds to the extent of the MS Lagrange radius for a fractional
mass equal to the value of $\mu=\MBH/\Mcl$, i.e. 0.05. Deep in this
region, the evolution is approximately homologous. The stellar BHs
concentrate in the center over a timescale $\sim \tento{(1-2)}{-3}\,\Tfp
\approx 4-8\,$Gyr. At the same time, the other stars slightly expand 
out of the center but the total density profile stays nearly
constant. 

During this first phase, the BHs come to dominate the central mass
density by forming a cusp around the MBH. This can be seen in
Figure~\ref{fig:DensStandModel}. We note that, at late times, the cusp
exponent becomes compatible with $\gamma=1.75$, but the lighter
objects form a profile with $\gamma<1.5$, flatter than the \citet{BW77} exponent.
However, it must be stressed that, for this model, the
stellar BHs never contribute more than $\sim 15$\,\% of the {\em number}
density in any region. Therefore they do not become a strictly
dominant species, in the sense that they still experience most of
their interactions with lighter objects. This is different from the
situation studied by
\citet{BW77}, who only considered larger values for the number
fraction $f_{\rm heavy}$ of massive objects (their smaller value being
$f_{\rm heavy}=1/16$ while we have $f_{\rm heavy}=f_{\rm BH}\simeq
0.002$) and smaller mass ratios $q=m_{\rm heavy}/m_{\rm light}$ (they
have $q\le 10$ while we have $q\approx 20-30$). Because our particle
number is not large enough to treat the system on a star-by-star
basis, it is still possible that, in a real MW-like nucleus, there
would be a region very close to the central MBH in which the stellar
BHs are numerically dominant and a clean \citet{BW77} cusp could
form. Our results strongly suggest that the radius of this region is
at least 100 times smaller than $\Rinfl$.

All other stellar species react to the segregation of the stellar BHs
by expanding away from the center. This evolution is very similar for
all objects of mass significantly lower than that of the BHs, with the
NSs showing slightly less expansion than the MS and WD stars. However,
to the resolution limit of our simulations, the density profiles show
no conspicuous central depletion, such as a flattening or even a dip
(as suggested by \citealt{CG02} for pulsars around {\SgrA}). Such a
density decrease is apparent only in comparison with the initial
conditions. It is very unlikely that this density decrease can be 
revealed by observations in the Galactic center
as a tell-tale indication of the presence of a cusp of stellar
BHs. Also, MS stars of different masses react essentially the same way, 
as can be seen in Figure~\ref{fig:LagRadMSStars},
and end up having the same density profiles. 

The fact that the stellar BH population is the main driving cause for the 
evolution of the central parts of our nucleus models becomes clear by running 
a simulation without any BHs (see Figure \ref{fig:LagRadNoBHs}). The most 
obvious difference is that the overall evolution, now driven by the mass 
segregation of NSs, is of order $5-10$ times slower, reflecting a correspondingly 
longer dynamical-friction time scale. The NSs are fully segregated only after 
of order 30--40\,Gyr. Consequently, even a clear-cut
observation that old visible stars form a  $\gamma\lesssim 1.5$ at the
center of the MW could not be interpreted as (indirect) evidence
for the existence of a population of invisible BHs following a steeper
profile: if BHs are not present, the system evolves too slowly to reach a 
relaxed state over 5--10\,Gyr and the observed distribution may
still reflect some ``initial conditions'' impose, for example, by a merger 
with another nucleus or by a large starburst due to massive gas inflow.

We note that the choice of $\eta=1.5$ as initial condition is rather arbitrary. 
It is mostly motivated by the observational constrains on the present-day stellar
distribution around {\SgrA}. We have considered models with $\eta$ in the range 
$1.2$ ($\gamma=1.8$) to $2.25$ ($\gamma=0.75$) to
assess the importance of the initial density profile on the late-time 
structure and evolution of our models. In
Figure~\ref{fig:LagRad2Eta}, we compare the evolution
of two models that share the same physics and most initial conditions,
including the total mass, the mass of the MBH, the stellar population
and (approximately) the enclosed stellar mass within
1\,pc, but different central profiles, namely $\eta=2.25$,
corresponding to a shallow cusp, and our usual $\eta=1.5$. The
$\eta=2.25$ model shows more evolution in the first $6-8$\,Gyr as it
``catches up'' with the $\eta=1.5$ case. After $t\sim 8$\,Gyr,
however, both nuclei have similar structures. In both cases, the BHs
dominate the mass density inside $R\simeq 0.3\,\pc$ (where their
density is $\simeq \tento{2}{5}\Mpccube$) at $t\simeq 10$\,Gyr. At
that time, the BHs and MS stars form cusps with $\gamma=1.7-1.8$ and
$\gamma=1.3-1.4$, respectively (for $R\lesssim 0.15\,\pc$) for both
simulations. In other terms, in the region of influence of the MBH, a
period of time of order $\trlx(\Rinfl)$ (which translates into
$7-10$\,Gyr for our MW-like models) seems enough to erase the details
of the ``initial conditions''.
\begin{figure}
  \centerline{\resizebox{0.87\hsize}{!}{%
    \includegraphics[bb=37 157 566 692, clip]{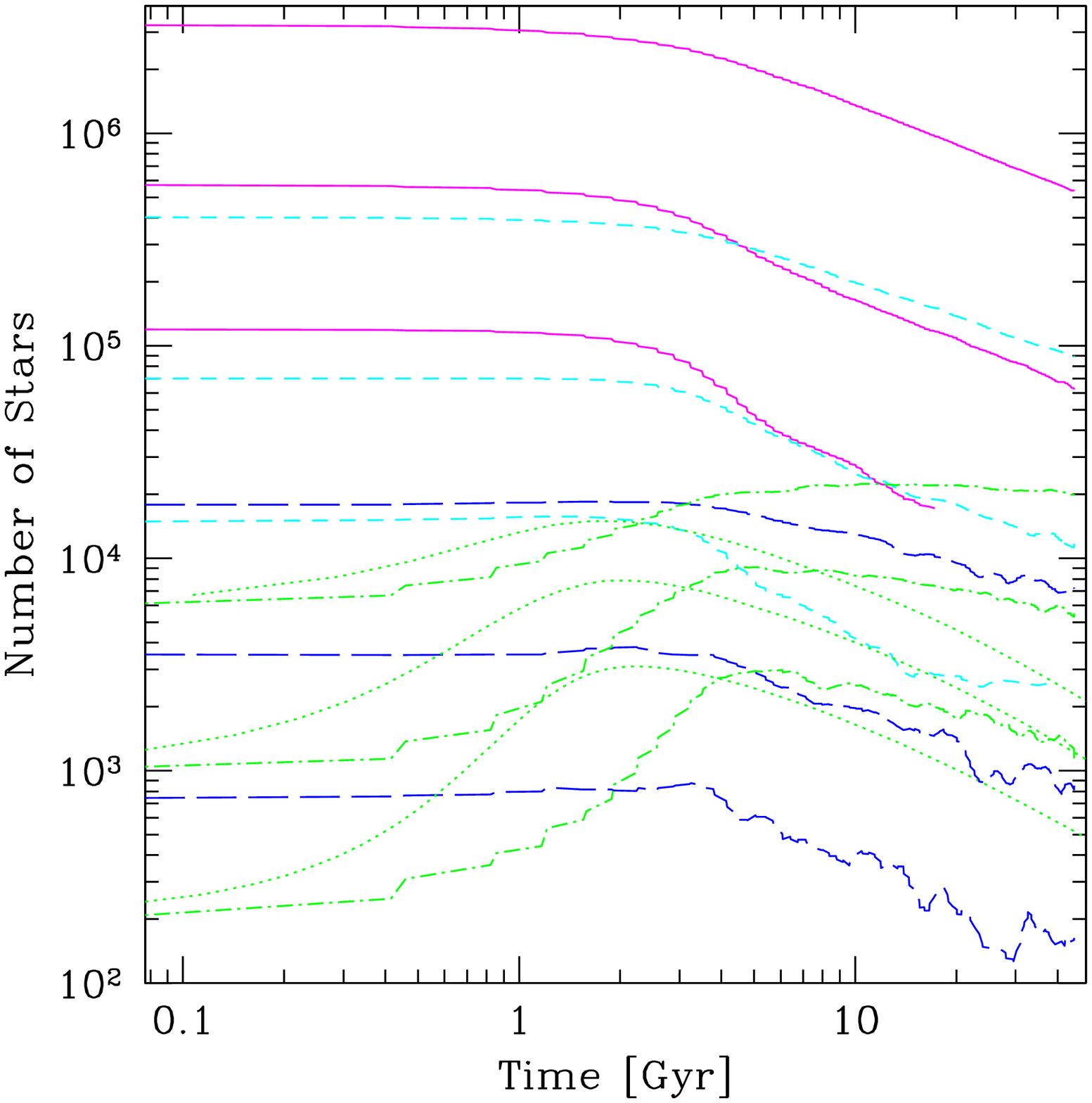}
    }}
  \caption{Number of stars of different types within $0.1$, $0.3$ and
  $1\,\pc$ from the center for model {\tt GN25}. Solid lines
  indicate MS stars, short dashes WDs, long dashes NSs and dash-dotted
  lines stellar BHs. The dotted lines are the result of the
  application of the analytical formula for dynamical friction
  (equations \ref{eq:dynfric1} and \ref{eq:dynfric2}) for the BHs,
  assuming a static background defined by the $\eta$-model and average
  stellar mass of the initial conditions. BHs that would have reached
  $R=0$ by dynamical friction are considered accreted by the MBH and
  not counted.}
  \label{fig:NobjCloseMBHStand}
\end{figure}
The initial conditions of model {\tt GN25} were chosen to be
compatible with the overall mass distribution in the {\SgrA} cluster,
as constrained by observations (see Figure~\ref{fig:MenclMW}). In
Figure~\ref{fig:MenclStandModel} we see that, despite mass segregation
and the slight expansion of the lighter stars, the enclosed mass
profile is still an acceptable fit to the {\SgrA} data after 10\,Gyr
of evolution. This is primarily because the evolution amounts mostly
to redistributing the various stellar types while keeping the total
density nearly constant. It is evident that the observations of the
current mass profile do not provide a strong constraint on initial
nucleus properties, as long as they match the stellar mass enclosed within
$R\simeq 2-3\,$pc. For the chosen initial conditions, the overall
expansion of the cluster occurs on a time scale longer than the Hubble
time but, as we will see, smaller nuclei expand significantly over a
few Gyrs, owing to their shorter relaxation time (see
Figure~\ref{fig:LagRadSmallNucl}).

In Figure~\ref{fig:NobjCloseMBHStand} we present the number of stars
of various types within distances of 0.01, 0.1 and 1\,pc of the MBH,
as a function of time. It is again evident that it takes
$3-5\,$Gyr for
the stellar BHs to concentrate in the inner pc. For a variety of
$\eta$ values and stellar populations, we find that between 20\,000
and 30\,000 of them populate this region after 5\,Gyr. Without mass
segregation their number would be of order 4-5 times lower. These
numbers bracket the estimate of 25\,000 obtained by
\citet{MEG00}. Similarly, for a stellar population similar to our case
S, \citet{Morris93} found that some $\tento{3.6}{4}$ BHs would
dominate the stellar mass density in the inner $0.8\,\pc$ (see line 4
of his Table~1). This agreement could be taken as proof that the
dynamical friction formula, used by \citet{Morris93} and
\citet{MEG00}, captures the process of mass segregation quite
accurately. However we think that this agreement is actually rather
fortuitous. In Figure~\ref{fig:NobjCloseMBHStand} we plot the
predictions of the dynamical friction formalism, assuming circular
orbits and a static background corresponding to the initial stellar
distribution. BHs that reach $R=0$ are assumed to merge with the MBH
and are not counted. Applied to our initial nucleus model, this
computation overestimates the speed and magnitude of mass
segregation. In particular, it leads to too many BHs being accreted by
the MBH and, consequently they lead to a fast {\it decline} in the number of
BHs populating in the central region after $t\simeq 2\,$Gyr. For
instance, from this simple treatment, one would expect only of order
7000 of them to inhabit the inner pc at $t=10\,$Gyr.  As expected,
this formalism also fails to reproduce the {\em structure} of the
central BH concentration by allowing BHs to sink in all the way down
to $R\approx 0$ and not taking into account their mutual
interactions. Clearly, once the BHs dominate the mass density in some
region, they start exchanging energy with each other at an important
rate, a process which cannot lead to an overall contraction. Finally,
based on the simple dynamical friction argument, one would erroneously
expect all stars significantly more massive than the average,
including the NSs, to segregate to smaller radii; this is clearly not
seen in the numerical simulations. We note that using the local,
self-consistent velocity distribution for an $\eta-$model instead of
relying on a Maxwellian approximation to compute the dynamical
friction coefficient makes a negligible
difference.
\begin{figure}
\centerline{\includegraphics{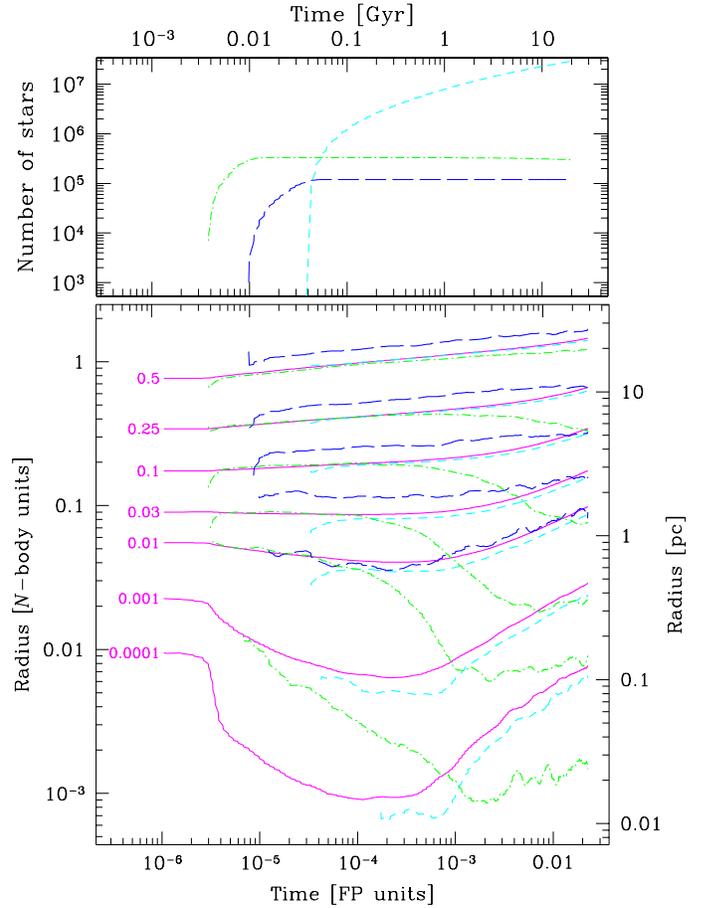}}
%
  \caption{Evolution of a model with stellar evolution and growth of
  the central MBH from a seed ({\tt GN78}). The top panel
  shows the total number of stellar remnants in the nucleus. The total number of stars
  is $\tento{2.13}{8}$. In the bottom panel, we show the evolution of
  the Lagrange radii for the various stellar species. A Lagrange
  sphere is specified by a fraction of the {\em instantaneous} total
  mass of stars of the corresponding species. The MBH grows from a
  seed of $\sim 1000\,\Msun$ to $\tento{3.95}{6}\,\Msun$ at
  $t=10\,$Gyr. Most of this increase comes from the accretion of a
  fraction $0.0653$ of the gas emitted by stellar evolution.}
  \label{fig:LagRadEvolStell}
\end{figure}

\begin{figure}
  \centerline{\resizebox{0.87\hsize}{!}{%
    \includegraphics[bb=37 157 566 692, clip]{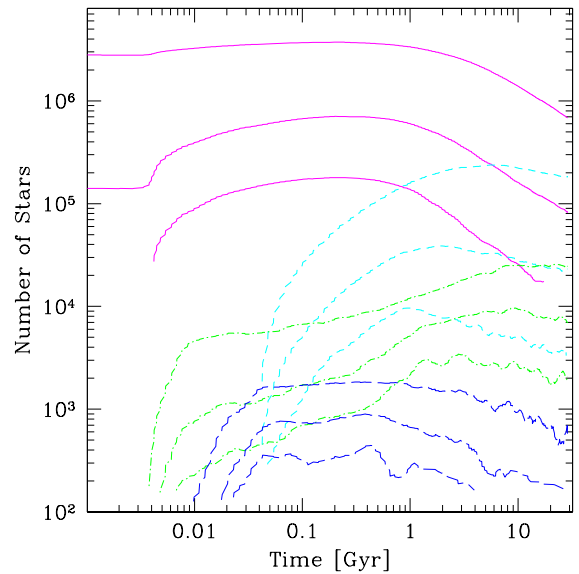}
    }}
  \caption{Number of stars of different types within $0.1$, $0.3$ and
  $1\,\pc$ from the center for model {\tt GN78}.}
  \label{fig:NobjCloseMBHEvolStell}
\end{figure}
So far we have focused on our standard {\SgrA} model. Except for mass
segregation, its initial conditions were set to reflect the state of
the MW nucleus {\em at the present epoch} in the sense that the
enclosed mass profile (interior to $\sim 3-5$\,pc) matches the
observational constrains and that the stellar population has a uniform
age of 10\,Gyr. Further stellar evolution was not included. Such a
model, chosen for its simplicity, is obviously not very realistic, not
even entirely self-consistent. In particular, during the $\sim 10$ Gyr
over which we allow dynamical evolution to proceed, the evolution of
stars with ZAMS mass above $\sim 1\,\Msun$ should be accounted for in
principle. Also, the MBH may have significantly grown during such a
long period. By considering a very different model, {\tt GN78}, {\em
tuned} to yield a {\SgrA}-like enclosed mass profile after 10\,Gyr of
evolution, we show that the conclusions about the mass-segregation
(and rates of interactions between stellar objects and the MBH, see
next subsection) are largely insensitive to our assumptions about the
past history of the nucleus, within the framework of the assumptions
common to both models (spherically symmetry, evolution in isolation,
etc.). This conclusion also applies to the other, less radical,
variations of the {\SgrA} model that we have considered but do not
discuss in detail. To help identify models that are applicable to the
{\SgrA} cluster, in Table~\ref{table:Mencl}, we indicate the enclosed
stellar mass $M_{\rm encl}(R)$ for $R=1$ and 3\,pc after 5 and 10 Gyr
of evolution. Observations indicate that, for {\SgrA}, $M_{\rm
encl}(1\,\pc)\simeq\tento{0.5-1}{6}\,\Msun$ and $M_{\rm
encl}(3\,\pc)\simeq\tento{0.5-1}{7}\,\Msun$ (see
Fig.~\ref{fig:MenclMW}). We note, that, because the density of
$\eta-$models decreases steeply for $R\ge R_{\rm b}$ and we cannot
afford large values of $R_{\rm b}$, lest the central resolution become
insufficient, it is difficult to put enough mass within 3\,pc of the
center.

Model {\tt GN78} is started as a cluster with $\eta=3$, i.e., no
initial central density cusp, containing a ``seed'' BH at its center,
$\MBH(0)=10^{-5}\Mcl(0)$ (because $\gamma=0$ but the velocity
dispersion is isotropic, the few particles initially in the influence
region are not in exact dynamical equilibrium). All stars are on the
ZAMS at $t=0$; as the simulation proceeds, they are turned into
remnants at the end of their MS lifetime, according to prescription F
of Table~\ref{table:remnants}. Natal kicks are imparted to NSs and
stellar BHs. We set $\Nstar=\tento{2.13}{8}$,
$\Mcl(0)=\tento{1.24}{8}\,\Msun$ and make the ad hoc assumption that
$6.53$\,\% of the gas emitted by stellar evolution is instantaneously
accreted my the MBH, in order to get, at $t=10\,Gyr$, $\MBH\simeq
\tento{3.5}{6}\,\Msun$ and $\MBH/\Mcl\simeq 0.05$, similar to the 
parameters of most other models. As tidal disruptions and coalescence
also contribute to the growth of the MBH, we obtain
$\MBH=\tento{3.95}{6}\,\Msun$ at $t=10\,$Gyr. Because the central
parts of the cluster strongly contract in the initial phase (see
below), we had to simulate clusters with different initial sizes to
find a value that yield a good fit to the observed enclosed mass
profile, namely $\Rnb(0)=16.2\,$pc.

We show the evolution of the structure of this model in
Fig.~\ref{fig:LagRadEvolStell} and plot in
Fig.~\ref{fig:NobjCloseMBHEvolStell} the number of stars in the
vicinity of the MBH. Nearly 90\,\% of neutron stars receive natal
kicks strong enough to escape from the nucleus. A strong and
relatively fast contraction of the inner regions starts at $t\simeq
3$\,Myr, which goes on, although at a much reduced rate until $t\simeq
100$\,Myr. This reflects the adiabatic contraction of the stellar
orbits, nearly unaffected by relaxation on such a short timescale, in
response to the growth of the MBH as it accretes the gas shed by
massive stars turning into BHs and NSs. At $t\simeq 10\,$Myr the MS
stars have formed inside $R=0.1\,$pc a profile compatible with the
cusp $\rho \propto R^{-2}$ predicted by the theory for an initial
distribution with $\eta=3$
\citep{QHS95,FB02b}. The later evolution of the nucleus is again 
dominated by relaxation. The system of BHs reaches its highest
concentration after some 2\,Gyr. After that time the structure and
evolution are essentially the same as that of the standard model.

Our assumption about the fraction of the mass
lost in stellar winds accreted by the MBH is ad-hoc. At early times it leads
to highly super-Eddington growth (see bottom panel of
Fig.~\ref{fig:dMdt}). It would be more physical to assume that the gas
accumulates at the center until the Eddington-fed MBH can accommodate
it but this would only introduce negligible changes in the results as
long as this central gas reservoir is seen as a point mass by the
stellar system
\citep{FB02b}. In any case, the fate of the interstellar gas in a 
galactic nucleus is a complex issue
\citep{DDC87a,DDC87b,CM97,CM99a,WBP99,CNSdM05}, well beyond the scope of this
study. Because the MBH acquires the bulk of its mass on a timescale
much shorter than the relaxation time but significantly longer than
the orbital time of the stars affected by its growth, the results of
our model apply to any situation of MBH growth respecting this
hierarchy of timescales, such as gas infall triggered by a galactic
merger \citep[e.g.,][]{BH91,BH96}.
\begin{figure}
\centerline{\includegraphics[scale=1.1]{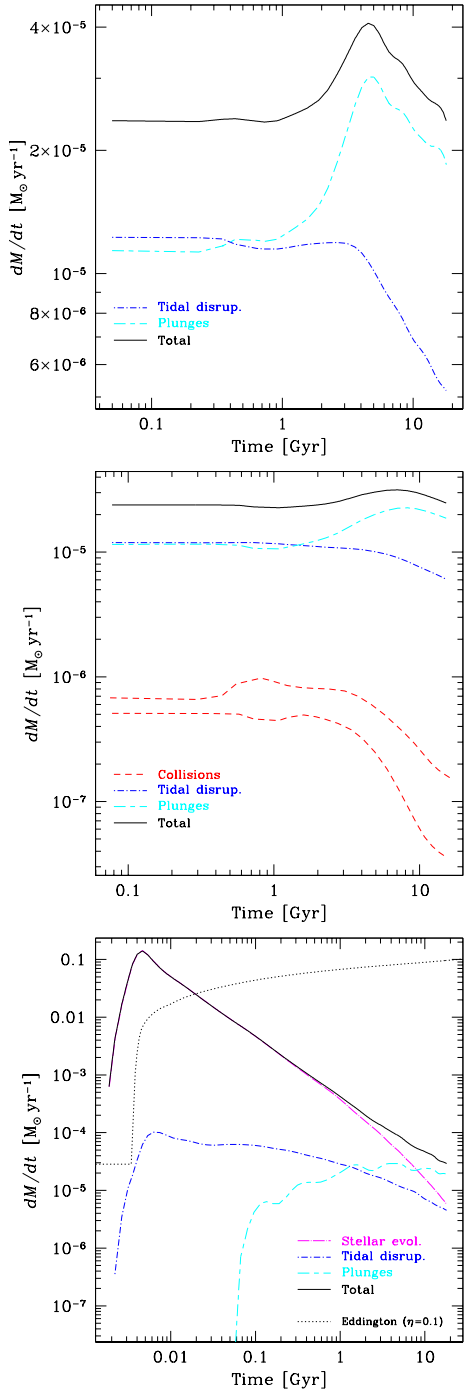}}
%
  \caption{Rate of accretion onto the MBH from various channels. The
  top panel is for our standard model {\tt GN25}. The middle panel
  is for simulations that include stellar collisions, {\tt GN45}
  and {\tt GN46}. The lower (thick) curve for the contribution of
  collisions corresponds to run {\tt GN45} in which the collisions between MS
  stars and compact objects were neglected; the upper (thin) curve is
  for run {\tt GN46} during which we assumed that such collisions always lead to
  complete destruction of the MS star. 100\,\% of the gas emitted in
  collisions is accreted by the MBH. The lower panel is for simulation
  {\tt GN78} which started with a seed BH ($\MBH\simeq 1000\,\Msun$) and
  stars on the ZAMS. Simple stellar evolution is included with a fixed
  fraction of the gas emitted when a MS star turns into a remnant
  being accreted by the MBH (see text). For this run, the
  Eddington accretion rate (with 10\,\% conversion factor) is also
  plotted but the growth of the MBH was not limited by it.}
  \label{fig:dMdt}
\end{figure}

\begin{figure}
\centerline{\includegraphics[scale=1.1]{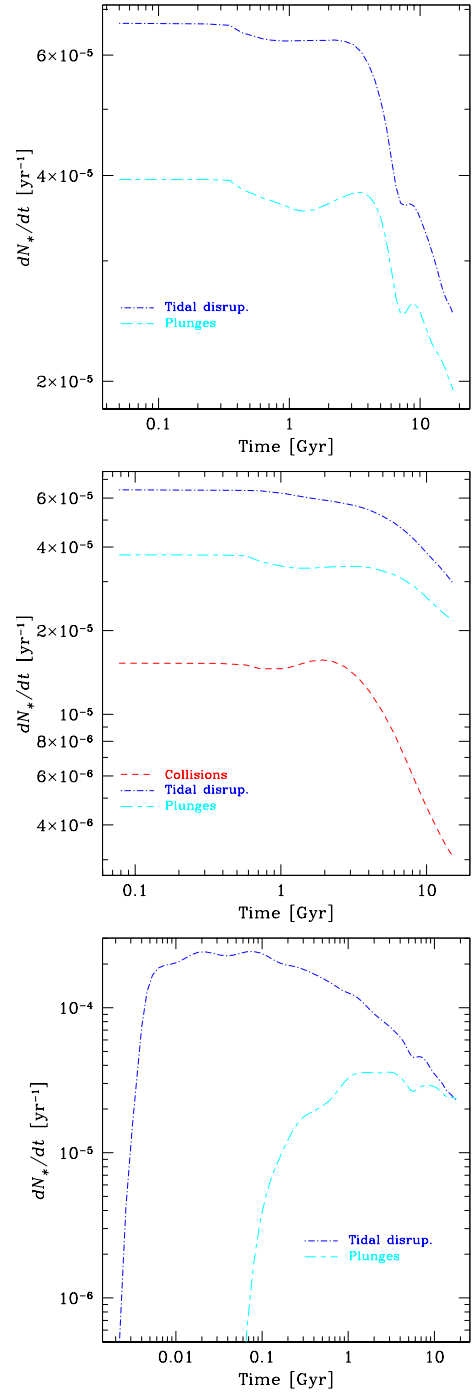}}
%
  \caption{Event rates for the same simulations as in Fig.~\ref{fig:dMdt}.}
  \label{fig:dNdt}
\end{figure}
To conclude the presentation of {\SgrA}-type models, we take a look at
the results for the rates of disruptive events. In
Fig.~\ref{fig:dMdt}, we plot the accretion rates onto the MBH, for our
standard, high-resolution model ({\tt GN25}), for two
lower-resolution models that include stellar collisions ({\tt GN45}
and {\tt GN46}) and for our ``alternative'' {\SgrA} run with
stellar evolution and progressive MBH growth ({\tt GN78}). We plot
the contributions of tidal disruptions (half of the mass of the star
is accreted), coalescences, collisions (100\,\% of the gas liberated
is accreted) and stellar evolution (for {\tt GN78}). The mass lost
in collisions between MS stars is determined from our SPH results
(\citealt{FB05,FRB05}; see \S~\ref{sec:MCcode}). As for collisions
between a MS star and a compact remnant, we considered two extreme
assumptions: either we neglected them altogether (but counted their
number), as in {\tt GN45}, or the MS star was considered to be
entirely destroyed in the process and the CR was left unaffected, as
in {\tt GN46} (in another run, we assumed half of the mass of the
MS star was accreted onto the CR).

In Fig.~\ref{fig:dNdt}, we show the  {\it number} rate of tidal
disruptions, coalescences and collisions for the same simulations.
Irrespective of the details of the models, we find that around
$t=10$\,Gyr, the tidal disruption rate is $|{dN}/{dt}|_{\rm td}\simeq
\tento{3-4}{-5}\,\peryr$, in good agreement with previous estimates 
for nuclei of similar structure \citep{Rees88,MT99,SU99}. Coalescences
are less frequent by some 20-30\,\% but dominate the mass accretion
rate owing to the important contribution of stellar BHs. In the models
without stellar evolution, mass segregation is responsible for the
significant increase in the coalescence rate taking place between
$t\simeq 2\,$Gyr and $t\simeq 6\,$Gyr. The decline in the rates at
later times is the consequence of the overall expansion of the
nucleus. The contribution of collisions to the growth of the MBH never
exceeds $10^{-6}\,\Msun\peryr$. At $t=10\,$Gyr, it is around
$\tento{5}{-7}\,\Msun\peryr$ if CR-MS collisions are neglected and
some 4 times higher if these events are disruptive. As we will see in
\S~\ref{subsec:role_las_coll}, collisions have also only little 
influence on the structure of the galactic nucleus, as far as it can be resolved by our simulations.


\subsubsection{Models for Nuclei of Different Masses}

\begin{figure}
  \resizebox{\hsize}{!}{%
    \includegraphics[bb=23 154 583 697, clip]{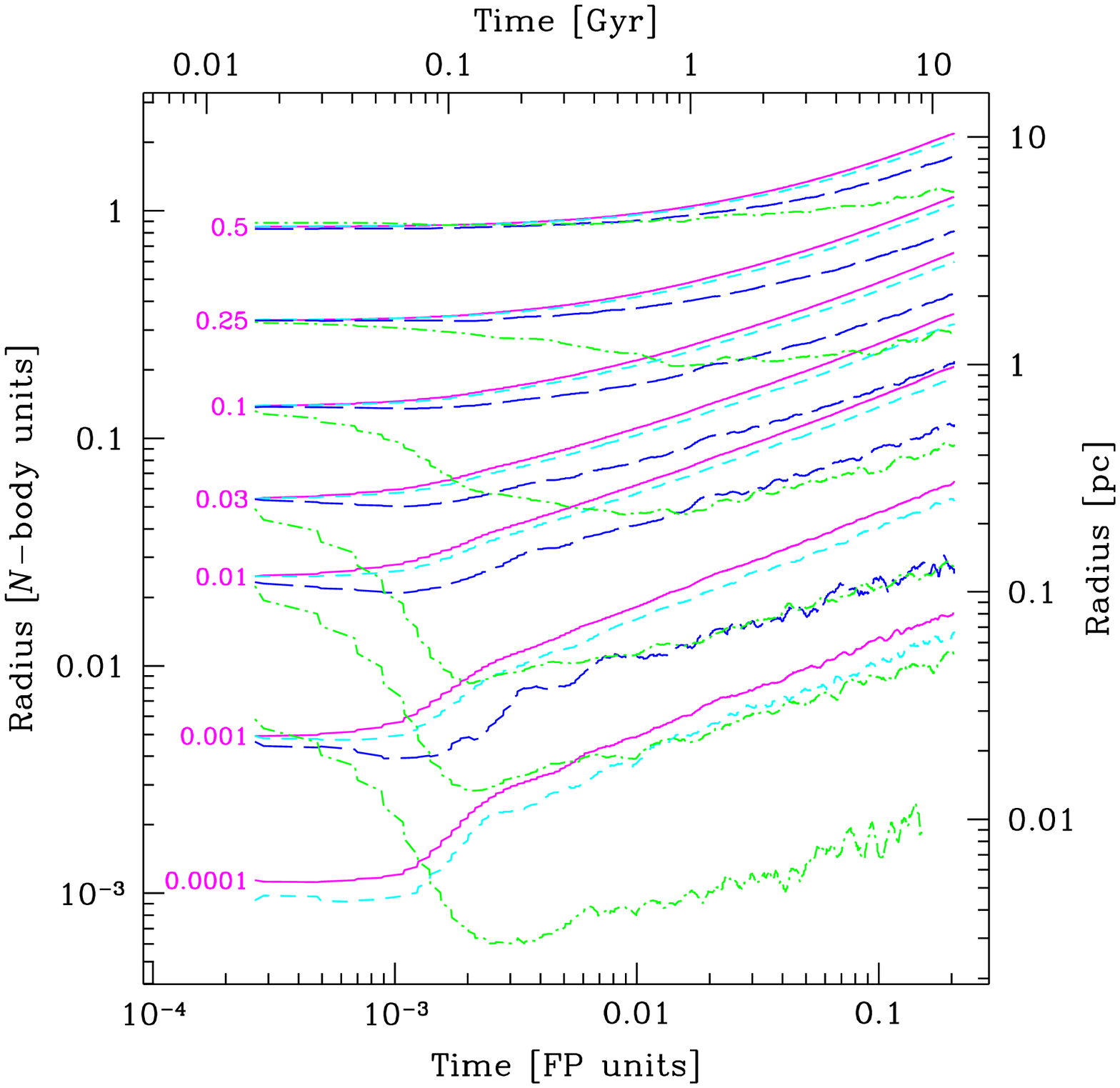}
    }
  \caption{Evolution of Lagrange radii for a small galactic nucleus.
  The initial conditions for this model ({\tt GN55}) are
  $\eta=1.5$, $\mu=0.05$, a stellar population of type F,
  $\MBH=10^5\,\Msun$ and $\Rnb=4.73\,\pc$.}
  \label{fig:LagRadSmallNucl}
\end{figure}

\begin{figure*}
  \centerline{\resizebox{0.87\hsize}{!}{%
    \includegraphics{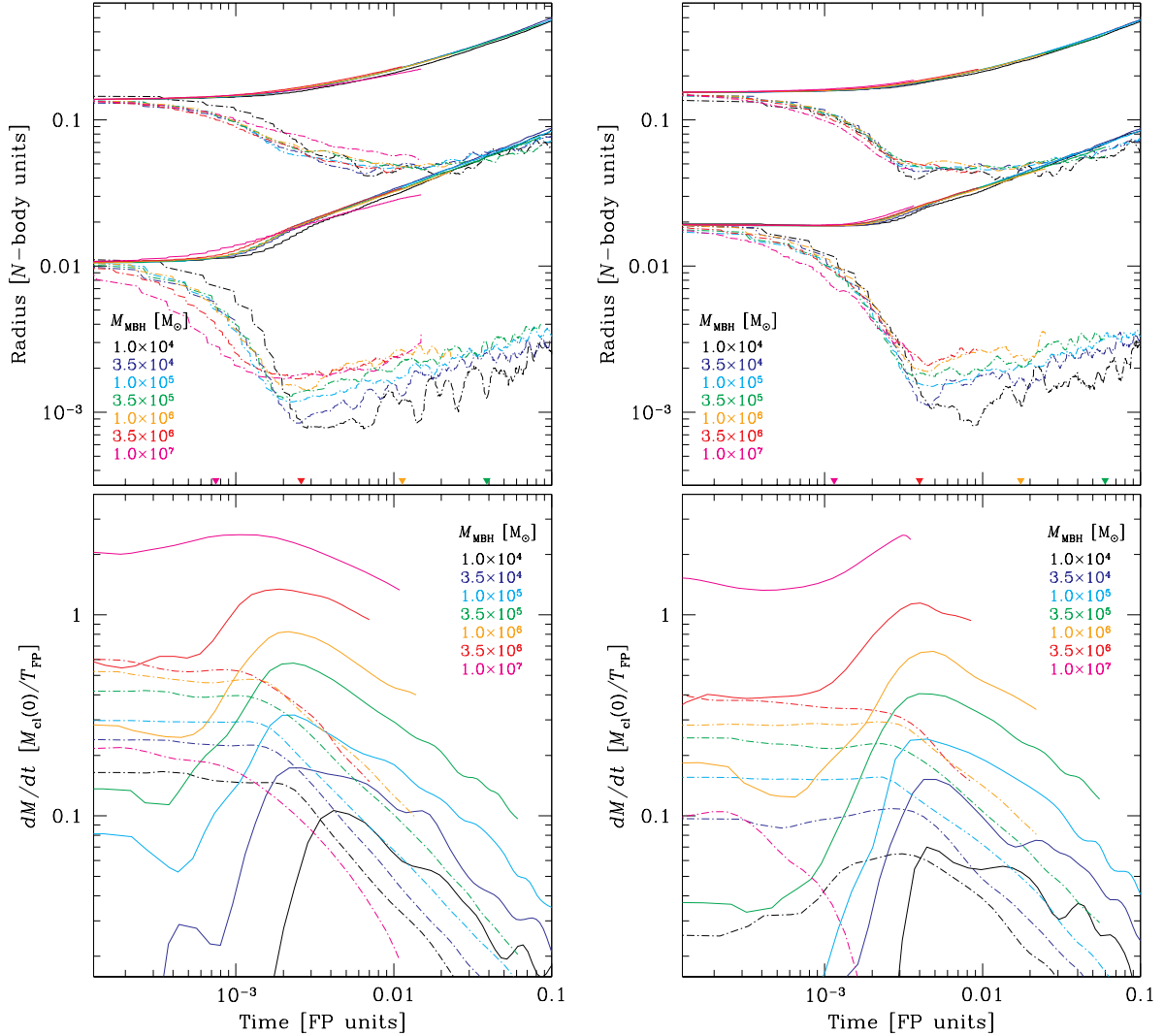}
    }}
  \caption{Evolution of galactic nuclei of different sizes. The left
  panels are for models with $\eta=1.5$, the right ones for
  $\eta=2.0$. For all models, $\mu\equiv\MBH(0)/\Mcl(0)=0.05$. We
  consider MBH masses ranging from $10^4$ and $10^7\,\Msun$ and scale
  the initial size of the cluster according to
  Eq.~\ref{eq:size_scaling}, i.e., $\Rnb \propto \sqrt{\MBH}$. In the
  top panels, we show the evolution of the Lagrange radii of
  fractional masses 0.1 and 0.003 for MS stars (solid lines) and
  stellar BHs (dot-dash). The triangles on the lower x-axis indicate 
  10\,Gyr for the models in which this corresponds to less than 0.1
  in Fokker-Planck time units.
  In the bottom panels, we plot the accretion
  rate onto the MBH. Solid lines indicate the contribution of
  coalescences and the dot-dashed lines that of tidal disruptions
  (with 50\,\% of the mass of each disrupted star beeing accreted).\newline
The models with $\MBH=10^4\,\Msun$
  ($\tento{3.5}{4}\,\Msun$) is made up of only
  $\Npart=\Nstar=\tento{6.1}{5}$ ($\tento{2.13}{6}$) particles and the
  0.003 Lagrange radius for BHs is determined with 3 (10) particles
  only, hence the large-amplitude oscillations. We used $\tento{4}{6}$
  ($<\Nstar$) particles for all other simulations plotted here.}
  \label{fig:comp_models}
\end{figure*}

\begin{figure}
  \resizebox{\hsize}{!}{%
    \includegraphics[bb=31 146 567 692,clip]{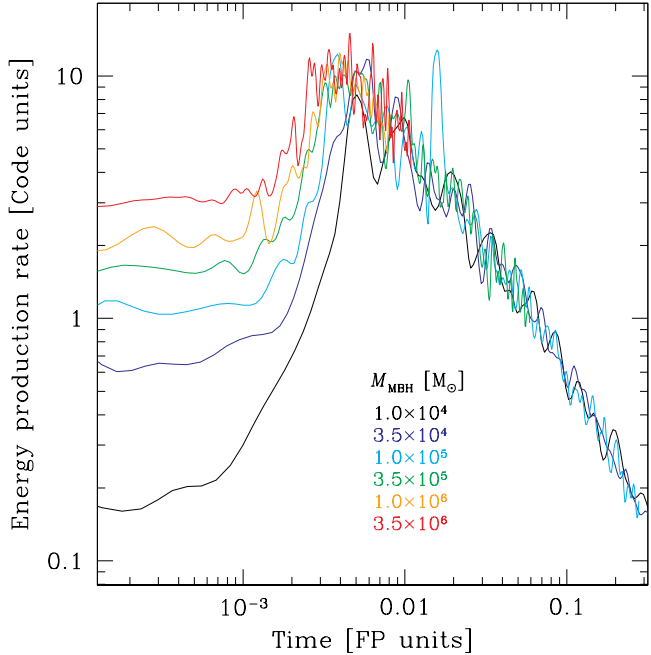}
    }
  \caption{Comparison between the rates of energy production due to
  tidal disruptions and coalescences with the MBH for nucleus models
  with $\eta=2.0$. The data is for same models as in the right panels
  of Fig.~\ref{fig:comp_models}. The units for the energy production
  rate is the $N-$body energy unit per Fokker-Planck time, i.e.,
  $G\Mcl^2\Rnb^{-1}\Tfp^{-1}$. The oscillations in the curves are
  essentially numerical noise. The different models show very similar
  energy production rates in the expansion phase.}
  \label{fig:comp_models_dEdt}
\end{figure}

We now explore how nuclei less or more massive than our standard
{\SgrA} case evolve. We recall that, for a given $\eta$ value, we
restrict ourselves to a one-parameter family of models by keeping
$\mu\equiv\MBH/\Mcl$ fixed and imposing $\Rnb \propto \sqrt{\MBH}$, a
scaling is inspired by the $M-\sigma$ relation (see
\S~\ref{sec:ini_nucl_models}). Hence, the model is specified by $\eta$ 
and $\MBH(0)$. We have considered $\MBH(0)$ values ranging from $10^4$
to $10^7\,\Msun$ and two values of $\eta$: 1.5 and 2.0.

We show the evolution of the model with $\MBH(0)=10^5\,\Msun$ and
$\eta=1.5$ (model {\tt GN55}) in
Fig.~\ref{fig:LagRadSmallNucl}. The most obvious feature is a faster
evolution, when measured in years, than the {\SgrA} nucleus, which
simply reflects a shorter relaxation time. This results in significant
expansion of the cluster central parts over a Hubble time. For
instance, the half-mass radius which showed hardly any change in the
{\SgrA} case, expands by a factor of about 2. In the models with
$\MBH(0)=10^4\,\Msun$, the whole cluster is expanding at $t=10\,$Gyr,
with the half-mass radius of the $\eta=1.5$ model increasing form 1.3
to 15.3\,pc. The expansion proceeds like $R(t)\propto t^\beta$ with
$\beta\simeq 2/3$ at large radii, as predicted for the self-similar
expansion powered by a central energy source
\citep{Henon65,Shapiro77,McMLC81,Goodman84, ASFS04}. The central parts appear 
to expand slower with $\beta\simeq 1/2$, a relation not yet
explicitly reported in the literature, to the best of our knowledge,
but consistent with the results of recent single-mass simulations with
the gas-dynamical model of cluster dynamics \citep{ASFS04} and
{\NBFOUR}
\citep{BME04a}. Our experimentations with the gas model indicate that this 
is a loss-cone effect and that the transition from $\beta\simeq 1/2$
to $2/3$ occurs around the critical radius. When tidal disruptions
through loss-cone diffusion are prevented (i.e., stars are
destroyed only when their energy becomes smaller than $-G\MBH/\Rtd$),
the whole cluster expands like $R(t)\propto t^{2/3}$.

A straightforward consequence of this strong expansion of small nuclei
is that one would have to start with initial conditions much more
compact to recover our assumed $\Rnb\propto \MBH^{1/2}$ relation
between different nuclei in the present-day universe. However, this
relation results from a naive application of the $M-\sigma$
relation. Observationally, $\sigma$ is a luminosity-weighted value
integrated over the light-of-sight and averaged over an aperture
covering a region much larger than the one dynamically influenced by
the MBH or by relaxation \citep{TremaineEtAl02}. In fact, in no other
case than the Milky Way is the region affected by relaxation actually
resolved by observations \citep{Merritt05}. Furthermore, the
$M-\sigma$ relation for $\MBH\lesssim 10^7\,\MBH$ is poorly
constrained anyway. In future studies of the evolution of
small galactic nuclei, a larger variety of models should considered by
allowing initial stellar clusters more and less dense than in our
single-parameter families.

In Fig.~\ref{fig:comp_models} we make a direct comparison between the
models of different masses. We plot the evolution of the 0.3 and
10\,\% Lagrange radii for MS stars and stellars BHs and accretion
rates onto the central MBH (through tidal disruptions and
coalescences). As we have seen, the distributions of WDs and NSs
evolve similarly as that of the MS stars; they are only slightly more
concentrated towards the center. For these plots, we have used
``natural'' units to stress the similarities between the various
models. Time is expressed in $\Tfp$, radius in $\Rnb$ and mass in
$\Mcl(0)$. From the Lagrange radii evolution, one sees that models of
different masses show a very similar structure at the same value of
$t/\Tfp$, which reflects the fact that evolution is driven by
relaxation. Significant differences are only visible at small
radii. They are the consequences of the ``central boundary
conditions'' imposed by tidal disruptions and coalescences. Unlike
relaxation, these processes introduce physical length scales in the
system: $\Rtd$ and $\RS$. The structure can only be independent
of the size and mass of the model at distances larger than the
corresponding critical radii.

For the $\eta=1.5$ series, the 0.3\,\% radius of the BHs contracts
slightly faster at early times for more massive, larger nuclei with
$\MBH\le\tento{3.5}{6}\,\Msun$. This seems to be the consequence of a
bigger growth of the central MBH in the early evolution phase during
which the stellar BHs segregate to the center ($t\lesssim
\tento{(1-3)}{-3}\,\Tfp$). In natural units, when the mass of the system 
is increased, the dynamical time at a given radius decreases. For a
fixed aperture of the loss cone, this would yield a higher accretion
rate in the full-loss-cone regime (at large radii) and a larger
critical radius while the empty loss-cone rate is unchanged. In
fact, as we use a $R\propto M^{1/2}$ scaling, the size of the loss
cone, at a fixed $R/\Rnb$ value, varies approximately like $\ThLC^2
\propto \MBH^{-1/6}$ for tidal disruptions ($\Rtd\propto \MBH^{1/3}$)
and like $\ThLC^2 \propto
\MBH^{1/2}$ for coalescences ($\Rplunge\propto \MBH$). All this indicates 
that the coalescence rate should increase with $\MBH$ in our families
of models, as indeed is the case. The situation for tidal disruptions
is more complicated, also because, as $\MBH$ increases, a larger and
larger fraction of MS stars are compact enough to withstand tidal
forces down to the last stable orbit around the MBH. For instance,
with $\MBH=10^7\,\Msun$, only MS stars more massive than $\approx
0.6\,\Msun$ can be tidally disrupted on non-plunge orbits, i.e., have
$\Rtd>\Rplunge$.

The ``segregation phase'' ends earlier and the concentration of
stellar BHs is less pronounced in the massive nuclei. This is likely a
result of the larger critical radius which yields an approximately
equal energy production rate (in ``natural'' units such as $N-$body
energy unit per $\Tfp$) for a lower stellar density in the inner
regions. At the same time a larger critical radius explains the larger
accretion rate, as stars are absorbed by the MBH when they are on
wider orbits. To drift from large distances to these orbits the
accreted stars have to dissipate less orbital energy and contribute
less heating towards the stellar system. In
Fig.~\ref{fig:comp_models_dEdt}, we verify that the energy production
rate through tidal disruptions and coalescences with the MBH is indeed
nearly the same for the different models with $\eta=2.0$ during the
expansion phase. As first realized by \citet{Henon75}, during the
gravothermal expansion of a cluster the conditions in the central
regions where energy is produced are indirectly controlled by the
large-scale structure. The latter determines how much energy is
transported outwards by 2-body relaxation to drive the expansion; this
``luminosity'' has to be balanced by central energy production through
disruptions and coalescences, in way similar to 
hardening of binaries which powers the post-collapse expansion of globular
clusters \citep[][ amongst others]{Shapiro77,ILB83,Heggie84,HH03}.

\subsubsection{Role of large-angle scatterings and collisions}
\label{subsec:role_las_coll}
\begin{figure}
\centerline{\includegraphics[scale=.9]{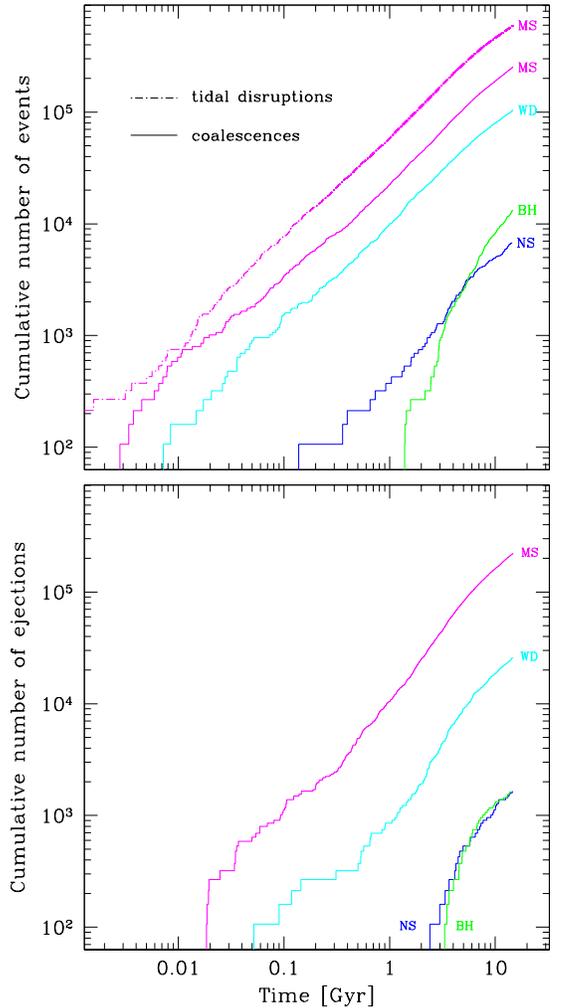}}
  \caption{Cumulative number of events in simulation {\tt GN54}. In
  the top panel, we show the numbers of tidal disruptions of MS stars
  and coalescences (i.e., plunges through the horizon of the MBH) for
  all stellar species. On the bottom panel, we plot the numbers of
  ejections from the nucleus, due to large-angle scatterings. Note
  that the number of ejections is always significantly smaller than
  that of coalescences with the MBH.}
  \label{fig:event_numbers}
\end{figure}
The discussion of direct collisions and large angle scatterings in
galactic nuclei would warrant another, specific paper. Here we only
consider the overall effects of these ``strong encounters'' between
stars on the structure and evolution of galactic nuclei.

In a few of our models, large-angle scatterings where introduced to
test whether a significant number of stellar objects are ejected from
the cusp through this mechanism. In Fig.~\ref{fig:event_numbers}, we
compare the number of ejections to that of coalescences and tidal
disruptions. For all stellar types, the number of ejections turn out
to be smaller by a factor of a few. In particular, stellar BHs are
nearly 10 times more likely to merge with the MBH than to be
ejected. \citet{LT80} argued that it was some 30-70\,\% as likely for
a cusp star to be ejected from the cluster as to be absorbed by the
MBH. Our results do not confirm this estimate but the nuclei we
consider are very different from the globular cluster model adopted by
\citeauthor{LT80}, in which the cusp is embedded in a large
constant-density stellar core. These authors also argue that the
probability of ejection from the core (as opposed to from the cluster)
is $3-10$ times that of absorption, so there is a possibility that
large-angle scatterings would reduce the rate of coalescences and/or
disruptions significantly even though they do not lead to numerous
ejections. For our {\SgrA}-type model, we find the number of BH-MBH
coalescences to be reduced by some $35-40$\,\% by the effects of
large-angle scatterings. Other numbers of events are much less
affected. We find no appreciable influence on the density profiles
around the MBH at $t\approx 10$\,Gyr. 

Collisions between stars are also found to have very little, if any,
impact on our results concerning mass segregation and rates of
coalescence and tidal disruptions. In most runs with collisions, we
only considered collisions between MS stars. Encounters
featuring one or two COs were counted but the mass or
trajectories of the stars were left unchanged. For a {\SgrA}-type
model, we find of order $\tento{6}{4}$ MS-MS collisions in 10\,Gyr
(the number of collisions between {\it particles} in the simulation is
smaller by a factor $\Nstar/\Npart=53.3$). The numbers of MS-WD, MS-NS
and MS-BH for the same period are about $\tento{3}{5}$, $\tento{2}{5}$
and $\tento{2-4}{4}$ respectively. Not surprisingly, collisions between
compact stars are extremely rare and their number is therefore very
uncertain given the resolution of our simulations. We find a number of
WD-NS and WD-BH events of the order of 100-1000 (corresponding to only
a few particle-particle encounters) and no collisions between 2
compact stars.

Despite the high velocities reached
in the central regions (the median value of the relative velocity at
''infinity'' for colliding stars is about $500\,\kms$), these
encounters are not highly disruptive. Collisions very rarely result in
mergers but the complete destruction of a MS star requires of order
20-30 collisions if encounters with CRs are neglected; on average only
about 4\,\% of the stellar mass is lost when two MS stars collide (see also
\citealt{FGR04c}). We are not able to 
detect any significant effect of collisions on the central density
profile of MS stars, down to a few $10^{-3}\,\pc$ of the MBH, even
when we assume that collisions with a CR result in the complete
disruption of the MS star. This is somewhat surprising in view of
estimates of the collision time such as done in
Fig.~\ref{fig:TscalesMW}, which suggest that inside $\sim 0.01\,\pc$
of the MBH, a MS star should suffer from $\sim 10$ collisions in
10\,Gyr. Strong central collisional depletion has been found by
\citet{MCD91} in their Fokker-Planck simulations but the nucleus models 
for which this effect was reported were much more collisional by
construction than those used here. To determine whether collisions can
have an observable effect on the stellar distribution at the Galactic
center, simulations with a much higher resolution are probably required.

\section{Summary and Outlook}
\label{sec:concl}
\subsection{Summary of simulations and results}

In this paper we report on our stellar dynamical simulations of
multi-mass models of galactic nuclei. Our main goal was to investigate
how stellar objects of different masses distribute themselves around a
central massive black hole (MBH), in response to relaxation, a process
known as mass segregation. 

This work is based on the use of a Monte-Carlo code which allows to
follow the secular evolution of spherical stellar cluster in dynamical
equilibrium over Gyrs. We performed about 90 different simulations, to
investigate the effects of various physical ingredients, assumptions
about their treatment, of the initial nucleus structure and to perform
some limited parameter-space exploration. For most models,
$\tento{4}{6}$ particles where used, requiring a few days of computing
time on a single-CPU PC. A few cases were computed with $10^6$ or
$\tento{8}{6}$ particles to establish that our results are not
strongly affected by the limitations in numerical resolution. The
latter number is close to the maximum number of particles that can fit
in the memory of standard 32-bit Linux PC.

All runs included the effects of the gravity of a central MBH, of the
self gravity of the stars, of 2-body relaxation, treated in the
Chandrasekhar (diffusive) approximation, and of the tidal disruption
of MS stars at the Roche limit around the MBH as well as direct
coalescence with the MBH for stars too compact to be tidally
disrupted. In most cases, stellar evolution was not included
explicitly; instead the stellar population consists, from the
beginning of the simulation of a mixture of MS and compact remnants
corresponding to a single star formation episode that took place
10\,Gyr ago. In a few models, explicit stellar evolution was included
with all stars starting on the MS and turning into compact remnants
(CRs) at the end of their MS life time. For simplicity, giants were
not considered because, as far as mass segregation is concerned, only
the mass of the star matters and the evolution of the stellar
distribution, being a relaxational process requires timescales much
longer than the duration of the giant phase. Stellar collisions and
large-angle gravitational deflections (not accounted for in the
diffusive treatment of relaxation) were considered in a small number
of models. We made no attempt to determine whether a given star-MBH
coalescence would occur as a gradual inspiral or a direct plunge
through the horizon of the MBH. This question, of central importance
for future low-frequency gravitational wave detectors such as LISA or
BBO, has to be addressed in future work with more appropriate
numerical methods.

All our runs are started as $\eta$-models. They have a central
power-law density cusp, $\rho \propto R^{\eta-3}$ and steeper
``cut-off'' at large radii, $\rho \propto R^{-4}$. In most cases, we
used parameters (mass of the MBH, stellar density around it, etc.)
corresponding to the stellar cluster around {\SgrA} at the center of
our Galaxy. We did not try to reproduce the very peculiar spatial and
age distribution of the bright IR stars observed within 1\,pc of
{\SgrA}. In this work, we adopt the position that these stars, useful
as they are as probes of the gravitational potential, are not
representative of the overall stellar population at the Galactic
center, assumed to be much older and therefore amenable to our
treatment. The usefulness of this assumption is that it defines a well-posed problem which consists an interesting limiting case. Clearly other situations have to be considered in future studies. For instance, an exciting
scenario, in sharp opposition to our simplifying assumptions, is that
the {\SgrA} cluster and its central MBH have been formed progressively
by the infall of rich stellar clusters, some of them containing
intermediate-mass black holes \citep{HM03,KFM04,GR05}. In this picture, which attempts to explain the existence of the very young, unrelaxed stellar populations and assumes the present epoch is not a special one, a stellar cluster should inspiral into the Galactic center every few Myr (but see \citealt{NS05} for arguments opposing this view and suggesting the young massive stars have formed {\it in situ} in an accretion disk). Such infalls would build up a mixed-age stellar population and reshuffle the orbits of stars already present in the nucleus quite significantly, and therefore yield a different
mass-segregation structure.

Based on our ``standard'' {\SgrA} models and a somewhat naive
application of the $M-\sigma$ relation, we have considered two
families of galactic nucleus models of different masses, for $\MBH$ in
the range $10^4-10^7\,\Msun$. One family has $\eta=2.0$, the other
$\eta=1.5$. The interval in $\MBH$ was chosen mostly to cover the
values that should yield gravitational wave signals in LISA band when
a compact star inspiral into the MBH. We embarked in the present study
as a first step towards more robust determinations of the rate and
characteristic of such extreme-mass ratio inspirals (EMRIs). This range
of models also covers systems that are both large enough (in terms of
the number of stars) to be amenable to treatment with the Monte-Carlo
method and small enough for relaxational effects to play a significant
role over some 10\,Gyr.

To ensure that the Monte-Carlo code, based as it is on a number of simplifying assumptions, yield correct results, we performed a number of comparisons with simulations performed with highly accurate (but much more computationally demanding) direct-summation {\NBFOUR} code. In particular, we performed a new {\NBFOUR} simulation of a two-component
model with a central massive object using 64\,000 particles. On the
GRAPE hardware at disposal not more than $\sim 10^4$ particles can be
used; hence it is not yet possible to simulate a system with a
realistic mass function using direct $N-$body but this 2-component toy
model demonstrated for the first time in a direct fashion that the MC
code treats mass segregation around a MBH very satisfactorily.

For the {\SgrA} models, our main results are the following. In all
cases, the stellar BHs, being the most massive objects (with a fixed
mass of $10\,\Msun$ or a range of masses, depending on the model),
segregate to the central regions. This segregation takes about 5\,Gyr
to complete. The nucleus then enters a second evolutionary phase which
is characterized by the overall expansion of the central regions,
powered by the accretion of stellar mass (of very negative energy)
onto the MBH. Although all species participate in the expansion, mass
segregation continues in a relative fashion, as the system of BHs
expands slower than the other components. The structure of the nucleus
at distances from the center larger than a few pc is left unaffected
by relaxation over a Hubble time.

BHs dominate the mass density within $\sim 0.2\,\pc$ of
the MBH but we do not find them to be more numerous than MS stars in
any region we can resolve (down to a few mpc, at $t=10\,$Gyr). Estimating the exponent for the density cusp the BHs form, $\rho
\propto R^{-\gamma}$, is difficult because of numerical noise, but, 
in most cases, $\gamma$ is compatible with the Bahcall-Wolf value
$\gamma=1.75$. In contrast, the less massive objects, such as MS
stars, form a cusp with $\gamma$ generally in the range $1.3-1.4$
which is significantly lower than the value of $1.5$ predicted by
\citet{BW77}. This is also found in the 2-component $N-$body
simulation. After 5-10\,Gyr of evolution,
we find of order $\tento{2-3}{3}$, $\tento{6-8}{3}$ and $\tento{2}{4}$
stellar BHs within 0.1, 0.3 and 1\,pc of the center,
respectively. About $10^4$ BHs coalesce with the MBH during a Hubble
time. Using the formalism of the dynamical friction for objects
on circular orbits in a fixed stellar background is an easy
alternative for estimating the concentration of massive objects in the
central regions. However, for stellar BHs, although this approach
offers a qualitatively correct picture, it overpredicts the
effectiveness of mass segregation. In the case of a model with
$\eta=1.5$ (whose relaxation time does not increase towards the
center), this yields too large a number of BHs accreted by the MBH
and {\it too few} being present within the inner 1\,pc after some
10\,Gyr.

All types of objects lighter than the BHs, including the NSs, are
pushed away from the central regions. Using the observed distribution
of these object to infer the presence of segregated BHs does not seem
to be possible though because, in the absence of BHs, it would take
the NSs more than 10\,Gyr to form a Bahcall-Cusp of their own, even
they would not receive any natal kick.

These results are not significantly affected by stellar collisions,
large-angle scatterings, the initial $\eta$ value. We also considered
three different prescriptions for the masses (and types) of compact
remnants and found no strong variations in the simulation
outcomes. Most interestingly, an alternative model in which stellar
evolution was included and the central MBH was grown from an IMBH seed
(by accretion of an {\it ad hoc} fraction of the mass lost by stars
when they turn into COs) yields basically the same structure of mass
segregation (and same rates of coalescences end tidal disruptions) at
$t\simeq 10\,$Gyr. These findings suggest that our main results are
not very sensitive to the special ``initial conditions'' used, as long
as they are fine-tuned to produce at $t=10\,Gyr$ a given MBH mass and
stellar mass within $\sim 1\,\pc$ of the MBH. However, it would be
instructive to consider a larger variety of models in future work,
including some with extended period of stellar formation. Our present
assumption of a single burst of stellar formation maximizes the number
fraction of stellar BHs and the time available for mass segregation.

When large-angle scatterings (not accounted for in the standard
diffusive treatment of relaxation) are explicitly included
(essentially as a special case of collisions), they are found to
have only little impact on the rate of tidal disruptions or
coalescences with the MBH. A stellar BH is about 10 times more likely
to by swallowed by the MBH than to be ejected from the nucleus. In
contrast with this, in their multi-mass $N-$body simulations,
\citet{BME04b} find that all stellar BHs except one are ejected from
the cluster and ascribe this result to strong interactions with
objects (generally another stellar BH) deeply bound to the IMBH. These
interactions are likely to be ``resonant'', i.e., the three objects
(including the IMBH) form a strongly interacting, chaotic
configuration for many orbital time until one of the lighter objects
is ejected (Baumgardt, personal communication; see e.g.,
\citealt{Hut93}). In principle, this mechanism can be included into 
the MC code by extending the loss-cone treatment used for tidal
disruptions and coalescences to interactions with the binary
consisting of the MBH and the most bound stellar object and resorting
to explicit integration of 3-body motion when a close interaction
between the binary and a third object is deemed to occur. 

However this process is probably of little importance in galactic
nuclei as a rough analysis suggests. Let's write $\Sigma_{\rm ej}$ and
$\Sigma_{\rm pl}$ to denote the cross sections for strong interaction
with the innermost stellar object (followed by ejection from the
nucleus) and for direct plunge through the horizon of the MBH,
respectively. Then assuming that the scaling laws for three-body
interactions established by
\citet{HHMcM96} apply all the way to mass ratios as extreme as considered 
here, we estimate $\Sigma_{\rm ej}/\Sigma_{\rm pl} \approx
(a/\RS)(m/\MBH)$ where $a$ is the semimajor axis of the stellar object
deeply bound to the MBH, $\RS$ is the Schwarzschild radius of the MBH
and $m$ is typical mass of stellar objects. Now, for the interaction
to be resonant, the inner binary has to be well separated from the
other objects in the cusp, $a<R_{1\ast}$ where $R_{1\ast}$ is the
radius containing (on average) one stellar object. Assuming a
power-law density cusp $n\propto R^{-\gamma}$ inside the influence
radius $\Rinfl$ of the MBH, one finds $R_{1\ast} \approx \Rinfl
(m/\MBH)^{1/(3-\gamma)}$. Therefore if the stellar velocity dispersion
outside $\Rinfl$ is $\sigma$, one finds $\Sigma_{\rm ej}/\Sigma_{\rm pl}
\approx (c/\sigma)^2(m/\MBH)^{(4-\gamma)/(3-\gamma)}$. For a $\sigma=20\,\kms$ 
globular cluster containing an $10^3\,\Msun$ IMBH with $\gamma=1.5$,
this ratio is of order $10^5$. But this is reduced to $10^{-2}$ for a
galactic nucleus with $\MBH=10^6\,\Msun$ and $\sigma=200\,\kms$.

%
\begin{figure}
  \centerline{\resizebox{0.87\hsize}{!}{%
  \includegraphics[bb=19 144 571 690, clip]{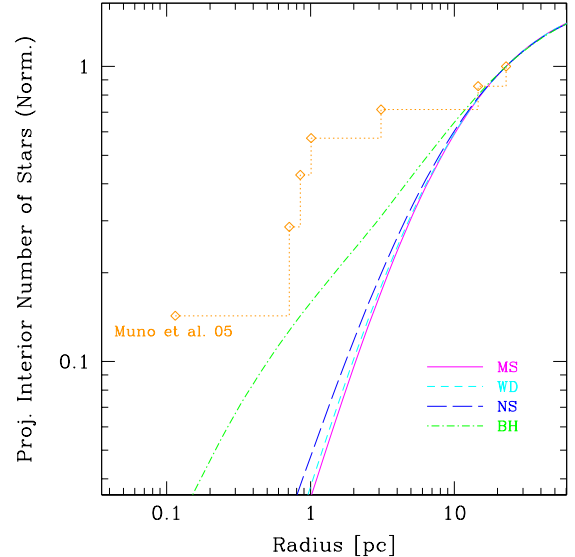}}}
  \caption{Comparison between the distribution of transient X-ray
  sources found by \citet{MunoEtAl05} at the Galactic center and the
  results of one of our simulations ({\tt GN25}, at
  $t=9.19\,$Gyr). The observational data are represented by diamonds
  connected by dotted lines. The smooth curves, one for each stellar
  species are the simulation data. Plotted is the number of sources
  whose sky position projects within a given distance of the center of
  the nucleus. This number is normalized to 1 at $R_{\rm
  norm}=23\,\pc$. A distance to the Galactic center of 8\,kpc has been
  assumed. The 7 seven transients are more concentrated around {\SgrA}
  than any stellar component in the simulation. See text for an
  assessment of the statistical significance of this result.}
  \label{fig:Muno05}
\end{figure}
\subsection{Astrophysical applications, including future work}

Although we have not attempted a realistic modeling of the Galactic
center, it is tempting to apply our results to one specific
observation of the {\SgrA} region. Using the {\em Chandra} X-ray
space-born observatory, \citet{MunoEtAl05} have detected 7 transient
sources which appear to be much more concentrated around {\SgrA} than
the overall stellar population. Here we examine whether this may be a
direct consequence of mass segregation, if these sources are all
stellar BHs accreting from a lower-mass companion. We make the strong
assumption that these binaries are not formed or affected by
interactions with other stars such as 3-body binary formation, partner
exchange, ionization, etc. Instead we consider they just react to
2-body relaxation as point objects with a total mass approximated by
the mass of the stellar BH.

In Fig.~\ref{fig:Muno05} we perform a graphical comparison between the
observed distribution of X-ray transients and the distribution of the
various species, most importantly the BHs, in our high-resolution
simulation {\tt GN25} at $t=9.29\,$Gyr. Clearly, the transients are
more centrally concentrated than the BHs in the simulation but given
the small number of observed sources the plot itself is not sufficient
to rule out our naive model for their distribution. If we pick up at
random 7 sources with projected distance from the center smaller or
equal to 23\,pc following our ``theoretical'' BH profile, we find that
their distribution is at least as concentrated as the observed one (in
the sense of the Kolmogorov-Smirnov test) in some 15\,\% of the
cases. It is therefore at this point not possible to exclude that the
transients owe their peaked profile purely to mass segregation but
this seems somewhat unlikely.

As pointed out by \citet{MunoEtAl05}, the rate of binary interactions
should also increase steeply towards the center and this probably
combines with mass segregation to produce the observed
distribution. In comparison with the situation in a globular cluster
with a well-defined core velocity distribution, the problem of binary
dynamics in the vicinity of a MBH is complicated by the fact that
there is no clear-cut definition of the hard-soft transition. The
Keplerian velocity dispersion increases virtually without bound when
one approaches the center. This may affect a binary on an orbit of
relatively large semimajor axis $a$ around the MBH because 2-body
relaxation will cause the orbit to reach down to a value $\Rperi =
(1-e)a \ll a$ over a timescale of order $\trlx\ln(1/(1-e))$
\citep[e.g.,][]{FR76}. Therefore, a binary may be disrupted even if it
is hard relative to the local velocity dispersion at the position
where it spends most of its time, i.e., at distances of order $a$ from
the MBH. 

The most extreme type of dynamical interaction a binary can experience
is the tidal separation of its members if its orbit brings it within
$\sim a_{\rm bin}(\MBH/m_{\rm bin})^{1/3}$ of the MBH. Here $a_{\rm
bin}$ is the semimajor axis of the binary itself and $m_{\rm bin}$ its
mass. This process is of great interest by itself both as a way to
create ``hypervelocity stars'' and to deposit a star on a tight orbit
around the MBH \citep{Hills88,YT03,GPZS05,MFHL05,Pfahl05}.

Given a model of the stellar distribution, one could estimate an
average lifetime for a binary of known properties and semimajor axis
(assumed fixed), accounting for the low$-\Rperi$ excursions caused by
relaxation. However the complex question of binary dynamics in a
galactic center would be better treated through self-consistent
stellar dynamical simulations of the sort presented here but including
binary processes. H\'enon-style Monte-Carlo code are particularly well
suited for following the evolution of large systems with a significant
fraction of binaries whose interaction can be computed accurately by
direct $3-$ and $4-$body integration, as has already been realized in
the context of young and globular clusters \citep{GS03,FGR06,GFR06}.

Concerning the prediction of EMRI rates and properties, the
determination of how 2-body relaxation shapes the stellar distribution
around the MBH is only a first --crucial-- step. A robust estimate of
the fraction of stars that eventually inspiral into LISA band, rather
than plunge directly through the horizon while still on a wide orbit
will probably require the development of a specific code. For stars on
very eccentric orbits, one needs to follow the combined effects of GW
emission and relaxation on a timescale significantly shorter than
allowed by the present {\MESSY} code where the time steps are a
function of the distance from the center and cannot depend explicitly
on orbital parameters, lest conservation of energy become
impossible. 

The work of \citet{HA05} (itself inspired by
\citealt{HB95}) indicates a promising avenue. The vast 
majority of EMRIs enter the GW regime when their orbit is confined
deep inside the region of influence of the MBH. Hence one could
develop a code specialized in the dynamics of stars on quasi-Keplerian
orbits around an MBH. \citeauthor{HA05} have followed the secular change of
eccentricity and semimajor axis of individual stars assuming a fixed
given stellar background to determine the diffusion coefficients for
2-body relaxation. A powerful development of their method would be to
evolve the stellar distribution self-consistently using a treatment of
relaxation borrowed from the H\'enon Monte-Carlo approach. One would
use individual time steps to better follow the evolution of stars on
high-eccentricity orbits until their fate (inspiral, plunge or,
possibly ejection) makes no more doubt. A Monte-Carlo code could
easily cope with the $10^6-10^7$ stars within the influence radius on
a star-by-star basis.

Recently \citet{HA06} have considered, for the first time in the
study of EMRIs, the role of ``resonant relaxation'', i.e., of the
random changes in eccentricity and orientation of the orbital planes
due to the non-vanishing but fluctuating torque exerted on an orbit by
the other orbits, each considered as an elliptical mass wire
\citep{RT96}. These authors find that resonant relaxation can increase 
the EMRI rate by of order a factor 10, an exciting result which is
calling for confirmation by other computation
techniques. Unfortunately, although strictly also a 2-body effect,
resonant relaxation is unlikely to be amenable to the type of local
2-body interactions at the core of the H\'enon Monte-Carlo
method.


\acknowledgments
We thank Holger Baumgardt for discussions and results of his $N-$body
simulations. PAS is indebted to him for invaluable help with the
modification of {\NBFOUR} during his visit to the Albert Einstein
Institut in August 2005. We are grateful to Tal Alexander, John
Fregeau, Atakan G\"urkan and Frederic Rasio for enlightening
discussions. MF received kind, patient and very valuable explanations
about Kolmogorov-Smirnov tests from Laurent Eyer.

We made extensive use of the computer cluster ``Typhoon''
at Northwestern University for the Monte-Carlo simulations, thanks to
the Beowulf wizardry of John Fregeau.

The $N-$body simulations of 2-component models were performed on the
{\em GRAPE} computers at the Astronomisches Rechen Institut in Heidelberg,
thanks to the support of Rainer Spurzem, Peter Berczik and Peter
Schwekendiek.

This research was supported by NASA ATP Grant NAG~5-13229. We started
writing the manuscript of this paper during the workshop ``LISA Data:
Analysis, Source and Science'' at the Aspen Center for Physics where
we benefited from the hospitality of the Center and from discussions
with many participants in the meeting.  The participation of MF in
this workshop was supported in part by NASA grant NNG05G106G. The work
of MF in Cambridge is funded through the PPARC rolling grant to the
theory group at the IoA. The work of PAS has been supported in the
framework of the Third Level Agreement between the DFG (Deutsche
Forschungsgemeinschaft) and the IAC (Instituto de Astrof\'\i sica de
Canarias).


\label{lastpage}

\begin{deluxetable}{llllllllllllllll}
\tabletypesize{\small}
\tablecaption{Prescriptions for the Compact Remnants
\label{table:remnants}}
\tablewidth{0pt}
\tablehead{
\colhead{}             &\multicolumn{4}{c}{White Dwarfs} &\colhead{} &\multicolumn{4}{c}{Neutron Stars} &\colhead{} &\multicolumn{3}{c}{Black Holes} &\colhead{} &\colhead{All}  \\ 
\cline{2-5} \cline{7-10} \cline{12-14} \cline{16-16}\\ 
\colhead{Prescription} &\colhead{$m^{\rm MS}_{max}$} &\colhead{$m$} &\colhead{$\langle m \rangle$} &\colhead{$f$} &\colhead{} &\colhead{$m^{\rm MS}_{max}$} &\colhead{$m$} &\colhead{$\langle m \rangle$} &\colhead{$f$} &\colhead{} &\colhead{$m$} &\colhead{$\langle m \rangle$} &\colhead{$f$} &\colhead{} & \colhead{$\langle m \rangle$}
}
\startdata
F:\,Fiducial & 8.0 & 0.6 & 0.6 & 0.110 && 22.0 & 1.4 & 1.4 & $\tento{4.8}{-3}$ && 10 & 10 &  $\tento{1.6}{-3}$ && 0.328 \\
BS:\,Belczynski $Z=0.02$ & 8.0 & 0.52--1.44 & 0.66 & 0.087 && 22.0 & 1.30--2.91 & 1.55 & $\tento{4.6}{-3}$ && 2.98--11.26 & 7.36 & $\tento{1.7}{-3}$ && 0.339 \\
BP:\,Belczynski $Z=10^{-4}$ & 6.43 & 0.57--1.43 & 0.77 & 0.108 && 22.0 & 1.30--3.03 & 1.51 & $\tento{6.3}{-3}$ && 3.23--27.02 & 15.02 & $\tento{2.1}{-3}$ && 0.364 \\
\enddata
\tablecomments{
$m^{\rm MS}_{max}$ is the maximum MS mass leading to the formation of
a compact remnant (CR) of the given type, $m$ indicates the mass range
of the CRs; $\langle m \rangle$ and $f$ are the average mass and
number fraction of in the stellar population, for a Kroupa IMF
spanning $0.2-120\,\Msun$ at an age of 10 Gyr. Masses are in units of
$\Msun$. $Z$ is the metallicity. The last column indicates the 
average mass of the stellar population (including MS stars) at 10\,Gyr.
}
\end{deluxetable}

\begin{deluxetable}{llllllllll}
\tabletypesize{\scriptsize}
\tablecaption{Properties of Simulated Nuclei
\label{table:ini_cond}}
\tablewidth{0pt}
 
\tablehead{
\colhead{Name} & \multicolumn{2}{c}{Structure}       & \colhead{Stellar}            & \colhead{$\MBH(0)$}       & \colhead{$\Nstar(0)$} & \colhead{$\Npart$} & \colhead{$\Rnb(0)$} & \colhead{Physics$^{\rm (b)}$} & \colhead{Comments}\\
               & \colhead{$\eta$} &  \colhead{$\mu$} & \colhead{pop.$^{\rm (a)}$}   & \colhead{($10^6\,\Msun$)} & \colhead{($10^8$)}    & \colhead{($10^6$)} & \colhead{(pc)}      &                               &                   
}
\startdata
{\tt GN01} & 1.5 & 0.05 & F & 3.5 & 0.01 & 1 & 10.0 & TC & $f_{\delta t}=0.01$\\
{\tt GN02} & 1.5 & 0.05 & F & 3.5 & 0.01 & 1 & 10.0 & TC & \\
{\tt GN03} & 1.5 & 0.05 & F & 3.5 & 2.132 & 1 & 10.0 & TC & $f_{\delta t}=0.01$\\
{\tt GN04} & 1.5 & 0.05 & F & 3.5 & 2.132 & 1 & 10.0 & TC & \\
{\tt GN05} & 1.5 & 0.05 & F & 3.5 & 2.132 & 4 & 10.0 & TC & \\
{\tt GN06} & 1.5 & 0.05 & F & 3.5 & 2.132 & 1 & 28.0 & TC & $f_{\delta t}=0.01$\\
{\tt GN07} & 1.5 & 0.05 & F & 3.5 & 2.132 & 1 & 28.0 & TC & \\
{\tt GN08} & 1.5 & 0.05 & F & 3.5 & 2.132 & 4 & 28.0 & TC & \\
{\tt GN10} & 1.5 & 0.05 & F & 3.5 & 2.132 & 1 & 28.0 & TC,\,LA & $f_{\delta t}=0.01$, $f_{\rm la}=2$\\
{\tt GN11} & 1.5 & 0.05 & F & 3.5 & 2.132 & 1 & 28.0 & TC,\,LA & $f_{\delta t}=0.01$, $f_{\rm la}=4$\\
{\tt GN12} & 1.5 & 0.05 & F & 3.5 & 2.132 & 1 & 22.22 & TC & $f_{\delta t}=0.01$\\
{\tt GN13} & 1.8 & 0.05 & F & 3.5 & 2.132 & 4 & 22.1 & TC & $f_{\delta t}=0.01$\\
{\tt GN14} & 1.5 & 0.025 & F & 3.5 & 4.264 & 1 & 28.0 & TC & $f_{\delta t}=0.01$\\
{\tt GN15} & 1.5 & 0.05 & F* & 3.5 & 2.132 & 1 & 28.0 & TC & $f_{\delta t}=0.01$, no BHs\\
{\tt GN17} & 1.5 & 0.05 & F & 3.5 & 2.132 & 4 & 28.0 & TC & As {\tt GN08}, other random seq.\\
{\tt GN18} & 1.2 & 0.05 & F & 3.5 & 2.132 & 4 & 25.2 & TC & \\
{\tt GN19} & 2.0 & 0.05 & F & 3.5 & 2.132 & 4 & 21.0 & TC & \\
{\tt GN20} & 2.25 & 0.05 & F & 3.5 & 2.132 & 4 & 19.25 & TC & \\
{\tt GN21} & 1.5 & 0.05 & BS & 3.5 & 2.132 & 4 & 28.0 & TC & \\
{\tt GN22} & 1.5 & 0.05 &  N & 3.5 & 2.132 & 4 & 13.0 & TC & \\
{\tt GN23} & 1.5 & 0.05 & BP & 3.5 & 2.132 & 4 & 28.0 & TC & \\
{\tt GN25} & 1.5 & 0.05 & F & 3.5 & 2.132 & 8 & 28.0 & TC & \\
{\tt GN26} & 1.5 & 0.1  & F  & 7.0 & 2.132 & 4 & 17.64 & TC & Central dens.\ $2\times$ stand.\\
{\tt GN29} & 1.5 & 0.05 & F & 3.5 & 2.132 & 4 & 28.0 & TC & \\
{\tt GN30} & 1.5 & 0.1  & F & 3.5 & 1.066 & 4 & 17.64 & TC & Central dens.\ $=$ stand.\\
{\tt GN33} & 1.5 & 0.05 & BS & 3.5 & 2.132 & 4 & 28.0 & TC & \\
{\tt GN34} & 1.5 & 0.1  & F & 3.5 & 1.066 & 4 & 17.64 & TC & Central dens.\ $=$ stand.\\
{\tt GN36} & 1.3 & 0.03 & F & 2.6 & 2.640 & 6 & 35.2 & TC & Similar to \citet{Freitag03}\\
{\tt GN44} & 1.5 & 0.05 & F* & 3.5 & 3.156 & 8 & 28.0 & TC & IMF down to $0.01\,\Msun$\\
{\tt GN45} & 1.5 & 0.05 & F & 3.5 & 2.132 & 4 & 28.0 & TC,\,C & \\
{\tt GN46} & 1.5 & 0.05 & F & 3.5 & 2.132 & 4 & 28.0 & TC,\,C & Disruptive CO-MS coll.\\
{\tt GN48} & 1.5 & 0.0283 & F & 3.5 & 2.132 & 4 & 28.0 & TC,\,SE & $f_{\rm SE}=0.025$, $\mu\simeq 0.05$ at 10\,Gyr\\
{\tt GN49} & 1.5 & 0.05 & F* & 3.5 & 2.132 & 1 & 28.0 & TC & no BHs\\
{\tt GN50} & 1.5 & 0.0283 & F & 3.5 & 2.132 & 4 & 20.0 & TC,\,SE & $f_{\rm SE}=0.025$ \\
{\tt GN53} & 1.5 & 0.05 & F & 3.5 & 2.132 & 4 & 28.0 & TC,\,LA & $f_{\rm la}=4$ \\
{\tt GN54} & 1.5 & 0.05 & F & 3.5 & 2.132 & 4 & 28.0 & TC,\,LA & $f_{\rm la}=8$ \\
{\tt GN55} & 1.5 & 0.05 & F & 0.1 & 0.0609 & 4 & 4.732 & TC,\,C & \\
{\tt GN56} & 1.5 & 0.05 & F & 0.35 & 0.2132 & 4 & 8.854 & TC,\,C & \\
{\tt GN57} & 1.5 & 0.05 & F & 1.0 & 0.6090 & 4 & 14.97 & TC,\,C & \\
{\tt GN58} & 1.5 & 0.05 & F & 3.5 & 2.132 & 4 & 28.0 & TC,\,C & \\
{\tt GN59} & 1.5 & 0.05 & F & 10.0 & 6.090 & 4 & 47.33 & TC,\,C & \\
{\tt GN60} & 1.5 & 0.05 & F & 3.5 & 2.132 & 4 & 28 & TC,\,SE(NK) & $f_{\rm SE}=0.025$, SE but old population\\
{\tt GN62} & 2.0 & 0.05 & F & 0.1 & 0.0609 & 4 & 3.550 & TC,\,C & \\
{\tt GN63} & 2.0 & 0.05 & F & 0.35 & 0.2132 & 4 & 6.641 & TC,\,C & \\
{\tt GN64} & 2.0 & 0.05 & F & 1.0 & 0.6090 & 4 & 11.23 & TC,\,C & \\
{\tt GN65} & 2.0 & 0.05 & F & 3.5 & 2.132 & 4 & 21.0 & TC,\,C & \\
{\tt GN66} & 2.0 & 0.05 & F & 10.0 & 6.090 & 4 & 35.50 & TC,\,C & \\
{\tt GN67} & 1.5 & 0.0283 & Fu & 3.5 & 2.132 & 4 & 28.0 & TC,\,SE(NK) & $f_{\rm SE}=0.025$\\
{\tt GN68} & 1.5 & 0.0566 & Fu & 3.5 & 1.066 & 1 & 17.64 & TC,\,SE & \\
{\tt GN69} & 1.5 & 0.0566 & Fu & 3.5 & 1.066 & 1 & 17.64 & TC,\,SE(NK) & \\
{\tt GN70} & 1.5 & 0.05 & F & 3.5 & 2.132 & 4 & 28.0 & TC,\,C & 50\,\% MS mass accreted onto BH in MS-BH collisions\\
{\tt GN74} & 1.5 & 0.0566 & BSu & 3.5 & 1.066 & 1 & 17.64 & TC,\,SE & \\
{\tt GN75} & 1.5 & 0.0566 & BSu & 3.5 & 1.066 & 1 & 17.64 & TC,\,SE(NK) & \\
{\tt GN76} & 3.0 & $10^{-5}$ & Fu & $\sim 0$ & 2.132 & 4 & 32.4 & TC,\,SE(NK) & , 6.53\,\% mass from SE accreted by MBH\\
{\tt GN77} & 1.5 & 0.0566 & BSu & 3.5 & 1.066 & 1 & 10.0 & TC,\,SE(NK) & \\
{\tt GN78} & 3.0 & $10^{-5}$ & Fu & $\sim 0$ & 2.132 & 4 & 16.2 & TC,\,SE(NK) & , 6.53\,\% mass from SE accreted by MBH\\
{\tt GN79} & 1.5 & 0.0566 & BSu & 3.5 & 1.066 & 1 & 7.0 & TC,\,SE(NK) & \\
{\tt GN80} & 1.5 & 0.1 & BS & 3.5 & 1.066 & 1 & 17.64 & TC & \\
{\tt GN81} & 1.5 & 0.045 & BSu & 3.5 & 1.066 & 1 & 8.0 & TC,\,SE(NK) & \\
{\tt GN82} & 1.5 & 0.045 & BSu & 3.5 & 1.066 & 1 & 5.0 & TC,\,SE(NK) & \\
{\tt GN83} & 1.5 & 0.05 & F & 0.035 & 0.02133 & 2.133 & 2.8 & TC,\,C & \\
{\tt GN84} & 2.0 & 0.05 & F & 0.035 & 0.02133 & 2.133 & 2.1 & TC,\,C & \\
{\tt GN85} & 1.5 & 0.05 & F & 0.01 & 0.006094 & 0.6094 & 1.50 & TC,\,C & \\
{\tt GN86} & 2.0 & 0.05 & F & 0.01 & 0.006094 & 0.6094 & 1.12 & TC,\,C & \\
{\tt GN87} & 1.5 & 0.045 & BSu & 3.5 & 1.066 & 4 & 8.0 & TC,\,SE(NK) & \\
{\tt GN88} & 2.0 & 0.05 & F & 0.035 & 0.02133 & 2.133 & 0.5 & TC,\,C & \\
{\tt GN90} & 1.5 & 0.1 & F* & 3.5 & 1.066 & 4 & 15.0 & TC,\,C,\,SE & Initial age of stellar pop.\ 5\,Gyr\\

\enddata

\tablecomments{
$\Nstar(0)$ is the initial number of stars (in $10^8$). $\Npart$ is the
number of particles (in $10^6$, generally $\Npart\ll\Nstar$). $\Rnb$ is
the $N-body$ length scale. If not indicated otherwise, the time
step parameter is $f_{\delta t}=0.04$ and the Coulomb logarithm is
$\ln\Lambda = \ln \GCoulomb
\Nstar$ with $\GCoulomb=0.01$. 50\,\% of the mass of tidally disrupted stars is accreted onto the MBH.
\newline
$^{\rm (a)}$ F: ``Fiducial''; BS: from Belczynski, solar metallicity; BP: from Belczynski, metal-poor (see text and Table~\ref{table:remnants}). An ``u'' indicates an unevolved IMF. An ``*'' indicates a pecularity explained in the ``Comments'' column.\newline
$^{\rm (b)}$ C: collisions; LA: large-angle scatterings; SE: stellar evolution (NK: natal kicks); TC: tidal disruptions and coalescences with MBH
}

\end{deluxetable}

\begin{deluxetable}{lcrllcrll}
\tabletypesize{\scriptsize}
\tablecaption{Stellar Mass Enclosed by $R=1$\,pc and $R=3$\,pc at 5 and 10 Gyr
\label{table:Mencl}}
\tablewidth{0pt}
 
\tablehead{
\colhead{Name} && 
\colhead{Time} & \colhead{$M_{\rm encl}(1\,{\rm pc})$} & \colhead{$M_{\rm encl}(3\,{\rm pc})$} &&
\colhead{Time} & \colhead{$M_{\rm encl}(1\,{\rm pc})$} & \colhead{$M_{\rm encl}(3\,{\rm pc})$} \\
&&
\colhead{(Gyr)} & \colhead{($10^6\,\Msun$)} & \colhead{($10^6\,\Msun$)} &&
\colhead{(Gyr)} & \colhead{($10^6\,\Msun$)} & \colhead{($10^6\,\Msun$)}
}
\startdata
{\tt GN04} && 4.97 & 2.82 & 13.00 && 10.00 & 2.04 & 11.00 \\
{\tt GN06} && 5.03 & 0.71 & 4.40 && 10.02 & 0.31 & 3.58 \\
{\tt GN07} && 5.16 & 0.86 & 4.61 && 10.03 & 0.54 & 4.04 \\
{\tt GN08} && 5.01 & 0.83 & 4.55 && 9.98 & 0.45 & 3.85 \\
{\tt GN10} && 4.94 & 0.83 & 4.56 && 10.04 & 0.41 & 3.76 \\
{\tt GN11} && 5.02 & 0.76 & 4.50 && 10.02 & 0.42 & 3.86 \\
{\tt GN12} && 5.01 & 0.97 & 5.59 && 9.98 & 0.46 & 4.59 \\
{\tt GN13} && 5.04 & 0.73 & 5.25 && 9.99 & 0.34 & 4.31 \\
{\tt GN14} && 5.00 & 0.85 & 8.39 && 10.05 & 0.015 & 6.93 \\
{\tt GN15} && 6.43 & 1.21 & 5.24 && 7.67 & 1.18 & 5.15 \\
{\tt GN17} && 5.09 & 0.98 & 4.51 && 9.78 & 0.77 & 3.89 \\
{\tt GN18} && 5.04 & 1.25 & 5.45 && 10.03 & 0.95 & 4.61 \\
{\tt GN19} && 5.04 & 1.12 & 5.80 && 10.04 & 0.87 & 4.91 \\
{\tt GN20} && 4.95 & 1.16 & 6.28 && 10.08 & 0.92 & 5.33 \\
{\tt GN21} && 4.91 & 1.04 & 4.65 && 10.01 & 0.82 & 4.06 \\
{\tt GN22} && 4.88 & 2.70 & 11.00 && 9.94 & 2.11 & 9.38 \\
{\tt GN25} && 5.05 & 0.97 & 4.49 && 10.04 & 0.75 & 3.82 \\
{\tt GN26} && 4.90 & 2.05 & 8.14 && 9.97 & 1.64 & 7.14 \\
{\tt GN29} && 5.03 & 0.99 & 4.51 && 9.89 & 0.80 & 3.91 \\
{\tt GN30} && 5.09 & 0.93 & 3.86 && 10.14 & 0.71 & 3.26 \\
{\tt GN33} && 4.93 & 1.05 & 4.67 && 10.21 & 0.83 & 4.07 \\
{\tt GN34} && 4.96 & 0.95 & 3.89 && 9.96 & 0.74 & 3.32 \\
{\tt GN36} && 4.85 & 0.87 & 4.30 && 9.99 & 0.67 & 3.57 \\
{\tt GN44} && 5.01 & 0.98 & 4.54 && 10.03 & 0.76 & 3.89 \\
{\tt GN45} && 5.29 & 1.04 & 4.62 && 10.03 & 0.82 & 4.06 \\
{\tt GN46} && 5.11 & 1.05 & 4.65 && 9.69 & 0.85 & 4.11 \\
{\tt GN48} && 4.94 & 0.72 & 2.98 && 10.11 & 0.54 & 2.43 \\
{\tt GN49} && 6.43 & 1.19 & 5.00 && 7.67 & 1.16 & 4.89 \\
{\tt GN50} && 5.04 & 0.92 & 4.14 && 10.11 & 0.71 & 3.37 \\
{\tt GN53} && 4.87 & 1.04 & 4.67 && 10.12 & 0.81 & 3.97 \\
{\tt GN54} && 5.07 & 0.98 & 4.54 && 9.90 & 0.76 & 3.87 \\
{\tt GN55} && 4.98 & 0.065 & 0.38 && 10.01 & 0.041 & 0.25 \\
{\tt GN56} && 5.00 & 0.19 & 1.07 && 10.07 & 0.14 & 0.78 \\
{\tt GN57} && 5.00 & 0.44 & 2.24 && 10.04 & 0.33 & 1.78 \\
{\tt GN58} && 5.00 & 1.05 & 4.58 && 9.98 & 0.83 & 3.95 \\
{\tt GN59} && 5.03 & 1.64 & 7.30 && 10.58 & 1.56 & 6.88 \\
{\tt GN60} && 4.98 & 1.06 & 4.67 && 9.82 & 0.93 & 4.18 \\
{\tt GN62} && 5.01 & 0.079 & 0.46 && 9.98 & 0.048 & 0.29 \\
{\tt GN63} && 5.01 & 0.25 & 1.39 && 9.95 & 0.17 & 1.01 \\
{\tt GN64} && 4.98 & 0.54 & 2.95 && 10.04 & 0.40 & 2.30 \\
{\tt GN65} && 4.96 & 1.16 & 5.85 && 10.06 & 0.96 & 5.02 \\
{\tt GN66} && 5.14 & 1.30 & 8.06 && 9.91 & 1.38 & 7.88 \\
{\tt GN67} && 4.93 & 0.64 & 2.22 && 10.01 & 0.51 & 1.90 \\
{\tt GN68} && 5.06 & 0.74 & 2.81 && 9.87 & 0.57 & 2.32 \\
{\tt GN69} && 4.95 & 0.72 & 2.74 && 10.02 & 0.55 & 2.27 \\
{\tt GN70} && 5.09 & 1.06 & 4.68 && 9.71 & 0.85 & 4.10 \\
{\tt GN74} && 5.09 & 0.80 & 3.01 && 9.90 & 0.63 & 2.54 \\
{\tt GN75} && 5.02 & 0.77 & 2.94 && 10.16 & 0.59 & 2.44 \\
{\tt GN76} && 4.87 & 0.53 & 2.23 && 10.22 & 0.44 & 1.96 \\
{\tt GN77} && 5.01 & 1.20 & 4.83 && 10.09 & 0.91 & 3.96 \\
{\tt GN78} && 4.90 & 0.99 & 5.01 && 9.99 & 0.79 & 4.14 \\
{\tt GN79} && 4.93 & 1.62 & 6.51 && 10.04 & 1.17 & 5.20 \\
{\tt GN80} && 4.81 & 1.01 & 4.01 && 10.07 & 0.79 & 3.48 \\
{\tt GN81} && 4.99 & 1.25 & 5.49 && 10.00 & 0.94 & 4.41 \\
{\tt GN82} && 5.01 & 3.69 & 17.00 && 10.05 & 2.68 & 13 \\
{\tt GN84} && 5.01 & 0.025 & 0.15 && 10.02 & 0.015 & 0.087 \\
{\tt GN85} && 4.99 & 0.0055 & 0.033 && 10.02 & 0.0031 & 0.018 \\
{\tt GN86} && 5.00 & 0.0061 & 0.036 && 10.03 & 0.0032 & 0.019 \\
{\tt GN87} && 5.01 & 1.23 & 5.42 && 10.01 & 0.90 & 4.31 \\
{\tt GN88} && 5.00 & 0.0075 & 0.042 && 9.98 & 0.0037 & 0.022 \\
{\tt GN90} && 5.00 & 2.36 & 9.19 && 10.00 & 1.86 & 7.91 \\
\enddata
%

\end{deluxetable}

\end{document}